\documentclass{elsarticle}
\usepackage{hyperref}
\usepackage{amsmath}
\usepackage{listings}
\usepackage{amssymb}
\usepackage{multirow}
\usepackage{multicol}
\usepackage{latexsym}
\usepackage[linesnumbered,ruled,algosection,nofillcomment]{algorithm2e}
\usepackage{CJK}
\usepackage{epsfig}
\usepackage{xcolor}
\usepackage{graphicx}
\usepackage{algorithmic}

\allowdisplaybreaks[4]

\newcommand{\bigcircp}  {\mbox{$\rlap{\hskip 4pt -}{\bigcirc}$}}

\newcommand{\bitor}  {\mbox{$\scriptstyle{|}$}}

\newcommand{\DEF}       {\stackrel{\rm def}{=}}

\newtheorem{Lem}{Lemma}

\newtheorem{Thm}{Theorem}

\def\squareforqed{\hbox{\rlap{$\sqcap$}$\sqcup$}}
\def\qed{\ifmmode\squareforqed\else{\unskip\nobreak\hfil
\penalty50\hskip1em\null\nobreak\hfil\squareforqed
\parfillskip=0pt\finalhyphendemerits=0\endgraf}\fi}

\newcounter{statement}
\def\stmnum{\hbox to .01pt{}\rlap{\rm \hskip -\displaywidth\thestatement.}}
\def\stm{\refstepcounter{statement}\topsep 2pt \trivlist \item[]\leavevmode
\hbox to\linewidth\bgroup $ \displaystyle \hskip\leftmargini}
\def\endstm{$\hfil \displaywidth\linewidth\stmnum\egroup \endtrivlist}

\newenvironment{Proof}{
\begin{list}{}{\setlength{\topsep}{\jot}\setlength{\parsep}{\topsep}%
\addtolength{\parsep}{-0.3\parsep}\setlength{\leftmargin}{0pt}}%
\parindent 4ex
\item[]\setcounter{statement}{0}\textbf{Proof:}}{\end{list}}

\makeatletter
 \let\@copyrightspace\relax
\makeatother

\newif\ifdraft\drafttrue
\ifdraft

\newcommand{\tc}[1]{{\color{red}{[{#1}---Tian]}}}
\else

\newcommand{\tc}[1]{}
\fi
\begin{document}

\begin{frontmatter}

%\title{A Compiler MC and Its Application for MSVL}

\title{Translating Xd-C Programs to MSVL Programs}

\author{Meng Wang}
\author{Cong Tian\corref{cor1}}
\author{Nan Zhang\corref{cor1}}
\author[]{Zhenhua Duan\corref{cor1}}

\author{Chenguang Yao}
\cortext[cor1]{Corresponding authors: Cong Tian (ctian@mail.xidian.edu.cn), Nan Zhang (nanzhang@xidian.edu.cn) and Zhenhua Duan (zhhduan@mail.xidian.edu.cn). This research is supported by the National Natural Science Foundation of China under Grant Nos. 61420106004,
61732013 and 61751207.}

\address{ICTT and ISN Lab, Xidian University, Xi'an, 710071, China}

%\ead{kaiyangxidian@126.com, zhhduan@mail.xidian.edu.cn, ctian@mail.xidian.edu.cn}

%\author[mymainaddress]{Zhenhua Duan\corref{mycorrespondingauthor}}
%\ead{zhenhua\_duan@126.com}

%\author[mymainaddress]{Cong Tian}
%\ead{tico\_tools@163.com}

%\address[mymainaddress]{No.2 South Taibai Road, Xi'an, China}

\begin{abstract}
C language is one of the most popular languages for software systems. In order to verify safety, reliability and security properties of such systems written in C, a tool UMC4M for runtime verification at code level  based on Modeling, Simulation and Verification Language (MSVL) and its compiler MC is employed. To do so, a C program $P$ has to be translated to an MSVL program {\ttfamily M} and the negation of a desired property $Q$ is also translated to an MSVL program {\ttfamily M'}, then ``{\ttfamily M and M'}'' is compiled and executed armed with MC. Whether $P$ violates $Q$ is checked by evaluating whether there exists an acceptable execution of new MSVL program ``{\ttfamily M and M'}''. Therefore, how to translate a C program to an MSVL program is a critical issue. However, in general, C is of complicated structures with {\ttfamily goto} statement.
In this paper, we confine the syntax of C
 in a suitable subset called Xd-C without loss of expressiveness. Further, we present a translation algorithm from an Xd-C program to an MSVL program based on translation algorithms for expressions  and statements.
Moreover, the equivalences between expressions and statements involved in Xd-C and MSVL programs are inductively proved. Subsequently, the equivalence between the original Xd-C program and the translated MSVL program is also proved.  %  based on operational semantics of Xd-C and MSVL.
In addition, the proposed approach has been implemented by a tool called $C2M$.  A benchmark of experiments including 13 real-world Xd-C programs is conducted. The results show that $C2M$ works effectively.
%The details of implementation including the architecture design, lexical, syntactic
%and semantic analysis, as well as preprocessing and scheduling algorithms.
%A case study is given to show how the approach works.
\end{abstract}


\end{frontmatter}

%\linenumbers

\section{Introduction}
Software systems written in C language are more than 13\% \footnote{https://www.tiobe.com/tiobe-index/}, second popular one, in the world
since C can be used to implement a complex system in a flexible way.
In order to verify properties of safety, reliability and security of such systems, many researchers focus on model checking \cite{clarke1994model,clarke1999model}.
 For the purpose of employing conventional model checkers such as NuSMV \cite{cimatti2002nusmv} and SPIN \cite{holzmann1997model}, an abstract model has to be extracted from a C program, and a desired property is specified by an LTL \cite{pnueli1977temporal} or CTL \cite{Clarke1986Automatic} formula.
Then the model checkers check whether the abstract model satisfying the property is valid.
However, as software systems become larger and more complex, it is difficult to acquire a model which is consistent with the original program. %Also, any errors may cause inconsistencies between the abstract model and original program.

In recent years, verifying software systems at code level has attracted more attentions \cite{GHEORGHE201845,NATARAJAN201724,Ivancic2005Model,henzinger2002lazy,beyer2009software,dietsch2015fairness}.
Tools like, SLAM \cite{ball2001automatically}, BLAST \cite{beyer2006the}, CPAChecker \cite{beyer2011cpachecker} and CBMC \cite{kroening2014cbmc}, support only safety property verification. They insert assertions into C source code and
then the verification is carried out by checking the reachability of error labels. % 源码
In order to verify more temporal properties such as liveness,
Ultimate LTLAutomizer \cite{dietsch2015fairness} and T2 \cite{brockschmidt2016t2} extend the software model checking approach by
reducing the verification problem to fair termination checking.
 To do that, a program to be verified written in C is first translated to an intermediate form, and then a desired property can be verified based on automata-theoretic approach for verification. % 源码转换
However, all these tools suffer from the state-explosion problem. Further, since there are no execution details of programs, the verification result is not always accurate, that is,
sometimes false positives (i.e., potential errors may be reported where
there are none) or false negatives (i.e., errors are not reported) may be produced.

As a lightweight verification technique, runtime verification checks whether a run of a system to be verified satisfies a
given property by monitoring the execution of the system. It alleviates the state-explosion problem since a single execution path is checked each time. With this approach, a system to be verified is described in a conventional program while a desired property is expressed in a formal specification language. Therefore,
 extracting events from the executing system and sending them to monitors can generate a large runtime overhead.
A better solution is to implement the system and monitor within the
same logical framework.
The runtime verification tool UMC4M \cite{Wang2017Full,8531789} takes a program {\ttfamily M} written in a Modeling, Simulation and Verification Language (MSVL) \cite{duan1996extended,duan2005temporal,zhang2016mechanism,yang2018compiler} and a desired property $P$ specified by
a Propositional Projection Temporal Logic (PPTL) \cite{duan2014a,duan2008decision} formula as input, and
 converts temporal property verification as a dynamic program execution task.
 With this tool, the negation of the desired property is translated to an MSVL program {\ttfamily M'}, then whether {\ttfamily M} violates $P$ is checked by evaluating whether there exists an acceptable execution of
new MSVL program ``{\ttfamily M and M'}''. Therefore, verification of MSVL programs can be carried out in the same logical framework.

In order to verify C programs by means of UMC4M, they have to be rewritten to MSVL programs.
%However, the syntax and semantics of C programs are relatively complicated.
In this paper, we confine the syntax of C language in a suitable subset called Xd-C and propose an algorithm to automatically translate Xd-C programs to MSVL programs.
Xd-C features most of the data types and statements of C, including all arithmetic types, arrays, pointers and struct
types, and all statements except {\ttfamily goto} statement, while
MSVL data types and statements include all Xd-C data types and statements.
 Therefore, an Xd-C program can be translated to an equivalent MSVL program in an automatic way.
In fact, we can treat that all variables in the MSVL program are framed, and the translation is in one-to-one manner. %The operational semantics of each statement in C fragments and the related
%statement in MSVL is equivalent.
The time complexity of the translation is linear ($O(n)$), where $n$ is the number of statements in an Xd-C program.
To prove the equivalence between the original Xd-C and translated MSVL programs, we present the operational semantics of Xd-C and MSVL.
The operational semantics of Xd-C is similar to that of another subset of C
language proposed by Blazy et al.,  called Clight \cite{blazy2009mechanized}.
It is presented as a big-step operational semantics and characterizes both termination and divergence behaviors. Whereas in MSVL,
the evaluation rules for left-value and right-value arithmetic expressions, and Boolean expressions are borrowed from \cite{yang2008operational}.
The semantic equivalence rules regarding a program, the
transition rules within a state and interval transition rules
are also formalized \cite{zhang2016mechanism,yang2008operational,wang2017msvl}.
Further, based on operational semantics of Xd-C and MSVL, we prove equivalences between expressions and statements by means of structural induction and rule induction, respectively.

The contributions of this paper are three-fold:
\begin{itemize}
  \item [(1)] We present algorithms to translate declarations, expressions and statements from Xd-C to MSVL.
   An example $bzip2$ \cite{SPEC}, a compression program to compress and decompress
input files, is used to show how the algorithms work.

  \item[(2)] The equivalence between an original Xd-C program and the translated MSVL program is proved based on operational semantics of Xd-C and MSVL.
  \item [(3)] We have implemented a translator $C2M$ and  conducted a benchmark of experiments including 13 real-world Xd-C programs. %integrated it in $MSV$ \cite{zhang2016model} so that verifying C programs can be supported.
\end{itemize}

The remainder of this paper is organized as follows. In Section \ref{Xd-C}, Xd-C is
briefly introduced. Further, MSVL is introduced in Section \ref{MSVL}.
Section \ref{trans} presents an algorithm for translating Xd-C programs to MSVL programs. Subsequently, in Section \ref{proof}, the equivalences between expressions and statements in Xd-C and MSVL are proved, respectively.  Moreover, the equivalence between Xd-C and translated MSVL programs is also proved. An implementation of the proposed approach is presented and the evaluation is conducted in Section \ref{imp}.
Section \ref{conc} concludes the paper.

\section{The Restricted C Fragment: Xd-C} \label{Xd-C}
 The restricted C fragment called Xd-C is confined in a subset of ANSI-C (C89 standard version). It consists of often-used types, expressions and statements of C language. Xd-C is similar to Clight \cite{blazy2009mechanized} but more than Clight.

\subsection{Types}
The supported types in Xd-C include arithmetic types (char, int, float and double in various sizes and signedness),
pointer, void pointer, function pointer and struct types. However, union type, static local variables and type qualifiers such as const, restrict and volatile are not
allowed in Xd-C. As storage-class specifiers, typedef definitions have been expanded away during parsing and type-checking. The syntax of Xd-C types is given as follows:\\
{\small
$$\begin{array}{lrl}
\mbox{Signedness}& sign&::= signed\mid unsigned\\
\mbox{int~ length}&   len&::=short\mid long\\
\mbox{Types}& \tau&::=int \mid sign~ int \mid len ~int \mid sign~ len ~int \\
&&~~\mid float\mid double \mid long~ double \mid char \mid sign ~char \\
&&~~ \mid structself \mid voidp \mid functp \mid \tau*
%\mid \tau ~funct(\tau^*)
%\mbox{Field ~lists}:& body&::=(\tau~ id;)^+
\end{array}
$$
}
Self-defined Types:
{\small
$$\begin{array}{ll}
structself&::= struct ~id_1 \{(\tau~ id_2;)^+\} | struct ~id_1\\
functp&::= [\tau|void] ((\tau,)^*\tau)* \mid [\tau|void] ()* \\
voidp&::= void*
%\mid \tau ~funct(\tau^*)
%\mbox{Field ~lists}:& body&::=(\tau~ id;)^+
\end{array}
$$
}\\
where
$struct ~id_1 \{(\tau~ id_2;)^+\}$ defines a structure $id_1$ consisting of body $(\tau~ id_2;)^+$; $[\tau|void] ((\tau,)^*\tau)* $ defines a function pointer with each parameter of type $\tau$ and a return value of type $\tau$ or $void$. $[\tau|void] ()*$ defines a function pointer with no parameter. Note that $id$ (possibly with subscriptions) is a string (name) consisting of characters and digits with a character as its head.

%\subsection{Data structures}
% of the function arguments $\tau^*$ and the return type $\tau$.

%Functions types specify the number and types of the function arguments ($\tau^*$) and the type of the function result ($\tau$). Xd-C struct type carries a local identifier $id$ and the list $body$ of their fields (names and types).

\subsection{Expressions}
The expression $e$ in Xd-C is inductively defined as follows:
\[
\begin{array}{rl}
    e ::= & c \mid le \mid \& le %\mide\mathit{++} \mid e\mathit{--} \mid \mathit{++}e\mid \mathit{--}e
    \mid (\tau) ~e \mid x(e_{1},...,e_{k}) \mid op_1 ~e
            \mid e_1 ~op_2 ~e_2  \mid e_1?e_2:e_3\\% \mid (e)\\ %\mid sizeof(\tau)
    x::= &id \mid id[e] \mid id[e_1][e_2] \mid  le.x \mid le\rightarrow x\\
    le::=&x|*e\\
    op_1::=&+ \mid - \mid \mbox{\~{}} \mid~ ! ~~~~~~~~op_2::=~ aop \mid bop \mid rop \mid eop \mid lop \\
    aop::=& + \mid - \mid * \mid / \mid \% ~~~~~~~~~~~~ bop::=~  << \mid >> \mid \& \mid \bitor \mid \hat{}\\
    rop::=&  < \mid > \mid <= \mid >=  ~~~~~~~~~~~~~~~~eop::=~ == \mid != \\
    lop::=& \&\& \mid \bitor\bitor
\end{array}
\]\noindent
where $c$ is an arbitrary constant, $id$ a variable, $id[e]$ the $e{th}$ element of array $id$ (counting from 0), $id[e_1][e_2]$ the element in row $e_1$ and column $e_2$, $le.x$ member $x$ of structural variable $le$ and $le\rightarrow x$ member $x$ of the structural variable that $le$ points to. $\& le$ takes the address of $le$ and $* e$ is the pointer dereferencing.
$x(e_{1},...,e_{k})$ is a function call with arguments $e_{1},...,e_{k}$ and it does not change the memory state.
%For an assignment $x=e\mathit{++}$, the value of $x$ is the current value of $e$, while the value of $e$ is changed to $e+1$. However, a single statement $e\mathit{++}$ means the value of $e$ is $e+1$. The same explanation can be given to $e\mathit{--}$.
%The result of $x=e++$ (resp. $e--$) is the value of $e$  and after the result is
%obtained, the value of $e$ is incremented (resp. decremented),
%Whereas $\mathit{++}e$ (resp. $\mathit{--}e$) represents
%the current value of $e$ plus (resp. minus) one.
%$\tau$ is a type name and
$(\tau) ~e$ represents the type cast of $e$ namely converting the value of $e$ to the value in type $\tau$. The type of $e$ before the type cast is a non-pointer type.
$op_1~ e$ is a unary expression including $+e$, $-e$, \~{}$e$ and $!e$. $op_2$ represents an binary operator including arithmetic operators $aop$ ($+$, $-$, $*$, $/$ and $\%$), bitwise operators $bop$ ($<<$, $>>$, $\&$, $\bitor$ and $\hat{}$~), relational operators $rop$ ($<$, $>$, $<=$ and  $>=$), equality operators $eop$ ($==$ and $!=$) and logical operators $lop$ ($\&\&$ and $\bitor\bitor$).
  Both $e_1$ and $e_2$ in $e_1~rop~e_2$ are of non-pointer types.
 $e_1?e_2:e_3$ is a conditional expression indicating that the result is $e_2$ if $e_1$ is not equal to 0, and $e_3$ otherwise.
%However it must satisfy that the function does not change any memory units or any external variables whose scopes are not limited to the function.

%The expression $le$ that can occur in left-value position can inductively defined as follows:
% $$le::=x|*e$$

%Moreover, assignments in Xd-C are presented as statements and cannot appear within expressions.
%The operational semantics of expressions is given in Appendix A. $\langle e, \sigma \rangle$ denotes the value of expression $e$ at state $\sigma$. More details can be found in \cite{blazy2009mechanized}.

\subsection{Statements}
The following are the elementary statements in Xd-C:
\[
\begin{array}{llll}
\mbox{Statements:}& cs::= ; &\mbox{null}\\
%& |e;& \mbox{expression} \\
& |le\mathit{++};& \mbox{post increment} \\
& |le\mathit{--};& \mbox{post decrement} \\
& |le = e; & \mbox{assignment} \\
   % & | le~ aop= ~e;~ &\mbox{compound assignment} \\
  %  &| le ~bop=~ e; \\
 %   &|\mbox{\ttfamily $s_1;s_2$} &\mbox{sequence}\\
    &|\mbox{\ttfamily if($e$)\{$cs_1$\}else\{$cs_2$\}}&\mbox{conditional} \\
    &|\mbox{\ttfamily switch($e$)\{$sw$\}}& \mbox{\ttfamily switch}\\
     &|\mbox{\ttfamily while($e$)\{$cs$\}}& \mbox{{\ttfamily while} loop} \\
     &|\mbox{\ttfamily do\{$cs$\}while($e$)$;$}&\mbox{{\ttfamily do} loop}\\
     &|\mbox{\ttfamily for($cs_1; e; cs_2$)\{$cs$\}}&\mbox{{\ttfamily for} loop}\\
  %  &|\mbox{\ttfamily f($e_{1},\ldots, e_{n}$)}&\mbox{function call}\\
    &|\mbox{\ttfamily continue}; &\mbox{next iteration of the current loop}\\
     &|\mbox{\ttfamily break}; &\mbox{exit from the current loop}\\
      &|\mbox{\ttfamily return $e$}; &\mbox{return from the current function}\\
      &|\mbox{\ttfamily return}; \\
      &|cs_1;cs_2 & \mbox{sequence} \\
      &|x(e_1,...,e_k);& \mbox{function call}\\
      %&|funct\{s\} &\mbox{function definition}\\
%&|\mbox{\ttfamily case $n:$ $s;sw$}\\
\mbox{Switch cases:} & sw::= sw_1| sw_2;sw_1 \\
&sw_1::= \mbox{\ttfamily default}: cs; &\mbox{default case}\\
%& \mbox{\ttfamily default}: s; &\mbox{default case}\\
&sw_2::= \mbox{\ttfamily case $n:$ $cs;sw_2$} &\mbox{labeled case}\\
%&|\mbox{\ttfamily case $n:$ $s;sw$} &\mbox{labeled case}
\end{array}
\]
%\[
% ie::=\left\{
% \begin{array}{ll}
% %\{(ie;)^*\}& \mbox{if the type of $id$ is structure;}\\
% \{(e)^*\}& \mbox{if the type of $id$ is structure;}\\
% \{(\{(ie,)^*ie\},)^*\{(ie,)^*ie\}\}& \mbox{if the type of $id$ is two dimensional array;}\\
% e &\mbox{otherwise}
% \end{array}\right .
%\]
A null statement performs no operations.
%An expression statement ($e;$) is evaluated as a void expression for its side effects.
A post increment statement $le\mathit{++}$ means that the value of $le$ is changed to $le+1$ while a post decrement statement $le\mathit{--}$ indicates that the value of $le$ is changed to $le-1$.
 In an assignment statement ``$le =e;$'', the value of $e$ %is converted to the type of the assignment expression and
replaces the value stored in the location designated by $le$.
%As previously mentioned, assignment $e_1 = e_2;$ of an r-value $e_2$ to an l-value $e_1$ is treated as a statement.
%Continuous assignments such as $e_1=e_2=e_3;$ are not supported.
%A compound assignment of the form $le~ aop= e;$ (or $le ~bop=e;$) can be expressed by the simple assignment
%$le= le~aop~e;$ (or $le=le~bop~ e;$).
%Sequence statement \mbox{\ttfamily $s_1;s_2$} means that $s_1$ and $s_2$ executes sequentially.
In a conditional statement \mbox{\ttfamily if($e$)\{$cs_1$\}else\{$cs_2$\}}, $cs_1$ is executed if expression $e$ compares unequal to 0, and $cs_2$ is executed otherwise. In a {\ttfamily switch} statement \mbox{\ttfamily switch($e$)$\{sw\}$}, $e$ is the controlling expression, and the expression of each case label shall be an integer constant expression.
% and no two of the case constant expressions in the same switch statement have the same value. There may be at most one default label in a switch statement.
There are three kinds of iteration statements in Xd-C including {\ttfamily while} loop, {\ttfamily do} loop and {\ttfamily for} loop statements. An iteration statement causes the body of the loop to repeatedly execute until controlling expression $e$ equals 0.
%The loop condition must exclude the incrementation and decrementation expressions such as $e\mathit{++}$, $\mathit{++}e$, $e\mathit{--}$ and $\mathit{--}e$. % 不支持增量与减量算符
In a {\ttfamily while} loop statement {\ttfamily while($e$)\{$cs$\}}, the evaluation of $e$ takes place before each execution of $cs$  while in a {\ttfamily do} loop statement ``{\ttfamily do\{$cs$\}while($e$)$;$}'', the evaluation of $e$ takes place after each execution of $cs$. In a {\ttfamily for}  loop statement {\ttfamily for($cs_1; e; cs_2$)\{$cs$\}}, $cs_1$ executes once at the beginning of the first iteration, $e$ is the condition of the loop, $cs_2$ executes at the end of each iteration, and $cs$ is the body of the loop.
Jump statements including ``{\ttfamily continue$;$}'', ``{\ttfamily break$;$}'', ``{\ttfamily return $e;$}'' and ``{\ttfamily return$;$}'' are supported in Xd-C, but not the {\ttfamily goto} statement. A {\ttfamily continue} statement shall appear only in the body of a loop. A \mbox{\ttfamily break} statement terminates execution of the smallest enclosing \mbox{\ttfamily switch} or iteration statement. A \mbox{\ttfamily return} statement appears only in the body of a function.
%Note that, the return type of a function shall be void or an object type other than array type.
%The definition of a function that accepts a variable number of arguments is not supported in Xd-C.
%A Xd-C program is composed of a list of declarations of global variables, a list of functions and an identifier naming the entry point of the program (the main function).

An Xd-C program is composed of a list of declarations, a list of functions and a main function.
It can be defined as follows:
{\small\[\begin{array}{lrll}
\mbox{Array}& array&::= &\tau~ id[n] \mid \tau ~id[m][n] \mid \tau ~id[n]=\{(e,)^*e\}\\
&&&\mid \tau ~id[]=\{(e,)^*e\} \mid   \tau ~id[m][n]=\{(e,)^*e\}\\
&&&\mid \tau ~id[m][n]=\{(\{(e,)^*e\},)^*\{(e,)^*e\}\}\\
\mbox{Structure}&   structure&::=& struct ~id_1 \{(\tau~ id_2;)^+\}\\ %| struct ~id_1\\
\mbox{Variable list} & varlist&::=& id %~\mbox{variable}\\
\mid id=e       %\mbox{initialize $id$, except for $structself$}\\
\mid varlist,varlist %\mbox{variable list}\\
\\
\mbox{Declaration}& Pd&::=& \tau ~varlist \mid array \mid structure\\
\mbox{Parameter}& par&::=& ~ \mid (\tau~ id_2,)^*(\tau~ id_2)\\
\mbox{Function}& funct&::=&[\tau|void] ~id_1(par)\{(Pd;)^*cs\}\\
&&&\mid extern~ [\tau|void] ~id_1(par)\\
\mbox{Program}& P&::=&(Pd;)^* (funct;)^* \\
&&&int~ main(int~ argc, char \mathit{**argv})\{(Pd;)^*cs\}
\end{array}\]
}\noindent
where $\tau~ id[n]$ defines a one dimensional array $id$ having $n$ elements of type $\tau$ while $\tau ~id[m][n]$ defines a
two dimensional array $id$ having  $m\times n$ elements of type $\tau$; $id=e$ defines an initialization of $id$ except for $structself$;
$[\tau|void] ~id_1((\tau~\linebreak id_2,)^*(\tau~ id_2))\{(Pd;)^*cs\}$ defines a function $id_1$ with each parameter $id_2$ of type $\tau$ and a return value of type $\tau$ or $void$; $[\tau|void] ~id_1()\{(Pd;)^*cs\}$ defines a function $id_1$ with no parameter.

\textbf{Summary: }
As we can see, some constructs and facilities in ANSI-C (C89) are not supported in Xd-C.
In the following, we show a key negative list which Xd-C does not support.
 \begin{itemize} \setlength{\itemsep}{0pt}
   \item [(1)] {\ttfamily goto} statement;
   \item [(2)] $union$ structure;
   \item [(3)] %$e\mathit{++}$, $e\mathit{--}$,
   $\mathit{++}e$ and $\mathit{--}e$ expressions;
   \item [(4)] ($a=b,b=c,d=(f(x),0)$) comma statements;
   \item [(5)] $\mathit{op=}$ compound assignments where $op::=+\mid -\mid *\mid /\mid \% \mid >>\mid <<\mid \&\mid  {\scriptstyle{|}} \mid \hat{}$ ;
   \item [(6)] $struct~ A~a$; $a=\{(void()*)b, (void()*)c\}$ structure assignments;
   \item [(7)] $x=y=z$ continuous assignments;
   \item [(8)] $typedef$, $extern$, $static$, $auto$ and $register$ storage-class specifiers;
   \item [(9)] $const$ and $volatile$ type qualifiers;
   \item [(10)] local variables in a block;
   \item [(11)] nested cases in a {\ttfamily switch}  statement;
   \item [(12)] assignment expressions such as {\ttfamily if$((y=fun())==x)$};
   \item [(13)] function pointers pointing to external functions;
   \item [(14)] functions that accept a variable number of arguments.
 \end{itemize}

In fact, the constructs and facilities in the above negative list except for {\ttfamily goto} statement can be implemented by Xd-C although the implementation might be tedious. Therefore, Xd-C is a reasonable subset of ANSI-C (C89) in practice.

%The definition of a function that accepts a variable number of arguments is not supported in Xd-C. %Note that $id$ (possibly with subscriptions) is a string (name) consisting of characters and digits with a character as its head.

\subsection{Operational semantics of Xd-C}
The operational semantics of expressions and statements in Xd-C is borrowed from \cite{blazy2009mechanized} and given in Appendix A.
The formal semantics of a large subset of C language called Clight is presented in \cite{blazy2009mechanized}.
Clight features most of the types and operators of C, including all arithmetic types, pointer, {\ttfamily struct} and {\ttfamily union} types, as well as all C control structures except {\ttfamily goto} statement, while Xd-C supports all types, expressions and statements in Clight except {\ttfamily union} type.%For convenience of proof, we rewrite the operational semantics in another form.

\begin{figure}[htb!]
  \centering
  \includegraphics[width=4.8in]{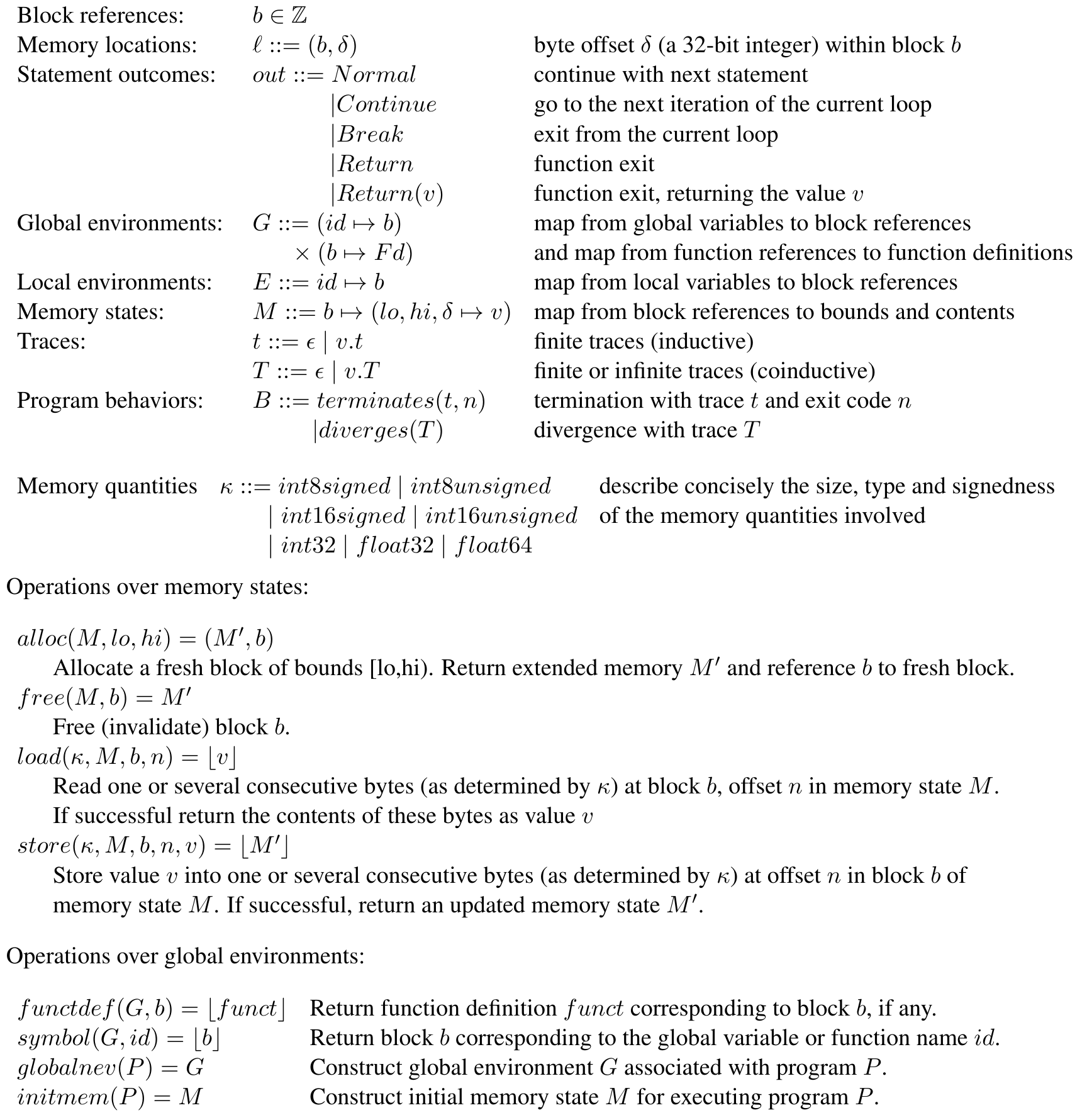}\\
  \caption{Semantic elements: block references, memory locations, statement outcomes, environments, memory states, etc}\label{semele}
\end{figure}

The semantic elements including block references, memory locations, statement outcomes, evaluation environments, memory states, traces, program behaviors and operations over memory states and global environments are defined in Figure \ref{semele}.
Memory location $\ell$ is a pair of a memory block reference $b$ and a byte offset $\delta$ within this block.
Statement outcome $out$ indicates how an execution terminates: either normally by running to completion or prematurely via a jump statement. Global environment $G$ maps program-global variables and function names to memory block references, as well as those references corresponding to function pointers to the definitions of functions. Local environment $E$ maps function scoped variables to their memory block reference. Memory state $M$ maps memory block references to bounds and contents. Each memory block has lower and upper bounds $lo$, $hi$.
$B$ describes the program behavior. % whether terminates with trace $t$ and exit code $n$ or diverges with trace $T$.
The basic operations over memory
states ($alloc$, $free$, $load$ and $store$) and  global environments ($\mathit{functdef}$, $\mathit{symbol}$, $\mathit{globalnev}$ and $\mathit{initmem}$) are also summarized. Note that,
for functions returning ``option'' types, $\lfloor x \rfloor$ corresponds to success with return value $x$, and $\emptyset$ to failure.

The semantics is defined by the following judgements:
{\small
\[
\begin{array}{ll}
G,E \vdash le, M \stackrel{l}{\Rightarrow} \ell &\mbox{(evaluation of left-value expressions)}\\
G,E \vdash e, M \Rightarrow v &\mbox{(evaluation of right-value expressions)}\\
G,E \vdash cs, M\stackrel{t}{\Rightarrow}  out, M' & \mbox{(evaluation of statements, terminating case)}\\
G \vdash funct(v_{args}), M \stackrel{t}{\Rightarrow} v, M'& \mbox{(evaluation of function invocations, terminating case)}\\
G,E \vdash cs, M\stackrel{T}{\Rightarrow}  \infty & \mbox{(evaluation of statements, diverging case)}\\
G \vdash funct(v_{args}), M \stackrel{T}{\Rightarrow} \infty & \mbox{(evaluation of function invocations, diverging case)}\\
\vdash P \Rightarrow B& \mbox{(execution of whole programs)}\\
\end{array}
\]
}

Each judgement relates a syntactic element and an initial memory state $M$ to the result of executing this syntactic element. For a left-value expression $le$, the result is a location $\ell$ while for a right-value expression $e$, the result is its value $v$. For the execution of a statement $cs$ in the terminating case, the result is a statement outcome $out$ and an updated memory state $M'$ while in the diverging case, the result is $\infty$.  For a function call $funct(v_{args})$ in the terminating case, the result is its value $v$ and a memory state $M'$ while in the diverging case, the result is $\infty$, where $v_{args}$ is a list of values of arguments.
For the  execution of a program $P$, it terminates with trace $t$ and exit code $n$ or diverges with trace $T$.

%For an expression in l-value position, the result is a location : a pair of a
%block identifier b and a byte offset δ within this block. For an expression in r-value
%position and for a function application, the result is a value v: the discriminated union
%of 32-bit integers, 64-bit floating-point numbers, locations (representing the value
%of pointers), and the special value undef representing the contents of uninitialized
%memory. Clight does not support assignment between struct or union, nor passing
%a struct or union by value to a function; therefore, struct and union values
%need not be represented.

%where  $G$ is a global environment and it maps global variables to block references as well as function references to function definitions; $E$ is a local environment and maps local variables to block references; $M$ is an initial memory state and maps block references to bounds and contents; $\ell$ is a location defined by a pair $(b, \delta)$,
%where $b$ is a block identifier and $\delta$ an offset within $b$;  $M'$ is an updated memory state and $B$ a program behavior (termination or divergence); $out$ indicates how an execution terminated: either normally by running to completion or prematurely via a jump statement.

\section{MSVL} \label{MSVL}
 MSVL is a Modeling, Simulation and Verification Language \cite{duan1996extended,duan2005temporal,zhang2016mechanism} which can be used to both model and execute a system. It is a subset of Projection Temporal logic (PTL). %programming language supporting not only conventional programming but also parallel programming.
 There are some statements in MSVL that cannot expressed by Xd-C, thus a suitable subset of MSVL can support Xd-C well. This section briefly introduces the suitable subset of MSVL  which is borrowed from \cite{duan1996extended,zhang2016mechanism,yang2008operational,wang2017msvl,duan2008framed,duan2004framed}.

\subsection{Syntax}
The left-value expression $la$, right-value expression $ra$ and Boolean expression $b$ in the subset of MSVL are inductively defined as follows:
{\small
\begin{align}
    \notag  v&::= id \mid id[ra] \mid id[ra_1][ra_2] \mid la.v \mid pt\rightarrow v\\
    \notag  h&::=v~~~~~~~~~~~g::=v\\
   \notag  pt&::= la \mid \&la\mid (\tau)~ra\mid ra_1+ra_2\mid ra_1-ra_1 \mid ext~g(ra_{1},\ldots,ra_{k}) \mid \\
   \notag  &~~~~~~ext~h(ra_{1},\ldots,ra_{k}, RVal)\\
    \notag  la&::= v\mid *pt \\%~~~~~~~~~~~~~~~~~~~~~~~~ pt::= la \mid \&la\mid (\tau)~ra\mid ra_1+ra_2\mid ra_1-ra_1\\
   % \notag la&::=lan\mid \bigcirc la\\
   \notag ra &::= c  \mid la \mid \&la \mid (\tau) ~ra \mid mop_1~ ra \mid ra_1 ~mop_2 ~ra_2  \mid if(b)then~ ra_1~else~ra_2 \mid \bigcircp ra \mid \\
 \notag    &~~~~~ext~g(ra_{1},\ldots,ra_{k})\mid ext~h(ra_{1},\ldots,ra_{k},RVal)\\
\notag  mop_1&::=+ \mid - \mid \mbox{\~{}}\\
\notag mop_2&::=+\mid -\mid*\mid / \mid \%\mid << \mid >> \mid \& \mid \bitor \mid \hat{}~\\
\notag    b &::= true \mid false \mid ra_{1}~mrop~ra_{2} %\mid ra_{1}<ra_{2}
%\mid a_{1}>a_{2} \mid ra_{1}<=ra_{2} \mid ra_{1}>=ra_{2} \\ \mid ra_{1}!=ra_{2}
  \mid \neg b \mid b_{1}~ \wedge ~b_{2}\mid b_{1} ~\vee ~b_{2}\\
 \notag mrop&::=<\mid >\mid<=\mid >=\mid =\mid !=
\end{align}
}\noindent
%{
%\begin{align}
%    e ::= & c \mid id \mid id[e] \mid id[e_1][e_2] \mid g(e_{1},...,e_{m}) \mid e.id \mid e\rightarrow id \mid e\mathit{++} \mid e\mathit{--} \mid \\
%           &\mathit{++}e\mid \mathit{--}e \mid (\tau) ~e \mid op_1 ~e \mid e_1 ~op_2 ~e_2  \mid e_1?e_2:e_3 \\ %\mid sizeof(\tau)
%    op_1::=& \& \mid * \mid + \mid - \mid \mbox{\~{}} \mid~ ! ~~~~~~~~op_2::=~ aop \mid bop \mid rop \mid eop \mid lop \\
%    aop::=& + \mid - \mid * \mid / \mid \% ~~~~~~~~~~~~ bop::=~  << \mid >> \mid \& \mid | \mid \hat{}\\
%    rop::=&  < \mid > \mid <= \mid >=  ~~~~~~~~~~~~~~~~eop::=~ == \mid != \\
%    lop::=& \&\& \mid ||
%\end{align}
%}\noindent
where $v$ denotes a variable and $pt$ an expression of a pointer type.  % $c$ is a constant, $m$ and $n$ integers, $x$ a variable,
The explanations of $c$, $la$, $\&la$  and $(\tau) ~ra$ are the same as explanations of $c$, $le$, $\&le$ and $(\tau) ~e$ in Xd-C, respectively.
$g(ra_{1},\ldots,ra_{k})$ is a function call of an external function while $ ext~h(ra_{1},\ldots, ra_{k},RVal)$ is an external function call of an MSVL user-defined function, where $RVal$ is the return value. %A state function contains no temporal operators.
%Note that a unary operation
%($op_1~a$, $op_1::= \& \mid * \mid + \mid - \mid \mbox{\~{}}$) can be treated as a state function $g(a)$; a binary operation ($a_1~op_2~a_2$, $op_2::=+\mid -\mid*\mid / \mid \%\mid << \mid >> \mid \& \mid \bitor \mid \hat{}~$) can be treated as a state function $g(a_1,a_2)$.
An external call of function $ext~h(ra_{1},\ldots,ra_{k},RVal)$ means that we concern only the return value of the function but not the interval over which the function is executed.
%$\bigcirc la$ stands for the value of $la$ at the next state and
$\bigcircp ra$  stands for the value of $ra$ at the previous state. %For the
We assume that all variables used are framed.
The following are the elementary statements in the subset of MSVL:
 \[
\begin{array}{llll}
 ms::=& \mbox{\ttfamily empty} &\mbox{Termination}\\
      &|\mbox{\ttfamily skip}& \mbox{Skip}\\
      &|\mbox{\ttfamily $la\Leftarrow ra$}&\mbox{Assignment}\\
      &|\mbox{\ttfamily $la:=ra$}&\mbox{Unit Assignment}\\
     &|\mbox{\ttfamily $ms_1$~and~$ms_2$}& \mbox{Conjunction}\\
     &|\mbox{\ttfamily next $ms$}& \mbox{Next}\\
     &|\mbox{\ttfamily $ms_1;ms_2$} &\mbox{Sequence}\\
    &|\mbox{\ttfamily if($b$)then\{$ms_1$\}else\{$ms_2$\}}&\mbox{Conditional}\\
    &|\mbox{\ttfamily while($b$)\{$ms$\}}&\mbox{While}\\
   % &|\mbox{\ttfamily return $a$}&\mbox{Return}\\
  %  &|\mbox{\ttfamily return}&\mbox{Return}\\
     &|\mbox{\ttfamily $h(ra_{1},\ldots,ra_{k})$}&\mbox{Function call}\\
    &|\mbox{\ttfamily ext $h(ra_{1},\ldots,ra_{k})$}&\mbox{External function call}
\end{array}
\]\noindent
%where $a$, {\ttfamily $a_1$} and $a_2$ are arithmetic expressions and {\ttfamily $b$} is a boolean expression. %and {\ttfamily $ms_1$} and {\ttfamily $ms_2$} are programs of the subset of MSVL.
The termination statement ``{\ttfamily empty}'' means that the current state is the final state of an interval. ``{\ttfamily skip}'' specifies one unit of time over an interval.
The assignment ``{\ttfamily $la\mathit{\Leftarrow}ra$}'' indicates that {\ttfamily $la$} is assigned the value of {\ttfamily $ra$} at the current state
while ``{\ttfamily $la:=ra$}'' means that the value of {\ttfamily $la$} at the next state equals the current value of {\ttfamily $ra$} and
the length of the interval is one unit of time.
The conjunction statement ``{\ttfamily $ms_1$ and $ms_2$}'' indicates that {\ttfamily $ms_1$} and {\ttfamily $ms_2$} are executed concurrently. ``{\ttfamily next $ms$}'' means that $ms$ will be executed at the next state.
``{\ttfamily $ms_1;ms_2$}'' means that {\ttfamily $ms_1$} is executed until
its termination from this time point then {\ttfamily $ms_2$} is executed or {\ttfamily $ms_1$} is infinitely executed.
``{\ttfamily $h(ra_{1},\ldots,ra_{k})$}'' and ``{\ttfamily ext $h(ra_{1},\ldots,ra_{k})$}'' are internal and external function calls, respectively.
The meanings of other statements are the same as Xd-C. Note that all the above statements are defined by PTL formulas in Appendix B.

In addition, data type $\tau$ in MSVL \cite{wang2017msvl} is defined  %including integer, float, char, pointer, array and struct
%have been defined and they are the types
the same as in Xd-C.
An MSVL program $P$ can be defined as follows:
{\small\[\begin{array}{lrll}
\mbox{Array}& marray&::= &\tau~ id[n] \mid \tau ~id[m][n] \mid \tau ~id[n]\Leftarrow\{(ra,)^*ra\}\\
&&&\mid \tau ~id[m][n]\Leftarrow\{(ra,)^*ra\}\\
&&&\mid \tau ~id[m][n]\Leftarrow\{(\{(ra,)^*ra\},)^*\{(ra,)^*ra\}\} \\
\mbox{Structure}&   mstructure&::=& struct ~id_1 \{(\tau~ id_2~ and)^*~\tau~ id_2\}\\ %| struct ~id_1\\
\mbox{Variable list} & mvarlist&::=& id %~\mbox{variable}\\
\mid id\Leftarrow ra       %\mbox{initialize $id$, except for $structself$}\\
\mid varlist,varlist %\mbox{variable list}\\
\\
\mbox{Declaration}& mPd&::=& \tau ~mvarlist \mid marray \mid mstructure \\
\mbox{Function body}& \phi&::=&(mPd;)^*ms\\
\mbox{Function}& mfunct&::=&function ~id_1((\tau~ id_2,)^* \tau~ RVal)\{\phi\}\\
&&&\mid function ~id_1((\tau~ id_2,)^* \tau~ id_2)\{\phi\}\\
&&&\mid function ~id_1()\{\phi\}\\
\mbox{Program}& P&::=&(mPd;)^* (mfunct;)^* ms
\end{array}\]
}\noindent
where in a structure definition $struct ~id_1 \{(\tau~ id_2~ and)^*~\tau~ id_2\}$, $and$ is used to connect each member $id_2$ of struct $id_1$; in a function fragment $function ~id_1((\tau~ id_2,)^* \tau\linebreak~ RVal)\{\phi\}$, $RVal$ is the return value;
$function ~id_1((\tau~ id_2,)^* \tau~ id_2)\{\phi\}$ and $function ~id_1()\{\phi\}$ define functions without a return value.

\subsection{Operational semantics}
The operational semantics of expressions and statements in the subset of MSVL is borrowed from \cite{zhang2016mechanism,yang2008operational,wang2017msvl}
\subsubsection{Notation}
%Operators such as $\wedge$ and $\vee$ are not sensitive to the order of the components in a program.
In order to directly express programs $p_1$, ..., $p_k$ connected by $\wedge$ and $\vee$, the following notations are defined.
{\small $$(1)\wedge\{p_1,...,p_k\}\DEF p_1 \wedge...\wedge p_k (k\geq 1)~~~(2) \vee\{p_1,...,p_k\}\DEF p_1\vee...\vee p_k (k\geq 1)$$}
Let $\mathcal{V}$ denote a set of variables and $D$ the set of all data in type $\tau$ needed by us.
A state $s$ is defined as a pair $(s^{l}, s^r)$, where $s^{l}$ is a mapping $\mathcal{V}\longrightarrow \mathbb{Z}\times N_0$ and $s^{r}$ a mapping $\mathcal{V}\longrightarrow D$. $Dom(s)$ is defined as $Dom(s)=Dom(s^l)=Dom(s^r)=\mathcal{V}$. The $i$th state $s_i = (s_i^l, s_i^r)$. $s_i^l(x)$ denotes the location of variable $x$ in type $\tau$ at state $s_i$ and $s_i^r(x)$ the value of $x$ in type $\tau$ at $s_i$. Note that
for an array $a[n]$ in type $\tau$, if $s_i^l(a)=(b,0)$, $s_i^l(a[j])=(b,j*sizeof(\tau))$ for $0\leq j<n$ and
for an array $a[m][n]$ in type $\tau$, if $s_i^l(a)=(b,0)$, $s_i^l(a[j][k])=(b,(j*n+k)*sizeof(\tau))$ for $0\leq j<m$ and $0\leq k<n$.  Let $\sigma= \langle s_0,...\rangle $ denote an interval.  Over an interval $\sigma$, if a variable $x$ is not released, $s_i^l(x)=s_{i+1}^l(x)$, for $i\geq 0$; otherwise, $x$ is removed from the state.

 We have two types of configurations, one for expressions, and the other for programs. A configuration of a program
$p$ is a quadruple $(p, \sigma_{i-1}, s_i, i)$, where $\sigma_{i-1}= \langle s_0,..., s_{i-1}\rangle (i > 0)$ records information of
all states, $s_i$ is the current state and $i$ counts the number of states in $\sigma_{i-1}$. Further, let the initial configuration be  $c_0=(p,\epsilon, s_0,0)$.
When a program $p$ is terminating, it is reduced to true and the state is written as $\emptyset$. So the final configuration is $c_f=(true, \sigma, \emptyset, |\sigma|+1)$ if it can terminate, otherwise, there will always be a transition from the current state to the next one.
% and $c_f=(p', \sigma_{i-1}, s_i, \omega)$, where $p'$ is a program, otherwise.
%Actually, after program $p$ is transformed from current state $s_i$ to next state $s_{i+1}$, $s_{i+1}$ becomes the new current state and $s_i$ is appended to $\sigma_{i_1}$ as a reduced state.
%So $\sigma_{i-1}$ is extended to $\sigma_i$. Therefore, the length of $\sigma$  is continuously increased with reduced states until the last state is appended to $\sigma$. In this way, the whole model of program $p$ can be obtained eventually.
Let $\rightarrowtail$ denote the congruence relation between configurations.
$c \stackrel{*}{\rightarrowtail} c'$ implies that $c$ is transformed to $c'$ by zero or several steps within a state.
Notation $\rightarrow$ is a binary relation  between two configurations with different states.  $c\stackrel{*}{\rightarrow}c'$ implies that $c$ is transformed to $c'$ after zero or several states and $c\stackrel{+}{\rightarrow}c'$ implies that $c$ is transformed to $c'$ after at least one state.

Similarly, for an arithmetic (or Boolean) expression $a$, the configuration is $(a,  \sigma_{i-1}, s_i, i)$. The evaluation of a left-value is of the form $(la, \sigma_{i-1}, s_i,i) \stackrel{l}{\Rightarrow} (b,\delta)$, which means that the location of $la$ is $(b,\delta)$ at state $s_i$. %,  or $(pt, \sigma_{i-1}, s_i, i) \stackrel{l}{\Rightarrow}  \&x$ which means the value of $pt$ is the address of a variable $x$ at state $s_i$.
The evaluation of a right-value is of the form $(ra, \sigma_{i-1}, s_i,i) \Downarrow n$, which means that the value of $ra$ is $n$ at state $s_i$.

For accessing the locations and values of variables at a state, the following notations are used.
Let $n$ be a value in $D$, $x$ and $y$ variables.
 %We use ${p}_x$ to assert whether an assignment has taken place to $x$ in the execution of a program. $\dot{p}_x$ denotes assignment flag $p_x$ or $\neg p_x$.
 $(s_i^l, s_i^r[n/x])$ means that the location of $x$ is not changed and the value of $x$ is changed to $n$, and other variables are not changed at $s_i$. Thus, we have,
$$(s_i^l, s_i^r[n/x])(y)=\left\{\begin{array}{ll}(s_i^l(y),n)& y=x \\ (s_i^l(y),s_i^r(y)) &y\neq x\end{array}\right .$$

%Projection function $\Pi_i$ is defined to obtain the components of the pair. For instance, $\Pi_1(s_i(x))=\Pi_1(m, \dot{p}_x)=m$ gets the first component from the pair.

%Further, for convenience, we use a simple notation $s_i[w]$ to mean that all variables and propositions appearing in state program $w$ at current state $s_i$ are instantiated with their values. As an example, $s_i[(m_1, \dot{p}_{x_1})/x_1][(m_2, \dot{p}_{x_2})/x_2]...[(m_n, \dot{p}_{x_n})/x_n]$ is abbreviated to $s_i[w]$ if $w\equiv (x_1=m_1 \wedge \dot{p}_{x_1})\wedge(x_2=m_2 \wedge \dot{p}_{x_2})...\wedge (x_n=m_n \wedge \dot{p}_{x_n})$.

 %$\Pi_1(s_i(x))$ is the value of  variable $x$ at $s_i$.

\subsubsection{Evaluation of expressions}
%The evaluation relation, $\Downarrow$, is concerned with how the overall results of expressions are obtained.
Rules in Tables \ref{semle} and \ref{semre} are evaluation rules of arithmetic expressions for left-values and right-values, respectively. $sizeof(\tau)$ returns the storage size of type $\tau$ and
$type(a)$ the type of expression $a$. $\mathit{field\_offset}(v, \varphi)$ returns the byte offset of the
field named $v$ in a struct whose field list is $\varphi$ or $\emptyset$ if $v$ does not appear in $\varphi$.
$ptr(b,\delta)$ denotes a pointer value pointing to $(b,\delta)$.
%The evaluation rules of arithmetic expressions are given in Table \ref{semae}.
 Rule L1 deals with variables, L2 and L3 elements of arrays, L4 and L5 members of structure variables and L6 pointer dereferencing.
 %We use $s_i[x]=(b,\delta)$ to denote the location of variable $x$, where $b$ is a memory block reference and $\delta$ a byte offset
%within this block.
 Rule R1 tackles with constants. R2 deals with expressions which can also appear in left-value position. R3 handles expressions with the address-of operator, R4 type cast operators, R5 - R8 arithmetic operators and R9 the previous ($\bigcircp$) operator.

  %expressions with next ($\bigcirc$) and previous ($\bigcircp$) operators. %Variables can refer to their previous values obtained from interval $\sigma$.
% The condition, $1 \leq n-m\leq i$, ensures that the evaluation of variables must
%be within the range of the interval. %Actually, rule A2 can be treated as a special case of rule A3 with $m=n$.
%Notice that the configuration $(\bigcirc ^m x, \sigma_{i-1}, s_i, i)(m \geq 1)$ is not permitted since we cannot evaluate expressions involving only future operators
%in the current state. Rules A5 - A8 handle the evaluation of expressions with arithmetic operators.
 %Rule A3 handles the evaluation of expressions with function call.
  Rules B1 - B6 in Table \ref{sembe} tackle with Boolean expressions. %They can easily be understood as those in conventional programming languages.

  \begin{table}[htb!]
  \small
\centering
\caption{Evaluation rules of arithmetic expressions for left-values}
\label{semle}
\begin{tabular}{|c|l|} \hline
L1& $(id, \sigma_{i-1}, s_i,i) \stackrel{l}{\Rightarrow} s_i^l(id)$\\[1.2ex]
L2& $\displaystyle\frac{(ra, \sigma_{i-1}, s_i, i) \Downarrow n~~~(id, \sigma_{i-1}, s_i,i) \stackrel{l}{\Rightarrow} (b,0)}{(id[ra], \sigma_{i-1},s_i,i)\stackrel{l}{\Rightarrow} (b,n*sizeof(\tau))}$, where $\tau$ is the type of $id[ra]$ \\[3ex]
L3& $\displaystyle\frac{(ra_1, \sigma_{i-1}, s_i, i) \Downarrow n_1~~~~~(ra_2, \sigma_{i-1}, s_i, i) \Downarrow n_2~~~~~(id, \sigma_{i-1}, s_i, i) \stackrel{l}{\Rightarrow} (b,0)}{(id[ra_1][ra_2], \sigma_{i-1},s_i,i)\stackrel{l}{\Rightarrow} (b,(n_1*n+n_2)*sizeof(\tau))}$,\\
 &where $n$ is the number of elements in each row of $id[ra_1][ra_2]$ and \\
 & $\tau$ is the type of $id[ra_1][ra_2]$ \\ %*(*(id + e1) + e2)
L4& $\displaystyle\frac{
\begin{array}{c}
(la, \sigma_{i-1}, s_i, i) \stackrel{l}{\Rightarrow} (b,\delta)~~~~~type(la)=struct ~id'\{\varphi\}\\
\mathit{field\_offset}(v,\varphi)=\lfloor \delta'\rfloor
\end{array}}{(la.v, \sigma_{i-1},s_i,i)\stackrel{l}{\Rightarrow} (b,\delta+\delta')}$ \\[3ex]
L5& $\displaystyle\frac{\begin{array}{c}
(pt, \sigma_{i-1}, s_i, i) \Downarrow ptr(b,\delta)~~~~~type(pt)=struct ~id'\{\varphi\}*\\
\mathit{field\_offset}(v,\varphi)=\lfloor \delta'\rfloor
\end{array}}{(pt\rightarrow v, \sigma_{i-1},s_i,i)\stackrel{l}{\Rightarrow} (b,\delta+\delta')}$ \\[2ex]
L6&$\displaystyle\frac{(pt, \sigma_{i-1}, s_i, i) \Downarrow ptr(b,\delta)}{(*pt, \sigma_{i-1},s_i,i)\stackrel{l}{\Rightarrow} (b,\delta)}$\\[2ex]
%A3& $(\bigcirc lan, \sigma_{i-1}, s_i, i) \stackrel{l}{\Rightarrow} s_{i-(n-m)}(x)$, if $1\leq n-m\leq i$\\
\hline
\end{tabular}
\end{table}

\begin{table}[htb!]
\small
\centering
\caption{Evaluation rules of arithmetic expressions for right-values}
\label{semre}
\begin{tabular}{|c|l|} \hline
R1& $(c, \sigma_{i-1}, s_i,i) \Downarrow c$ \\
%R2& $\displaystyle\frac{(la, \sigma_{i-1}, s_i, i) \stackrel{l}{\Rightarrow} x}{(la, \sigma_{i-1},s_i,i)\Downarrow s_i(x)}$\\
R2& $\displaystyle\frac{(la, \sigma_{i-1}, s_i, i) \stackrel{l}{\Rightarrow} s_i^l(x)}{(la, \sigma_{i-1},s_i,i)\Downarrow s_i^r(x)}$\\
R3& $\displaystyle\frac{(la, \sigma_{i-1}, s_i, i) \stackrel{l}{\Rightarrow} (b,\delta)}{(\&la, \sigma_{i-1},s_i,i)\Downarrow ptr(b,\delta)}$\\
R4& $\displaystyle\frac{(ra, \sigma_{i-1}, s_i, i) \Downarrow n_1}{((\tau) ~ra, \sigma_{i-1},s_i,i)\Downarrow n}$, where $n=(\tau) ~n_1$\\[2ex]
R5& $\displaystyle\frac{(ra_1, \sigma_{i-1}, s_i, i) \Downarrow n_1}{(mop_1~ra_1, \sigma_{i-1},s_i,i)\Downarrow n}$, where $n=mop_1~n_1$\\[2ex]
R6& $\displaystyle\frac{(ra_1, \sigma_{i-1}, s_i, i) \Downarrow n_1~~~~(ra_2, \sigma_{i-1}, s_i, i) \Downarrow n_2}{(ra_1~mop_2~ra_2, \sigma_{i-1},s_i,i)\Downarrow n}$, where $n=n_1~mop_2~n_2$\\[2ex]
R7& $\displaystyle\frac{(b, \sigma_{i-1}, s_i, i) \Downarrow true~~~~ (ra_1, \sigma_{i-1}, s_i, i) \Downarrow n_1}{(if(b)then~ra_1~else~ra_2, \sigma_{i-1},s_i,i)\Downarrow n_1}$\\[2ex]
R8& $\displaystyle\frac{(b, \sigma_{i-1}, s_i, i) \Downarrow false~~~~ (ra_2, \sigma_{i-1}, s_i, i) \Downarrow n_2}{(if(b)then~ra_1~else~ra_2, \sigma_{i-1},s_i,i)\Downarrow n_2}$\\[2ex]
R9& %$(\bigcircp v, \sigma_{i-1}, s_i, i) \Downarrow s_{i-1}(x)$\\
 $\displaystyle\frac{(ra, \sigma_{i-m-1}, s_{i-m}, i-m) {\Downarrow} n}{(\bigcircp^m ra, \sigma_{i-1},s_i,i)\Downarrow n}$, where $m\leq i$\\
%A9& $\displaystyle\frac{(e_1, \sigma_{i-1}, s_i, i) \Downarrow n_1, ..., (e_m, \sigma_{i-1}, s_i, i) \Downarrow n_m}{(g(e_1,...,e_m), \sigma_{i-1},s_i,i)\Downarrow n}$, where $n=g(n_1,...,n_m)$\\[2ex]
\hline
\end{tabular}
\end{table}

{\small
\begin{table}[htb!]
\small
\centering
\caption{Evaluation rules of Boolean expressions}
\label{sembe}
\begin{tabular}{|c|l|} \hline
B1& $(true, \sigma_{i-1}, s_i,i) \Downarrow true$ \\
B2& $(false, \sigma_{i-1}, s_i, i) \Downarrow false$ \\
B3& $\displaystyle\frac{(ra_1, \sigma_{i-1}, s_i, i) \Downarrow n_1~~~~(ra_2, \sigma_{i-1}, s_i, i) \Downarrow n_2}{(ra_1~mrop~ra_2, \sigma_{i-1},s_i,i)\Downarrow t}$~~~$t=\left\{\begin{array}{ll}true& \mbox{if}~n_1~mrop~n_2\\ false &\mbox{otherwise}\end{array}\right .$\\
%B4& $\displaystyle\frac{(e_1, \sigma_{i-1}, s_i, i) \Downarrow n_1, (e_2, \sigma_{i-1}, s_i, i) \Downarrow n_2}{(e_1<e_2, \sigma_{i-1},s_i,i)\Downarrow t}$~~~$t=\left\{\begin{array}{ll}true& \mbox{if}~n_0<n_1\\ false &\mbox{otherwise}\end{array}\right .$\\
B4& $\displaystyle\frac{(b, \sigma_{i-1}, s_i, i) \Downarrow true}{(\neg b, \sigma_{i-1},s_i,i)\Downarrow false}$~~~ $\displaystyle\frac{(b, \sigma_{i-1}, s_i, i) \Downarrow false}{(\neg b, \sigma_{i-1},s_i,i)\Downarrow true}$\\
B5& $\displaystyle\frac{(b_1, \sigma_{i-1}, s_i, i) \Downarrow t_1~~~~ (b_2, \sigma_{i-1}, s_i, i) \Downarrow t_2}{(b_1~\wedge~ b_2, \sigma_{i-1},s_i,i)\Downarrow t}$, $t=\left\{\begin{array}{ll}true& \mbox{if}~b_1=true \\ &\mbox{ and } b_2=true\\ false &\mbox{otherwise}\end{array}\right .$\\
B6& $\displaystyle\frac{(b_1, \sigma_{i-1}, s_i, i) \Downarrow t_1~~~~(b_2, \sigma_{i-1}, s_i, i) \Downarrow t_2}{(b_1~\vee~ b_2, \sigma_{i-1},s_i,i)\Downarrow t}$, $t=\left\{\begin{array}{ll}true& \mbox{if}~b_1=true \\ &\mbox{ or } b_2=true\\ false &\mbox{otherwise}\end{array}\right .$\\
\hline
\end{tabular}
\end{table}
}

\subsubsection{ State reduction}
The semantic equivalence rules regarding programs are formalized in Table \ref{sems}.
Rule SKIP is concerned with statement {\ttfamily skip} and $\bigcirc ${\ttfamily empty} specifies one unit of time over an interval.
UASS handles unit assignment statement $la:=ra$. $la$ is assigned by $ra$ at the next state and it takes one unit of time.
The conjunction statement {\ttfamily $ms_1$ and $ms_2$} can be expressed by $\wedge \{ms_1, ms_2\}$. We define {\ttfamily more} as
{\ttfamily more}$\DEF \bigcirc true$.
$\bigcirc ms$ implies that $ms$ will be executed at the next state and {\ttfamily more} means that the current interval is not yet over.
$\Box ms$ is handled by rule ALW depending on {\ttfamily more} or {\ttfamily empty} encountered in programs.
In order to keep consistence of operational semantics with the Xd-C sequential statement $r;ms_2$ in which $r$ could execute over an infinite interval, in this paper, we use weak chop `$;$' instead of strong chop `$;_s$' as most of time in MSVL programs.
 As a matter of fact, the two operators can be defined by each other. Formally, $r; ms_2\DEF (r;_s ms_2) \vee (r\wedge \Box \mbox{\ttfamily more})$ and $r;_s ms_2\DEF (r\wedge \lozenge \varepsilon; ms_2)$.
Thus, rule CHOP deals with $(r ; ms_2)$ in light of the structure of program $r$ in four forms. In
the case of $(r\equiv \wedge\{w, ms_1\})$, and $w$ being a state program or true, $(r ; ms_2)$ is reduced to $(\wedge \{w, ms_1 ; ms_2\})$; in the case of $(r\equiv \bigcirc ms_1)$, $(r ; ms_2)$
is transformed to $(\bigcirc(ms_1 ; ms_2))$; in the case of $(r \equiv \mbox{\ttfamily empty})$, $(r ; ms_2)$ is reduced to $ms_2$; and in case of $(r\equiv \Box \mbox{\ttfamily more})$, $(r ; ms_2)$ is reduced to $\Box \mbox{\ttfamily more}$.
Rule IF transforms the conditional statement to its equivalent program according to the definition. Rule WHL transforms the while statement to an equivalent conditional statement.

\begin{table}[htb!]
\small
\centering
\caption{Semantic equivalence rules of framed programs}
\label{sems}
\begin{tabular}{|c|l|} \hline
%Ass& $(Ass \Leftarrow) \wedge \{lbf(x),x\Leftarrow e\} \equiv x\Leftarrow e$ \\
SKIP& {\ttfamily skip} $\equiv \bigcirc ${\ttfamily empty} \\
UASS& If $(la,\sigma_{i-1},s_i,i)\stackrel{l}{\Rightarrow}s_i^l(x)$ and $(ra,\sigma_{i-1},s_i,i)\Downarrow n$  then\\
&$la:=ra \equiv \bigcirc (x\Leftarrow n \wedge ${\ttfamily empty}$)$\\
AND& {\ttfamily $ms_1$ and $ms_2$} $\equiv \wedge\{ ms_1, ms_2\}$\\
NEXT& {\ttfamily next} $ms \equiv \wedge \{\bigcirc ms, \Box \mbox{\ttfamily more}\}$ \\
ALW& (1){\ttfamily $\wedge \{\Box ms,$ empty$\}\equiv\wedge\{ms,$ empty$\}$} \\
&(2)$\wedge\{\Box ms, \mbox{\ttfamily ~more}\}\equiv \wedge\{ms, \bigcirc\Box ms\}$\\
CHOP& (1) $\wedge\{w,ms_1\};ms_2 \equiv \wedge\{w,ms_1;ms_2\}$\\
& (2) $ \bigcirc ms_1;ms_2 \equiv \bigcirc(ms_1;ms_2)$\\
 &(3) {\ttfamily empty}$;ms_2 \equiv ms_2$\\
 &(4) $\Box \mbox{\ttfamily more};ms_2\equiv \Box  \mbox{\ttfamily more}$\\
IF& {\ttfamily if$(b)$then$\{ms_1\}$else$\{ms_2\}$} $\equiv (b\wedge ms_1)\vee (\neg b \wedge ms_2)$\\
WHL& {\ttfamily while$(b)\{ms\}\equiv$ if$(b)$then$\{ms\wedge  \mbox{\ttfamily more};$while$(b)\{ms\}\}$else\{empty\}}\\
\hline
\end{tabular}
\end{table}
Semantic equivalence rules regarding true and false are listed in Table \ref{tf}. We use $p$ to represent an MSVL program.

\begin{table}[htb!]
\small
\centering
\caption{Semantic equivalence rules of truth values}
\label{tf}
\begin{tabular}{|c|l|c|l|c|l|} \hline
F1& $\wedge\{false,p\}\equiv false$ & F2 & $\vee\{p, false\}\equiv p$& F3 & $\wedge\{p, \neg p\}\equiv false$\\
T1& $\wedge\{p,true\}\equiv p$ & T2 & $\vee\{p, true\}\equiv true$& T3 & $\vee\{p, \neg p\}\equiv true$\\
\hline
\end{tabular}
\end{table}

The following rules are concerned with assignments.
\begin{itemize}
 \item [MIN1]If $\exists ~j$, $1\leq j\leq n$, $(la_j,\sigma_{i-1},s_i,i)\stackrel{l}{\Rightarrow} s_i^l(x_j)$ and $(ra_j,\sigma_{i-1},s_i,i)\Downarrow n_j$,
then\\
$(\wedge\{p,\wedge_{k=1}^{n} \{ la_k\Leftarrow ra_k\}\},\sigma_{i-1}, s_i, i)\rightarrowtail$\\
$(\wedge \{p, \wedge_{k=1,k\neq j}^{n} \{ la_k[n_j/x_j]\Leftarrow ra_k[n_j/x_j]\}\},\sigma_{i-1}, (s_i^l,s_i^r[n_j/x_j]), i )$.\\
If $la_j$ can be evaluated to the location of a variable $x_j$ and $ra_j$ a constant $n_j$, then conjunct $la_j\Leftarrow ra_j$ is eliminated from
the program in the configuration, where $x_j$ is set to $n_j$ at state $s_i$.
%It is really that $x_j\Leftarrow e_j$ is moved from the program to state $s_i$ in the configuration, where
$la_k[n_j/x_j]$ ($ra_k[n_j/x_j]$) means that variable $x_j$ is replaced by value $n_j$ in  $la_k$ ($ra_k$) for $1\leq k\leq n$ and $k\neq j$.
  \item [MIN2] If $(\bigcircp x,\sigma_{i-1},s_i,i)\Downarrow n (i\geq 1)$ and there is no state component $la\Leftarrow ra$ in $p$, where $(la,\sigma_{i-1},s_i,i)\stackrel{l}{\Rightarrow} s_i^l(x)$,
then\\
$(p,\sigma_{i-1}, s_i, i)\rightarrowtail$ $(p,\sigma_{i-1}, (s_i^l,s_i^r[n/x]), i )$.\\
If there is no assignment to variable $x$ at the current state,  $x$ keeps its previous value.
\end{itemize}
%
%\begin{table}[htb!]
%\centering
%\caption{Transition rules within a state}
%\label{trs}
%\begin{tabular}{|c|l|} \hline
%SUB-& $\wedge\{false,p\}\equiv false$\\
%T1& $\wedge\{p,true\}\equiv p$ & T2 & $\vee\{p, true\}\equiv true$& T3 & $\vee\{p, \neg p\}\equiv true$\\
%\hline
%\end{tabular}
%\end{table}

%\textbf{Interval reduction.}
Actually, once all of the variables involved in the current state have been set, the remained
subprogram is of the forms, $\bigcirc ms$ or {\ttfamily empty}. Rule TR1 in Table \ref{tr} deals with the former and rule TR2 the latter. Concretely, the transition of $(\bigcirc ms, \sigma_{i-1}, s_i, i)$ means that $ms$ will be executed at
next state $s_{i+1}$, and current state $s_i$ needs to be appended to $\sigma_{i-1}$. So $i$, the
number of states in $\sigma_{i-1}$, need plus one. The transition of $(\mbox{\ttfamily empty}, \sigma_{i-1}, s_i, i)$ means that $s_i$ is appended to $\sigma_{i-1}$ and the final configuration $(true, \sigma_{i-1} \cdot \langle s_i\rangle ,\emptyset, i+1)$
is reached. %Thus, the reduction process terminates.

\begin{table}[htb!]
\small
\centering
\caption{Interval transition rules}
\label{tr}
\begin{tabular}{|c|l|} \hline
TR1& $(\bigcirc ms, \sigma_{i-1}, s_i, i) \rightarrow (ms, \sigma_i, s_{i+1}, i+1)$ \\
TR2& $(\mbox{\ttfamily empty}, \sigma_{i-1}, s_i, i) \rightarrow (true, \sigma_i, \emptyset, i+1)$\\
\hline
\end{tabular}
\end{table}

\subsubsection{Type declaration statement}
The set of basic data types $\mathcal{T}_b$ is defined as follows:
$$
\begin{array}{ll}
\mathcal{T}_b\DEF &\{int, float, char, int\langle\rangle, float\langle\rangle, char\langle\rangle, int[n_0], int[n_0][n_1], \\
 &float[n_0],float[n_0][n_1], char[n_0], char[n_0][n_1]\}.
\end{array}$$
where $n_0,n_1\in N_0$.
Note that when an array is declared, the number of elements of the array needs to be specified.
For each basic data type $T\in \mathcal{T}_b$, a point type $T*$ is introduced and the set of the pointer types are $\mathcal{T}_{pb}=\{T*\mid T\in \mathcal{T}_b\}$. Let $\mathcal{T}_s$ denote a countable set of all possible names of struct types.
For each struct type $S\in\mathcal{T}_s$, $S*$ denotes the pointer type to $S$ and the set of pointer types to struct types is $\mathcal{T}_{ps}=\{S*\mid S\in \mathcal{T}_s\}$.
Let $\mathcal{T}^d\DEF \mathcal{T}_b \cup \mathcal{T}_{pb} \cup \mathcal{T}_s \cup \mathcal{T}_{ps}$ be the union of basic data types, struct types and their corresponding pointer types.
We define predicates $IS_T(\cdot)$, which means ``is of type $T$'', for each type $T\in \mathcal{T}^d$.
$$IS_T:T\longrightarrow B \mbox{ for } T\in \mathcal{T}^d$$
For every type $T\in \mathcal{T}^d$, the formula $IS_T(v)$ represents $v$ is a variable of type $T$.

%\begin{itemize}
%  \item [(1)] For each primitive type $T\in \{int, float, char\}$, $Is_T:T\rightarrow B$ with $(c,T)\mapsto true$,
%  \item [(2)] for each list type $T\langle\rangle\in \{int\langle\rangle, float\langle\rangle, char\langle\rangle\}$, $Is_{T\langle\rangle}: T\langle\rangle \rightarrow B$ with $(c,T\langle\rangle)\mapsto true$, and
%  \item [(3)] for each array type $T[n]\in \{int[1], int[2],...,float[1],...,char[1],...\}$, $Is_{T[n]}: T[]\rightarrow B$ with $(c,T[])\mapsto true$ iff $|c|=n$.
%  \item [(4)] for each pointer type $T\in \mathcal{T}_pb$, $Is_T:T\rightarrow B$ with $(v,T)\mapsto true$,
%\end{itemize}

Using these predicates, we define the type declaration statements as a derived PTL formula.
$$T ~x\DEF \Box IS_T(x), ~~~~~~~\mbox{for } T\in \mathcal{T}_b\cup \mathcal{T}_{pb}$$

Suppose a struct $S$
is defined with types of its members being $S_1*,..., S_n*\in \mathcal{T}_{ps} (n\geq 0)$.
Such a struct definition $S$ is called legal if each struct $S_i(1\leq i\leq n)$ is defined no later than the definition of $S$ and illegal otherwise.
We call a program with illegal struct definitions unhealthy.
To deal with unhealthy programs, a special proposition $\mu$ meaning ``unhealthy'' is introduced.
Moreover, in order to make sure that the values of a struct type should be consistent with the struct definition in terms of members and their types, for each struct $S$, a system variable $mem_S$ is used. %$mem_S$
We use notation $FPF(E_1,E_2)$ to denote the set of all finite partial functions from a set $E_1$ to another set $E_2$, i.e., all finite subsets of $E_1\times E_2$ that are partial functions. Formally,
{\small$$FPF(E_1,E_2)\DEF \{E\mid E\subset E_1\times E_2, E \mbox{ is finite and }(u,v), (u,v')\in E \mbox{ implies }v=v'\}.$$}\noindent
Variable $mem_S$ takes a special type $MEMTYPE \DEF FPF(\mathcal{V}, \mathcal{T}^d\backslash \mathcal{T}_s)$.
The interpretation of $mem_S$ is of the form $\{(a_1,T_1),...,(a_k, T_k)\}(k\geq 1)$. We define the struct definition as follows:
$$\begin{array}{l}
struct ~S\{T_1~a_1 ~and~... ~and ~T_k~ a_k\}\DEF \\
\Box(r_S\wedge mem_S=\{(a_1,T_1),...,(a_k,T_k)\})\wedge (r_{S_1}\wedge ... \wedge r_{S_n} \vee \mu)
\end{array}$$
where $\{S_1*,..., S_n*\}=\{T_1,..., T_k\}\cap \mathcal{T}_{ps}$ is the set of struct pointer types in $T_1,...,T_k$ and
$r_S$ (resp. $r_{S_i}$) denotes whether $S$ (resp. $S_i$) is defined or not.
%For each struct name $S$, we introduce a proposition to denote whether $S$ is defined or not.

We call a variable $x$ is consistent with struct $S$ in terms of members, if it has exactly the members and their types defined by $S$. To check the consistency of $x$ with $S$ in terms of members, we define a predicate $Con$ as follows:
$$\begin{array}{ll}
Con:& \bigcup\limits_{S\in \mathcal{T}_s} (S\times MEMTYPE\longrightarrow B)\\
&(v,S), \{(a_1,T_1),...,(a_k,T_k)\}\mapsto true, \mbox{if $v$ is of the form }\\
&\{(a_1,(v_1,T_1)),...,(a_k,(v_k,T_k))\};\\
&(v,S), \{(a_1,T_1),...,(a_k,T_k)\}\mapsto false, \mbox{otherwise.}
\end{array}
$$
$Con(x, mem_S)$ indicates $x$ is consistent with $S$ in terms of members.
To define variable declarations, we first extend the definition of unhealthy programs.
A variable declaration $S ~x$ or $S*x$ is called legal if a struct definition $struct ~S\{...\}$ is no later than the variable declaration and illegal otherwise. A program with illegal struct definitions or variable declarations is called unhealthy. Proposition $\mu$ is still used to mean ``unhealthy''. Then
the struct variable and struct pointer variable declarations can be specified as follows:
%We finalize the specifications of.
$$\begin{array}{rl}
S~x\DEF& r_S\wedge \Box(IS_S(x)\wedge Con(x,mem_S))\vee \neg r_S\wedge \mu,\\
S*~x\DEF& r_S \wedge \Box IS_{S*}(x)\vee \neg r_S\wedge \mu.
\end{array}
$$

\subsubsection{Function}\label{function}
Two kinds of functions can be used in MSVL: external functions, written in other programming
languages such as C, and user-defined functions written in MSVL.
There are also two kinds of function calls: external call and internal call. For an external call, the interval over which the callee function is executed is ignored, while for an internal call,
 the interval over which the callee function is executed is inserted and concatenated with
the main interval over which the caller function is executed. External functions can only be invoked as an external call while user-defined MSVL functions can be invoked as  either an external call or internal call.
As in C language, function calls can appear in expressions and statements in MSVL programs.
The evaluation rules of function calls in expressions and semantic equivalence rules of function calls in statements are given in the following parts.\\
\textbf{Evaluation rules of function calls in expressions}
All function calls appearing in expressions are external function calls. Such function could be an MSVL user-defined   or   external function. However it must satisfy that the function does not change any memory units or any external variables whose scopes are not limited to the function.
%For each expression $a$, the evaluation of $a$ related to interpretation $I=(\sigma, i,k,j)$, denoted by $I(a)$, is redefined based on the semantics of $PTL$.
Let $\sigma'=\langle s'_0,  . . . , s'_{|\sigma'|} \rangle$ be an interval over which the function is executed and $s'_0=s_i$.
\begin{itemize}
  \item [(1)] $f$ is an MSVL user-defined function %with parameters and a return value and
  defined as follows:
  $$function~f(\tau_1~ v_1, ..., \tau_k~ v_k, \tau~ RVal) \{\phi\}$$
   The following is the evaluation rule of function call $ext~f(ra_1,...,ra_k, RVal)$:
   $$\mbox{R10}~~~~~
   \displaystyle\frac{
   \begin{array}{l}(ra_1, \sigma_{i-1}, s_i, i) \Downarrow n_1, ..., (ra_k, \sigma_{i-1}, s_i, i) \Downarrow n_k,\\
    (\phi\wedge \bigwedge_{j=1}^{k} v_j\Leftarrow n_j, \epsilon, s'_0, 0)\stackrel{*}{\rightarrow}(true, \sigma', \emptyset, |\sigma'|+1)
   \end{array}}{(ext~f(ra_1,...,ra_k,RVal), \sigma_{i-1},s_i,i)\Downarrow s'^r_{|\sigma'|}(RVal)}$$
   %where $\phi[n_{j}/v_{j}(i=1..m)]$ denotes $\phi$ with $v_{j}$ replaced by $n_j$, for all $j=1..m$.
  % For an MSVL user-defined function with a return value but not parameters, the evaluation rule is similar. Hence, it is omitted here.
  \item [(2)] $g$ is an external function and the evaluation rule of $ext~g(ra_1,...,ra_k)$ is given as follows:
  $$\mbox{R11}~~~~~\displaystyle\frac{
   (ra_1, \sigma_{i-1}, s_i, i) \Downarrow n_1, ..., (ra_k, \sigma_{i-1}, s_i, i) \Downarrow n_k}{(ext~g(ra_1,...,ra_k), \sigma_{i-1},s_i,i)\Downarrow g(n_1,...,n_k)}$$
\end{itemize}
\textbf{Semantic equivalence rules of function calls in statements}\\
 (1) The execution of an internal call of a user-defined MSVL function in statements is actually substituted by the execution of the body of the function with the arguments. $f$ is an  MSVL user-defined function defined
 as follows:
  $$function~f((\tau_j~ v_j,)^* (\tau~ RVal)^?) \{mdcl; ms\}$$
  where $mdcl$ is a list of declarations $(mdcl = (mPd;)^*$);
  $(\tau~ RVal)^?$ denotes an optional occurrence of $\tau~ RVal$.
 The semantic equivalence rule of internal call statement $f((ra_j,)^*(RVal)^?)$ is given as follows:
$$\begin{array}{lll}
\mbox{FUN}&f((ra_j,)^*(RVal)^?)\equiv& ( (\tau_j~ v_j\Leftarrow ra_j\wedge)^* mdcl);ms; \\
&& \bigcirc (ext~mfree((v_j,)^*(RVal,)^?mdcl)\wedge \mbox{\ttfamily empty})
 \end{array}$$
where $mfree((v_j,)^*(RVal,)^?mdcl)$, which is an external function, releases the memory of variables $v_j$ and $RVal$, and variables declared in $mdcl$.\\
%For other kinds of MSVL user-defined functions, the semantic equivalence rule is similar. Hence, it is omitted here.\\
(2) For the execution of each external function call in an MSVL program, only the information upon the
beginning and ending points of the execution of the callee function is kept. \\ %Let $\rightarrowtail$ denote the congruence relation between configurations. $c \stackrel{*}{\rightarrowtail} c'$ implies that $c$ is transformed to $c'$ by several steps within a state.\\
 1) For any MSVL user-defined function, rule  EXT1  of
 the external function call $ext~ f ((ra_j,)^* (RVal)^?)$
is given as follows:
\begin{itemize}
  \item [EXT1.]
 If $\sigma'=\langle s'_0,..., s'_n\rangle$, $s'_0=s_i$ and $( ( (\tau_j~ v_j\Leftarrow ra_j\wedge)^*mdcl);ms,\epsilon,s'_0,0) \stackrel{+}{\rightarrow} (true, \sigma',\emptyset, n+1)$,
   then  $(\wedge\{\bigcirc p,ext~ f ((ra_j,)^*(RVal)^?)\},\sigma_{i-1}, s_i, i)$
$\rightarrow(p, \linebreak \sigma_{i}, s_{i+1}, i+1)$ and $s_{i+1}=s'_n$.
\end{itemize}
2) For any call of external functions, the function has been actually parameterized with arguments, which can be regarded as the initial state of executing the function. Each assignment in the
external function could be considered as the cause of state transitions. By executing the function, a finite state sequence can be generated and it is called a model of the function call. Then rules EXT2 and EXT3 of $ext~g(ra_1, . . . , ra_k)$ are given as follows:
\begin{itemize}
  \item [EXT2.] If for each $1\leq j\leq k$, $(ra_j,\sigma_{i-1},s_i,i)\Downarrow n_j$ and $\langle s_i \rangle$ is a model of  $g(n_1, . . . , n_k)$,
     then
   $(\wedge\{\bigcirc p, ext~ g (ra_1, . . . , ra_k)\},\sigma_{i-1}, s_i, i)$
$\rightarrow(p,\sigma_{i}, s_{i+1}, i+1)$  and  $s_{i+1}=s_i$. That means  $g (ra_1, . . . , ra_k)$ executes at a single state $s_i$, and state $s_i$ is inserted and concatenated with the main interval $\sigma_{i}$.

  \item [EXT3.] If for each $1\leq j\leq k$, $(ra_j,\sigma_{i-1},s_i,i)\Downarrow n_j$, $s'_0=s_i$ and
%one of the following two cases holds:
%\begin{itemize}
 % \item [(1)] $(\phi[n_i/v_i(i=1..m)],\sigma_{i-1},s_i,i) \stackrel{*}{\rightarrowtail} (\mbox{\ttfamily empty},\sigma_{i-1},s_i,i)$ and\\
 % $(\wedge\{\bigcirc p,ext~ f (ra_1, . . . , ra_m, RVal)\},\sigma_{i-1}, s_i, i)\rightarrow$
%$(p,\sigma_{i}, s_{i+1}, i+1)$
 % \item [(2)]
   $\sigma'=\langle s'_0,..., s'_n\rangle$ is a model of $g(n_1, . . . , n_k)$,
     then
   $(\wedge\{\bigcirc p,ext~ g (ra_1, . . . , ra_k)\}, \sigma_{i-1}, s_i, i)$
$\rightarrow(p,\sigma_{i}, s_{i+1}, i+1)$  and  $s_{i+1}=s'_n$. That means $g (ra_1, . . . , ra_k)$ executes over an interval $\sigma'$, and the final state  $s'_n$ over $\sigma'$ is inserted and concatenated with the main interval $\sigma_{i}$.
\end{itemize}

%\end{itemize}

%$$
%\begin{array}{l}
%I \models_c ext~g(a_1, . . . , a_n) \mbox{ iff}\\
%(1)~ j = k + 1; \mbox{ and}\\
%(2)~\mbox{one of the following two cases holds:}\\
%~~~~~(a)~ s_k = s_{k+1} \mbox{ and } \langle s_k \rangle \mbox{ is a model of } g(I[a_1], . . . ,I[a_n]), \mbox{ or}\\
%~~~~~(b)~ \mbox{there exists an (empty) interval } \sigma'' = \langle s''_1, . . . , s''_m\rangle (m \in N_0) \mbox{ such that}\\
%~~~~~~~~ \sigma' = \langle s_k \rangle \cdot \sigma'' \cdot \langle s_{k+1} \rangle\mbox{ and }
%\sigma'  \mbox{ is a model of } g(I[a_1], . . . ,I[a_n]).
%\end{array}
%$$

\section{Translation from Xd-C to MSVL} \label{trans}
In this section, an algorithm for translating an Xd-C program to an MSVL program is presented. Further, an example is given to show how the algorithm works.
%the semantics equivalence between the two programs is presented.
\subsection{Translation Algorithm}
An Xd-C program is composed of a list of declarations $(Pd;)^*$ and functions $(funct;)^*$ (including main function).
 Thus, as shown in Algorithm \ref{PrgmTr}, translating an Xd-C program to an MSVL program is actually
  translating declarations and functions from Xd-C to MSVL.

\begin{algorithm}[htb!]
\small
\caption{$PrgmTr(P)$}
\label{PrgmTr}
\renewcommand{\algorithmicrequire}{\textbf{Input：}}
\renewcommand\algorithmicensure {\textbf{Output：}}
\begin{algorithmic}[l]
\REQUIRE
an Xd-C program $P$ %&\mid array \mid structure
\ENSURE
an MSVL program $Q$
\STATE \textbf{begin function}
\STATE \quad \textbf{case}
\STATE  \quad\quad $P$ is $Pd$: \textbf{return} $DecTr(Pd)$;
\STATE  \quad\quad $P$ is $funct$: \textbf{return} $FuncTr(funct)$;
\STATE  \quad\quad $P$ is $P_1;P_2$: \textbf{return} $PrgmTr(P_1);PrgmTr(P_2)$;
\STATE \quad \textbf{end case}
\STATE \textbf{end function}
\end{algorithmic}
\end{algorithm}

Algorithm \ref{DT} translates Xd-C declarations to MSVL declarations. For a variable declaration $\tau ~varlist$, $varlist$ is translated to an MSVL variable list using $VLTr(varlist)$. For an array initialization $\tau ~id_c[]=\{(e,)^*e\}$, $count((e,)^*e)$ is used to count the number of elements in array $id_c$ and each element $e$ is translated to an MSVL expression by $ExTr(e)$.
For initialization of an array whose number of elements is specified, we just need to translate each element $e$ to an MSVL expression by $ExTr(e)$.
For a structure definition $struct ~id_{c1} \{(\tau~ id_{c2};)^* \tau~id_{c2}\}$, $and$ is used to replace `$;$'.
\begin{algorithm}[htb!]
\small
\caption{$DecTr(Pd)$}
\label{DT}
\renewcommand{\algorithmicrequire}{\textbf{Input：}}
\renewcommand\algorithmicensure {\textbf{Output：}}
\begin{algorithmic}[l]
\REQUIRE
a declaration $Pd$ %&\mid array \mid structure
\ENSURE
an MSVL fragment
\STATE \textbf{begin function}
\STATE \quad \textbf{case}
\STATE  \quad\quad $Pd$ is $\tau ~varlist$: \textbf{return} $\tau$ $VLTr(varlist)$ {\ttfamily and skip};
\STATE  \quad\quad $Pd$ is $\tau~id_c[n]$: \textbf{return} $\tau~id_m[n]$ {\ttfamily and skip};
\STATE  \quad\quad $Pd$ is $\tau~id_c[t][n]$: \textbf{return} $\tau~id_m[t][n]$ {\ttfamily and skip};
\STATE  \quad\quad $Pd$ is  $\tau ~id_c[n]=\{(e,)^*e\}$:
\STATE  \quad\quad ~~~~~~~\textbf{return} $\tau ~id_m[n]\Leftarrow\{(ExTr(e),)^*ExTr(e)\}$ {\ttfamily and skip};
\STATE  \quad\quad $Pd$ is $\tau ~id_c[]=\{(e,)^*e\} $:
\STATE  \quad\quad ~~~~~~~\textbf{return} $\tau ~id_m[count((e,)^*e)]\Leftarrow\{(ExTr(e),)^*ExTr(e)\}$ {\ttfamily and skip};
\STATE  \quad\quad $Pd$ is  $\tau ~id_c[t][n]=\{(e,)^*e\}$:
\STATE  \quad\quad ~~~~~~~\textbf{return} $\tau ~id_m[t][n]\Leftarrow\{(ExTr(e),)^*ExTr(e)\}$ {\ttfamily and skip};
\STATE  \quad\quad $Pd$ is  $\tau ~id_c[t][n]=\{(\{(e,)^*e\},)^*\{(e,)^*e\}\}$: \textbf{return} $\tau ~id_m[t][n]\Leftarrow$
\STATE  \quad\quad ~~~~~~~~~~$\{(\{(ExTr(e),)^*ExTr(e)\},)^*\{(ExTr(e),)^*ExTr(e)\}\}$ {\ttfamily and skip};
\STATE  \quad\quad $Pd$ is $struct ~id_{c1} \{(\tau~ id_{c2};)^* \tau~id_{c2};\}$:
 \STATE  \quad\quad ~~~~~~~\textbf{return} $struct ~id_{m1} \{(\tau~ id_{m2} ~and)^* \tau~id_{m2}\}$;
\STATE \quad \textbf{end case}
\STATE \textbf{end function}
\end{algorithmic}
\end{algorithm}

Algorithm \ref{VLTr} translates a variable list $Varlst$ from Xd-C to MSVL.  A variable $id_c$ is directly translated to variable $id_m$ in MSVL and a variable initialization $id_c=e$ is translated to $id_m\Leftarrow ExTr(e)$.
\begin{algorithm}
\small
\caption{$VLTr(Varlst)$}
\label{VLTr}
\renewcommand{\algorithmicrequire}{\textbf{Input：}}
\renewcommand\algorithmicensure {\textbf{Output：}}
\begin{algorithmic}[l]
\REQUIRE
a variable list $Varlst$ in Xd-C
\ENSURE
a variable list in MSVL
\STATE \textbf{begin function}
\STATE \quad \textbf{case}
\STATE  \quad\quad $Varlst$ is $id_c$: \textbf{return}  $id_m$;
\STATE  \quad\quad $Varlst$ is $id_c=e$: \textbf{return}  $id_m\Leftarrow ExTr(e)$;
\STATE  \quad\quad $Varlst$ is $varlist,varlist$: \textbf{return}  $VLTr(varlist), VLTr(varlist)$;
%\STATE  \quad\quad $Varlst$ is $v=e,varlist$: \textbf{return}  $v<==ExTr(e), VLTr(varlist)$;
\STATE \quad \textbf{end case}
\STATE \textbf{end function}
\end{algorithmic}
\end{algorithm}

Algorithm \ref{ExTr} shows how to translate each expression from Xd-C  to MSVL. A constant $c$ and a variable $id_c$ can directly be translated to $c$ and $id_m$ in MSVL while for other expressions such as $id_c[e]$, $x(e_{1},...,e_{k})$, $(\tau) ~e$, $op_1~e$ and $e_1~op~e_2$, sub-expressions $e_{1}$, $e_2$, $...$, $e_k$, $e$ and $x$ are translated to their corresponding MSVL expressions. Operations $\mathit{==}$, $\&\&$ and $\bitor\bitor$ are translated to $=$, $and$ and $or$, respectively.
Note that since a toolkit can only recognize strings of ordinary symbols, we replace $\neg$, $\wedge$ and $\vee$ by $!$, $and$ and $or$, respectively.
%$e_1 ~\mathit{==} ~e_2$ is translated to  $ ExTr(e_1) ~= ~ExTr(e_2)$, $e_1 ~\&\& ~e_2$ to $ExTr(e_1)~ and ~ExTr(e_2)$ and $e_1 ~|| ~e_2$ to $ExTr(e_1)~ and ~ExTr(e_2)$.
%Expressions $\mathit{++}e$, $\mathit{--}e$, $e\mathit{++}$ and $e\mathit{--}$ are translated to MSVL library function calls
%$AIncr(\&ExTr(e))$, $ADecr(\&ExTr(e))$, $BIncr(\&ExTr(e))$ and $BDecr(\&ExTr(e))$, respectively. Fig. \ref{libFunc} illustrates the implementations of these library functions.
Expression $e_1?e_2:e_3$ is translated to  { $if(ExTr(e_1))~then~ ExTr(e_2)~ else~ ExTr(e_3)$}.

\begin{algorithm}[htb!]
\small
\caption{$ExTr(E)$}
\label{ExTr}
\renewcommand{\algorithmicrequire}{\textbf{Input：}}
\renewcommand\algorithmicensure {\textbf{Output：}}
\begin{algorithmic}[l]
\REQUIRE
an expression $E$ in Xd-C
\ENSURE
an expression in MSVL
\STATE \textbf{begin function}
\STATE \quad \textbf{case}
\STATE  \quad\quad $E$ is $c$: \textbf{return} $c$;
\STATE  \quad\quad $E$ is $id_c$: \textbf{return} $id_m$;
\STATE  \quad\quad $E$ is $id_c[e]$: \textbf{return} $id_m[ExTr(e)]$;
\STATE  \quad\quad $E$ is $id_c[e_1][e_2]$: \textbf{return} $id_m[ExTr(e_1)][ExTr(e_2)]$;
\STATE  \quad\quad $E$ is $le.x$:\textbf{return} $ExTr(le).ExTr(x)$;
\STATE  \quad\quad $E$ is $le\rightarrow x$: \textbf{return} $ExTr(le)\rightarrow ExTr(x)$;
\STATE  \quad\quad $E$ is $*e$: \textbf{return} $*ExTr(e)$;
\STATE  \quad\quad $E$ is $\&le$: \textbf{return} $\&ExTr(le)$;
\STATE  \quad\quad $E$ is $x(e_{1},...,e_{k})$: if $x$ points to a user-defined function
\STATE  \quad\quad ~~~~~~~~~~~~~~~~~~~~~~~~~\textbf{return}  $ext~ExTr(x)(ExTr(e_1),...,ExTr(e_k),RVal)$;
\STATE  \quad\quad ~~~~~~~~~~~~~~~~~~~~~~~~else \textbf{return}  $ext~ExTr(x)(ExTr(e_1),...,ExTr(e_k))$;
%\STATE  \quad\quad $E$ is $\mathit{++}e$: \textbf{return}  $AIncr(\&ExTr(e))$;
%\STATE  \quad\quad $E$ is $\mathit{--}e $: \textbf{return}  $ADecr(\&ExTr(e))$;
%\STATE  \quad\quad $E$ is $e \mathit{++}$: \textbf{return} $BIncr(\&ExTr(e))$;
%\STATE  \quad\quad $E$ is $e\mathit{--}$: \textbf{return} $BDecr(\&ExTr(e))$;
\STATE  \quad\quad $E$ is $(\tau) ~e$: \textbf{return}  $(\tau) ~ExTr(e)$;
\STATE  \quad\quad $E$ is $op_1 ~e $: \textbf{return}  $op_1 ~ExTr(e)$;
\STATE  \quad\quad $E$ is $e_1 ~op ~e_2 ( op::= aop|bop|rop|!=)$: \textbf{return}  $ ExTr(e_1) ~op ~ExTr(e_2)$;
\STATE  \quad\quad $E$ is $e_1 ~\mathit{==} ~e_2$: \textbf{return}  $ ExTr(e_1) ~= ~ExTr(e_2)$;
\STATE  \quad\quad $E$ is $e_1 ~\&\& ~e_2$: \textbf{return}  $ExTr(e_1)~ and ~ExTr(e_2)$;
\STATE  \quad\quad $E$ is $e_1 ~\bitor\bitor ~e_2$: \textbf{return}  $ExTr(e_1)~ or ~ExTr(e_2)$;
\STATE  \quad\quad $E$ is $e_1?e_2:e_3$: \textbf{return}  {\ttfamily $if(ExTr(e_1))then~ ExTr(e_2)~ else~ ExTr(e_3)$};
\STATE \quad \textbf{end case}
\STATE \textbf{end function}
\end{algorithmic}
\end{algorithm}
%
% \begin{figure}[htb!]
%\centering
%\includegraphics[width=4.5in]{libFunc.png}
%\caption{Implementations of $AIncr$, $ADecr$, $BIncr$ and $BDecr$}
%\label{libFunc}
%\end{figure}

\begin{algorithm}[htb!]
\small
\caption{$FuncTr(funct)$}
\label{fTr}
\renewcommand{\algorithmicrequire}{\textbf{Input：}}
\renewcommand\algorithmicensure {\textbf{Output：}}
\begin{algorithmic}[l]
\REQUIRE
a function fragment $funct$ in Xd-C
\ENSURE
an MSVL fragment
\STATE \textbf{begin function}
\STATE \quad \textbf{case}
\STATE  \quad\quad $funct$ is $\tau_1 ~id_1((\tau_2~ id_2,)^*(\tau_2~ id_2))\{(Pd;)^*cs\}$:
\STATE  \quad\quad  ~~~~~~~~\textbf{return}  $function~ id_1((\tau_2~ id_2,)^*(\tau_2~ id_2), \tau_1~RVal)$
\STATE  \quad\quad  ~~~~~~~~ ~~~~~~~~~~ $\{(DecTr(Pd);)^*StmtTr(cs)\}$;
\STATE  \quad\quad $funct$ is $void ~id_1((\tau~ id_2,)^*(\tau~ id_2))\{(Pd;)^*cs\}$:
\STATE  \quad\quad  ~~~~~~~~\textbf{return}  $function~ id_1((\tau~ id_2,)^*(\tau~ id_2))\{(DecTr(Pd);)^*StmtTr(cs)\}$;
\STATE  \quad\quad $funct$ is $\tau ~id_1()\{(Pd;)^*cs\}$:
\STATE  \quad\quad  ~~~~~~~~\textbf{return}  $function~ id_1(\tau~ RVal)\{(DecTr(Pd);)^*StmtTr(cs)\}$;
\STATE  \quad\quad $funct$ is $void ~id_1()\{(Pd;)^*cs\}$:
\STATE  \quad\quad  ~~~~~~~~\textbf{return}  $function~ id_1()\{(DecTr(Pd);)^*StmtTr(cs)\}$;
\STATE \quad \textbf{end case}
\STATE \textbf{end function}
\end{algorithmic}
\end{algorithm}
A function $funct$ can be translated to an MSVL function using Algorithm \ref{fTr}.
It translates variable declarations and statements from Xd-C to MSVL by means of $DecTr$ and $StmtTr$, respectively.
Algorithm \ref{StmtTr} is presented to translate each Xd-C statement to an MSVL statement. %For variable declaration statement $\tau ~(varlist)$, $varlist$ is first translated into an MSVL variable list using Algorithm $VLTr$, then the statement is specified with the interval length 1.
A null statement is translated to MSVL statement {\ttfamily empty}.
Simple assignment $le=e$, post increment $le\mathit{++}$ and post decrement $le\mathit{--}$ are translated to MSVL unit assignment statements $ExTr(le):=ExTr(e)$, $ExTr(le):=ExTr(le)+1$ and $ExTr(le):=ExTr(le)-1$, respectively. A conditional statement is translated to MSVL conditional statement {\ttfamily if$(ExTr(e))$then$\{StmtTr(cs_1)\}$else$\{StmtTr(cs_2)\}$}. %For a sequential statement $s_1;s_2$, its sub-statements $s_1$ and $s_2$ are translated to MSVL statements using $StmtTr$.
 Sequential, {\ttfamily switch}, {\ttfamily while} loop, {\ttfamily do} loop and {\ttfamily for} loop statements are translated to MSVL statements by Algorithms $ChopTr$, $SwitchTr$,  $WhileTr$, $DoTr$ and $ForTr$, respectively. %The translation of do loop and for loop statements are performed with the help of $WhileTr(S)$.
%In order to translate the function definition statement $\tau ~fn(\linebreak para) \{s\}$ to an MSVL statement, $s$ should be translated into MSVL statement using $StmtTr(s)$.
In order to translate  {\ttfamily continue},  {\ttfamily break} and {\ttfamily return} in an Xd-C program to MSVL statements, variables $continue$, $break$ and $return$  are introduced as key variables in MSVL to handle {\ttfamily continue},  {\ttfamily break} and {\ttfamily return} statements in the Xd-C program. The translation of a function call statement ``$x(e_1,...,e_k);$'' is divided into three cases:
%In an MSVL program a variable $continue$ is used to denote whether a
\begin{itemize}
  \item [(1)] For a function call of a user-defined function without a return value, all sub-expressions $e_1$,...,$e_k$ and $x$ are translated to MSVL expressions by $ExTr$.
  \item [(2)] For a function call of a user-defined function with a return value,  an extra argument $RVal$ storing the return value of the function call is introduced.
  \item [(3)] For a function call of an external function, a key word $ext$ is added before the function call which represents the function call is an external call.
\end{itemize}

\begin{algorithm}[htb!]
\small
\caption{$StmtTr(S)$}
\label{StmtTr}
\renewcommand{\algorithmicrequire}{\textbf{Input：}}
\renewcommand\algorithmicensure {\textbf{Output：}}
\begin{algorithmic}[l]
\REQUIRE
an elementary statement $S$ in Xd-C
\ENSURE
an MSVL statement
\STATE \textbf{begin function}
\STATE \quad \textbf{case}
%\STATE  \quad\quad $S$ is $\tau ~varlist$;: \textbf{return}  $\tau ~VLTr(varlist)$ {\ttfamily and skip};
\STATE  \quad\quad $S$ is $;$: \textbf{return} {\ttfamily empty};
%\STATE  \quad\quad $S$ is $e;$: \textbf{return} $ExSTr(S)$;
\STATE  \quad\quad $S$ is $le\mathit{++};$: \textbf{return} $ExTr(le):=ExTr(le)+1$;
\STATE  \quad\quad $S$ is $le\mathit{--};$: \textbf{return} $ExTr(le):=ExTr(le)-1$;
\STATE  \quad\quad $S$ is $le=e;$: \textbf{return} $ExTr(le):=ExTr(e)$;
\STATE   \quad\quad $S$ is $cs_1;cs_2$: \textbf{return} $\mathit{ChopTr}(cs_1;cs_2)$;
%\STATE  \quad\quad $S$ is $le ~aop=~e;$: \textbf{return} $ExTr(le):=ExTr(le) ~aop ~ExTr(e)$;
%\STATE  \quad\quad $S$ is $le ~bop=~e;$: \textbf{return} $ExTr(le):=ExTr(le) ~bop ~ExTr(e)$;
\STATE  \quad\quad $S$ is {\ttfamily if($e$)\{$cs_1$\}else\{$cs_2$\}}:
\STATE  \quad\quad \quad~~ \textbf{return} {\ttfamily if($ExTr(e)$)then\{$StmtTr(cs_1)$\}else\{$StmtTr(cs_2)$\}};
\STATE  \quad\quad $S$ is {\ttfamily switch($e$)$\{sw\}$}: \textbf{return} $break:=0; SwitchTr(sw,e);break:=0$;
\STATE  \quad\quad $S$ is {\ttfamily while($e$)\{$cs$\}}: \textbf{return} $WhileTr(S)$;
\STATE  \quad\quad $S$ is {\ttfamily do\{$cs$\}while($e$)$;$}: \textbf{return} $DoTr(S)$;
\STATE  \quad\quad $S$ is {\ttfamily for($cs_1; e; cs_2$)\{$cs$\}}: \textbf{return} $ForTr(S)$;
\STATE  \quad\quad $S$ is {\ttfamily continue$;$}: \textbf{return} $continue:=1$;
\STATE  \quad\quad $S$ is {\ttfamily break$;$}: \textbf{return} $break:=1$;
\STATE  \quad\quad $S$ is {\ttfamily return $e;$}: \textbf{return} {\ttfamily $return:=1$ and $RVal:=ExTr(e)$};
\STATE  \quad\quad $S$ is {\ttfamily return$;$}: \textbf{return} {\ttfamily $return:=1$};
%if {\ttfamily return} occurs in $s_1$
%\STATE   \quad\quad \quad\quad \quad\quad \quad ~ \textbf{return} $StmtTr(s_1);if(return=0)then\{StmtTr(s_2)\}$;
%\STATE   \quad\quad \quad\quad \quad\quad ~  else \textbf{return} $StmtTr(s_1);StmtTr(s_2)$;
\STATE  \quad\quad $S$ is $x(e_{1},...,e_{k});$:
\STATE  \quad\quad ~~~~~if $x$ points to a user-defined function without a return value
\STATE  \quad\quad ~~~~~~~~~~\textbf{return}  $ExTr(x)(ExTr(e_1),...,ExTr(e_k))$;
\STATE  \quad\quad ~~~~~else if $x$ points to a user-defined function with a return value
\STATE  \quad\quad ~~~~~~~~~~~~~~\textbf{return}  $ExTr(x)(ExTr(e_1),...,ExTr(e_k),RVal)$;
\STATE  \quad\quad ~~~~~~~~~~~else \textbf{return}  $ext~ExTr(x)(ExTr(e_1),...,ExTr(e_k))$ {\ttfamily and skip};
\STATE \quad \textbf{end case}
\STATE \textbf{end function}
\end{algorithmic}
\end{algorithm}

%
%Algorithm \ref{ExSTr} translates each expression statement to an MSVL statement. For function call expression statement ($x(e_1,...,e_m);$), we just need to translate $x(e_1,...,\linebreak e_m)$ to an MSVL expression using Algorithm $ExTr(E)$.
%Since other expression statements have no side effects, they are translated to  {\ttfamily empty}.
%
%\begin{algorithm}
%\small
%\caption{$ExSTr(S)$}
%\label{ExSTr}
%\renewcommand{\algorithmicrequire}{\textbf{Input：}}
%\renewcommand\algorithmicensure {\textbf{Output：}}
%\begin{algorithmic}[l]
%\REQUIRE
%an expression statement $e;$ in Xd-C
%\ENSURE
%an MSVL statement
%\STATE \textbf{begin function}
%\STATE \quad \textbf{case}
%\STATE  \quad\quad $S$ is $x(e_1,...,e_m);$: \textbf{return}  $ExTr(E)$;
%\STATE  \quad\quad default: \textbf{return}  {\ttfamily empty};
%\STATE \quad \textbf{end case}
%\STATE \textbf{end function}
%\end{algorithmic}
%\end{algorithm}

A sequential statement ``$cs_1;cs_2$;'' is translated to an MSVL statement by Algorithm \ref{ChopTr} in four cases.
If there is no {\ttfamily break}, {\ttfamily return} or {\ttfamily continue} in $cs_1$, sub-statements $cs_1$ and $cs_2$ are translated to MSVL statements using $StmtTr$. If there is {\ttfamily break}, {\ttfamily return} or {\ttfamily continue} in $cs_1$, the sequential statement is translated to ``{\ttfamily $StmtTr(cs_1);$if$(break=0)$ then$\{StmtTr(cs_2)\}$ else$\{$empty$\}$}'', ``{\ttfamily $StmtTr(cs_1);$if$(return=0)$ then$\{StmtTr(cs_2)\}$ else$\{$empty$\}$}'' or ``{\ttfamily $StmtTr(\linebreak cs_1);$ if$(continue=0)$ then$\{StmtTr(cs_2)\}$ else$\{$empty$\}$}''.
\begin{algorithm}[htb!]
\small
\caption{$ChopTr(S)$}
\label{ChopTr}
\renewcommand{\algorithmicrequire}{\textbf{Input：}}
\renewcommand\algorithmicensure {\textbf{Output：}}
\begin{algorithmic}[l]
\REQUIRE
an Xd-C sequential statement $S$: $cs_1;cs_2$
\ENSURE
an MSVL statement
\STATE \textbf{begin function}
\STATE \quad \textbf{case}
\STATE  \quad\quad $cs_1$ contains no {\ttfamily break}, {\ttfamily return }or {\ttfamily continue}:
\STATE  \quad\quad\quad\textbf{return}  $StmtTr(cs_1);Stmt(cs_2)$;
\STATE  \quad\quad $cs_1$ contains {\ttfamily break}:
\STATE  \quad\quad\quad \textbf{return}  {\ttfamily $StmtTr(cs_1);$ if$(break=0)$then$\{StmtTr(cs_2)\}$else$\{$empty$\}$};
%\STATE  \quad\quad\quad \quad\quad\quad\quad\quad\quad \quad\quad~  {\ttfamily if$(break=0)$then$\{Stmt(s_2)\}\};$}
\STATE  \quad\quad $cs_1$ contains {\ttfamily return}:
\STATE  \quad\quad\quad \textbf{return}  {\ttfamily $StmtTr(cs_1);$ if$(return=0)$then$\{StmtTr(cs_2)\}$else$\{$empty$\}$};
\STATE  \quad\quad $cs_1$ contains {\ttfamily continue}:
\STATE  \quad\quad \quad \textbf{return} {\ttfamily $StmtTr(cs_1);$ if$(continue=0)$then$\{StmtTr(cs_2)\}$else$\{$empty$\}$};
\STATE \quad \textbf{end case}
\STATE \textbf{end function}
\end{algorithmic}
\end{algorithm}

Algorithm \ref{SwitchTr} translates each case of a {\ttfamily switch} statement to
an MSVL conditional statement. For a {\ttfamily switch} statement, if the value of $e$ is $n$ and no {\ttfamily break} or {\ttfamily return} statement occurs before it, {\ttfamily case $n:cs$} is chosen to execute; for the case following case $n$ (including the default case), if there is no {\ttfamily break} or {\ttfamily return} before it, the case is also executed.

Algorithm \ref{WhileTr} translates a {\ttfamily while} loop statement  {\ttfamily while$(e)\{cs\}$} to an MSVL statement according to whether there is {\ttfamily break}, {\ttfamily return} or {\ttfamily continue}. If there is no {\ttfamily break}, {\ttfamily return} or {\ttfamily continue} statement in $cs$, it is directly translated to a {\ttfamily while} statement in MSVL. If there is {\ttfamily break} or {\ttfamily return} in $cs$, the value of $break$ or $return$ should also be concerned in the condition of the {\ttfamily while} statement. If there is a {\ttfamily continue} statement in $cs$, the value of $continue$ should be set to 0 at the end of each loop.
The translation procedures of {\ttfamily do} loop and {\ttfamily for} loop statements are similar to the translation of {\ttfamily while} loop and are shown in Algorithm \ref{DoTr} and \ref{ForTr}, respectively.

\begin{algorithm}[htb!]
\small
\caption{$SwitchTr(SW, e)$}
\label{SwitchTr}
\renewcommand{\algorithmicrequire}{\textbf{Input：}}
\renewcommand\algorithmicensure {\textbf{Output：}}
\begin{algorithmic}[l]
\REQUIRE
{\ttfamily switch case} statement $SW$ and controlling expression $e$
\ENSURE
an MSVL statement
\STATE \textbf{begin function}
\STATE \quad \textbf{case}
\STATE  \quad\quad $SW$ is {\ttfamily default$:cs;$}:
\STATE  \quad\quad\quad\textbf{return} {\ttfamily if($switch=1~ and ~break=0 ~ and~ return=0$)}
 \STATE  \quad\qquad\qquad~~~~~{\ttfamily then$\{StmtTr(cs)\}$ else$\{$empty$\}$};
\STATE  \quad\quad $SW$ is {\ttfamily case $n:cs;sw$}:
\STATE  \quad\quad\quad   \textbf{return} {\ttfamily if$((ExTr(e)=n~or~switch=1)~ and ~break=0~and~$
 \STATE  \quad\quad\quad\quad \quad\quad  ~$return=0)$ then$\{StmtTr(cs)\}$ else$\{$empty$\}$$;SwitchTr(sw)$};
\STATE \quad \textbf{end case}
\STATE \textbf{end function}
\end{algorithmic}
\end{algorithm}

\begin{algorithm}[htb!]
\small
\caption{$WhileTr(S)$}
\label{WhileTr}
\renewcommand{\algorithmicrequire}{\textbf{Input：}}
\renewcommand\algorithmicensure {\textbf{Output：}}
\begin{algorithmic}[l]
\REQUIRE
an Xd-C {\ttfamily while} loop statement $S$: {\ttfamily while($e$)\{$cs$\}}
\ENSURE
an MSVL statement
\STATE \textbf{begin function}
\STATE \quad \textbf{case}
\STATE  \quad\quad $cs$ contains no {\ttfamily break}, {\ttfamily return} or {\ttfamily continue}:
\STATE  \quad\quad\quad\textbf{return}  {\ttfamily while$(ExTr(e))\{StmtTr(cs)\}$};
\STATE  \quad\quad $cs$ contains {\ttfamily break}:
\STATE  \quad\quad\quad \textbf{return}  {\ttfamily while$(ExTr(e) ~and ~break=0)\{StmtTr(cs)\};break:=0$};
\STATE  \quad\quad $cs$ contains {\ttfamily return}:
\STATE  \quad\quad\quad \textbf{return}  {\ttfamily while$(ExTr(e) ~and ~return=0)\{StmtTr(cs)\}$};
\STATE  \quad\quad $cs$ contains {\ttfamily continue}:
\STATE  \quad\quad\quad \textbf{return} {\ttfamily  while$(ExTr(e))\{StmtTr(cs);continue:=0\}$};
\STATE \quad \textbf{end case}
\STATE \textbf{end function}
\end{algorithmic}
\end{algorithm}

\begin{algorithm}[htb!]
\small
\caption{$DoTr(S)$}
\label{DoTr}
\renewcommand{\algorithmicrequire}{\textbf{Input：}}
\renewcommand\algorithmicensure {\textbf{Output：}}
\begin{algorithmic}[l]
\REQUIRE
an Xd-C {\ttfamily do} loop statement $S$: {\ttfamily do$\{cs\}$while$(e);$}
\ENSURE
an MSVL statement
\STATE \textbf{begin function}
\STATE \quad \textbf{case}
\STATE  \quad\quad $cs$ contains no {\ttfamily break}, {\ttfamily return} or {\ttfamily continue}:
\STATE  \quad\quad\quad\textbf{return}  $StmtTr(cs);${\ttfamily while$(ExTr(e))\{StmtTr(cs)\}$};
\STATE  \quad\quad $cs$ contains {\ttfamily break}:
\STATE  \quad\quad\quad \textbf{return}  {\ttfamily $StmtTr(cs);$}
\STATE  \quad\quad\quad\quad\quad\quad {\ttfamily while$(ExTr(e) ~and ~break=0)\{StmtTr(cs)\};$$break:=0$};
\STATE  \quad\quad $cs$ contains {\ttfamily return}:
\STATE  \quad\quad\quad \textbf{return}  {\ttfamily $StmtTr(cs);$while$(ExTr(e) ~and ~return=0)\{StmtTr(cs)\}$};
\STATE  \quad\quad $cs$ contains {\ttfamily continue}:
\STATE  \quad\quad\quad \textbf{return} {\ttfamily $StmtTr(cs);continue:=0;$}
\STATE  \quad\quad\quad\quad\quad\quad {\ttfamily while$(ExTr(e))\{StmtTr(cs);continue:=0\}$};
\STATE \quad \textbf{end case}
\STATE \textbf{end function}
\end{algorithmic}
\end{algorithm}

%{\ttfamily for($s_1; e; s_2$)\{$s$\}}: \textbf{return} $StmtTr(s_1);WhileTr(${\ttfamily while$(e)\{s;s_2\})$};

\begin{algorithm}[htb!]
\small
\caption{$ForTr(S)$}
\label{ForTr}
\renewcommand{\algorithmicrequire}{\textbf{Input：}}
\renewcommand\algorithmicensure {\textbf{Output：}}
\begin{algorithmic}[l]
\REQUIRE
an Xd-C {\ttfamily for} loop statement $S$: {\ttfamily for($cs_1; e; cs_2$)\{$cs$\}}
\ENSURE
an MSVL statement
\STATE \textbf{begin function}
\STATE \quad \textbf{case}
\STATE  \quad\quad $cs$ contains no {\ttfamily break}, {\ttfamily return} or {\ttfamily continue}:
\STATE  \quad\quad\quad\textbf{return}  $StmtTr(cs_1);${\ttfamily while$(ExTr(e))\{StmtTr(cs;cs_2)\}$};
\STATE  \quad\quad $cs$ contains {\ttfamily break}:
\STATE  \quad\quad\quad \textbf{return}  {\ttfamily $StmtTr(cs_1);$}
\STATE  \quad\quad\quad\quad\quad\quad ~~{\ttfamily while$(ExTr(e) ~and ~break=0)\{StmtTr(cs;cs_2)\};$$break:=0$};
\STATE  \quad\quad $cs$ contains {\ttfamily return}:
\STATE  \quad\quad\quad \textbf{return}  {\ttfamily $StmtTr(cs_1);$while$(ExTr(e) ~and ~return=0)\{StmtTr(cs;cs_2)\}$};
\STATE  \quad\quad $cs$ contains {\ttfamily continue}:
\STATE  \quad\quad\quad \textbf{return} {\ttfamily $StmtTr(cs_1);$}
\STATE  \quad\quad\quad\quad\quad\quad ~~{\ttfamily while$(ExTr(e))\{StmtTr(cs);StmtTr(cs_2);continue:=0\}$};
\STATE \quad \textbf{end case}
\STATE \textbf{end function}
\end{algorithmic}
\end{algorithm}

\subsection{An Example}
\begin{figure}[htb!]
\centering
\includegraphics[width=4.4in]{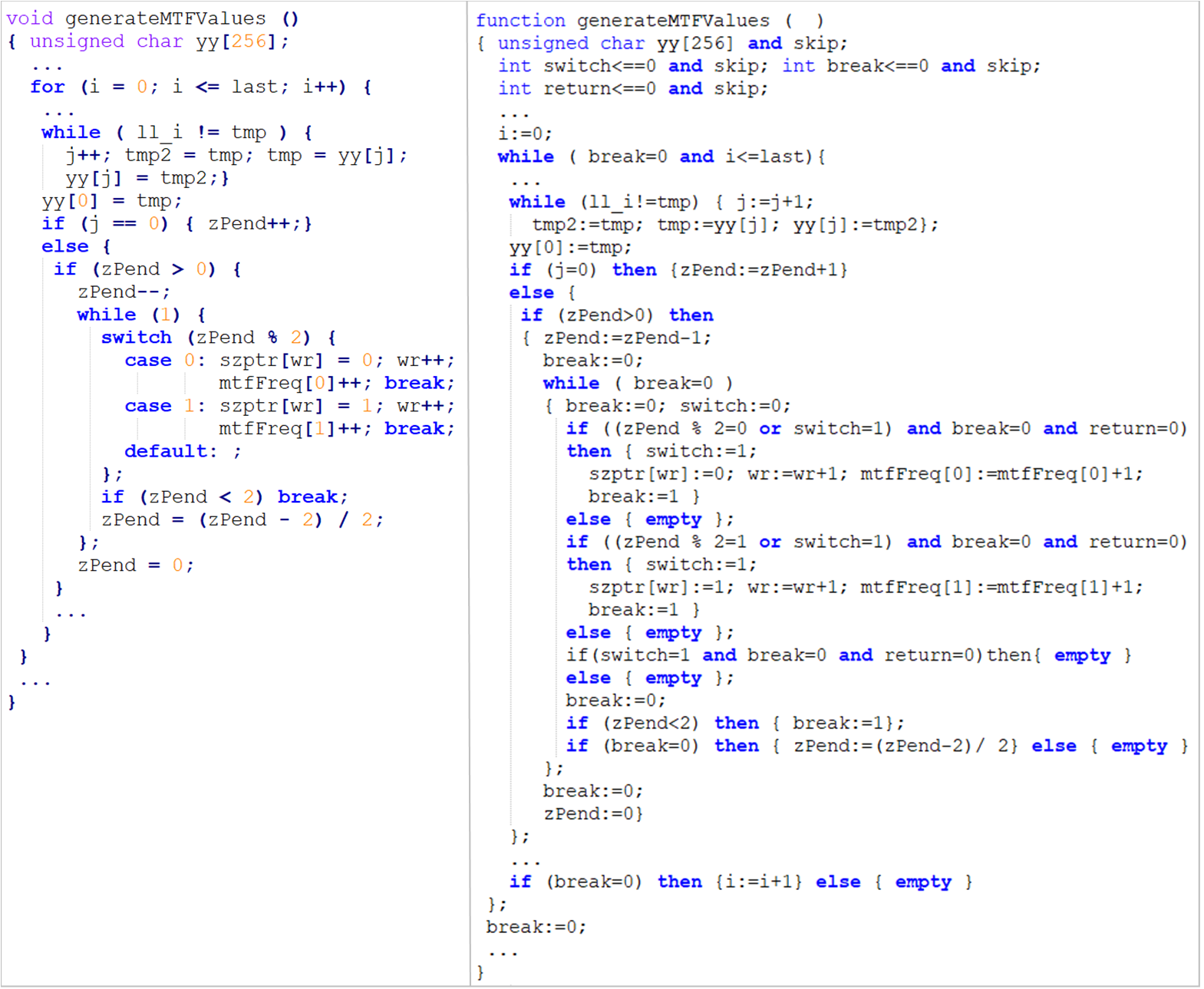}
\caption{Translating $generateMTFValues$ from Xd-C to MSVL}
\label{Case}
\end{figure}

In this section, an application $bzip2$ \cite{SPEC} is used to show how an Xd-C program is translated to an MSVL program.
$bzip2$ is a compression program to compress and decompress input files.
As shown in Fig.\ref{Case}, the left-hand side is the core of a function $generateMTFValues$ in $bzip2$, including most kinds of Xd-C statements and the right-hand side is the translated MSVL program by using the translation algorithms.
 Various kinds of Xd-C statements in the program are translated to their equivalent MSVL statements as follows:
\begin{itemize}
  \item [(1)] Function definition statement {\ttfamily void $generateMTFValues()\{...\}$} can directly be translated to MSVL function definition statement {\ttfamily function}\linebreak $generateMTFValues()\{...\}$.
  \item [(2)] Variable declaration statement ``{\ttfamily unsigned char $yy[256];$}'' is translated to {\ttfamily unsigned char $yy[256]$ and skip}.
  \item [(3)] Simple assignment statement ``{\ttfamily $i=0;$}'' is directly translated to MSVL unit assignment statement {\ttfamily $i:=0$}.
 % \item [(4)] Expression statement {\ttfamily $i++;$} is directly translated to an MSVL library function call $BIncr(\&i)$.
  \item [(4)]   {\ttfamily for$(i=0;i\mathit{<=}last;i\mathit{++})\{...\}$} is translated to ``{\ttfamily $i:=0;$ while$(break=0 ~and~i<=last)\{...;$if$(break=0)$then$\{i:=i+1\}$else$\{$empty$\} \};break:=0$}''.
  \item [(5)] {\ttfamily while$(ll_i \mathit{!=} tmp)\{...\}$} is directly translated to an MSVL {\ttfamily while} statement.
  \item [(6)] Conditional statement {\ttfamily if$(j \mathit{==} 0)\{...\}$else$\{...\}$} is translated to {\ttfamily if$(j=0)$ then$\{...\}$else$\{...\}$}.
  \item [(7)]  ``{\ttfamily break;}'' is translated to $break:=1$.
  \item [(8)]  {\ttfamily switch$(zPend \% 2)\{$case 1$:...;$ case 2$: ...;$ default$:~;\}$} is translated to the following MSVL statement:  \\
      {\ttfamily \small  $break:=0; ~switch:=0;$\\
         if$((zPend \% 2=0~ or~switch=1)~and~break=0~and~return=0)$\\
         then$\{switch:=1;...\}$ else$\{$empty$\};$\\
         if$((zPend \% 2=1~or~switch=1)~ and~break=0~and~return=0))$\\
         then$\{switch:=1;...\}$ else$\{$empty$\};$\\
         if$((switch=1~ and~break=0~and~return=0))$then$\{$empty$\}$ else$\{$empty$\};$\\
         $break:=0$}.
\end{itemize}
\section{Proof of Equivalence} \label{proof}
Since the types in MSVL are the same as in Xd-C fragments, we only need to prove the equivalence between expressions and statements involved in Xd-C and MSVL programs.

In order to account for differences in allocation patterns between Xd-C and
MSVL programs, a function $\alpha$ is used to denote a memory injection \cite{BlazyDL06}. It is an injective function from Xd-C block reference $b$ to $(b',\delta')$, which means that block $b$ in the Xd-C memory state corresponds to block $b'$ and offset $\delta'$ in the MSVL memory state.

Based on a memory injection $\alpha$, an equivalence relation between an Xd-C value $v$ and an MSVL
value $n$, written by $\alpha \vdash v\sim n$,  is defined as follows:
%$$
%\begin{array}{l}
%\alpha \vdash c\sim c~~~~~
%\displaystyle\frac{\alpha(b)=(b',\delta)~~~~i'=i+\delta}{\alpha \vdash ptr(b,i)\sim ptr(b',i')}~~~~~
%\displaystyle\frac{\alpha\vdash v_1\sim n_1~~...~~\alpha\vdash v_m\sim n_m}{\alpha \vdash (v_1,...,v_m)\sim (n_1,...,n_m)}
%\end{array}$$
$$
\begin{array}{ll}
(1)&\alpha \vdash c\sim c, \mbox{ where } v=n=c.\\
(2)&\alpha \vdash ptr(b,i)\sim ptr(b',i'), \mbox{ where } v=ptr(b,i) \mbox{ and } n=ptr(b',i'), \mbox{ if and} \\
&\mbox{only if there exists $\delta\in N_0$ such that }
\alpha(b)=(b',\delta) \mbox{~and } i'=i+\delta. \\
%\alpha \vdash (v_1,...,v_m)\sim (n_1,...,n_m)  \mbox{ if and only if } \alpha\vdash v_1\sim n_1,...,\alpha\vdash v_m\sim n_m.\\
% \alpha \vdash ptr(b,i)\sim ptr(b',i') \mbox{ if and only if there exists $\delta\in N_0$ such that } \alpha(b)=(b',\delta)\\
 %\mbox{and } i'=i+\delta. \\
\end{array}$$
 Rule (1) means that a constant $c$ in Xd-C is equivalent to $c$ in MSVL. For example, $\alpha\vdash 1\sim 1$ and $\alpha \vdash 1.2\sim 1.2$.
Rule (2) means that a pointer value $ptr(b,j)$ in Xd-C is equivalent to a pointer value $ptr(b',j')$ in MSVL.
For example, we assume the location of a variable $x$ in an Xd-C program is $(\mbox{0xffff0000},0)$. If $\alpha(\mbox{0xffff0000})=(\mbox{0xffffaaaa},8)$ which indicates that an Xd-C block $\mbox{0xffff0000}$ corresponds to an MVSL block 0xffffaaaa and offset 8, we have $\alpha\vdash ptr(\mbox{0xffff0000},0)\sim ptr(\mbox{0xffffaaaa},8)$.

%In the following, we sometimes use $P\cong Q$ to denote $\vdash P\leftrightarrow Q$ and
%$P\supset Q$ to denote $\vdash P\rightarrow Q$.

\begin{Lem}\label{adre}
For a given $\alpha$, any $b,b'\in \mathbb{Z}$ and $i,i',j,j'\in N_0$, if $\alpha \vdash ptr(b,i)\sim ptr(b',i')$ and $j=j'$, then $\alpha \vdash ptr(b,i+j)\sim ptr(b',i'+j')$.
\end{Lem}

\begin{Proof}
\[
\begin{array}{llll}
(1)&&\alpha \vdash ptr(b,i)\sim ptr(b',i')\wedge j=j'&\\
(2)&\Longrightarrow& \alpha(b)=(b',\delta) \wedge i'=i+\delta \wedge j=j'&\\
(3)&\Longrightarrow& \alpha(b)=(b',\delta) \wedge i'+j'=i+j+\delta& \\
(4)&\Longrightarrow& \alpha \vdash ptr(b,i+j)\sim ptr(b',i'+j')&
\end{array}
\]

\end{Proof}

\rightline{$\square$}

Let $\mathbb{M}$ denote the set of all memory states $M$ in Xd-C programs and $\mathbb{S}$ the set of all states $s$ in MSVL programs. % $\mathbb{G}$ the set of all global environments $G$, $\mathbb{E}$ the set of all local environments $E$ and $O$ the set of all statement outcomes $out$ in Xd-C programs.
 The equivalences of states, expressions and statements are respectively defined as follows.

\newtheorem{definition}{Definition}
\begin{definition} \label{se} (State Equivalence)
For a given memory injection $\alpha$, an Xd-C memory state $M$ is equivalent to an MSVL state $s$, denoted by $\alpha\vdash M\sim s$, if and only if the following condition holds:\\
%For any variable $x$ in the Xd-C and MSVL programs, $b\in \mathbb{Z}$, $j\in N_0$ and $v\in D$, if $(G,E\vdash x, M \stackrel{l}{\Rightarrow} (b,j))$ and $(G,E \vdash x, M \Rightarrow v)$ in the Xd-C program, then  $\alpha\vdash ptr(b,j)\sim ptr(s^l(x))$ and $\alpha\vdash v\sim s^r(x)$.\\
For any variable $x_c\in Dom(G\cup E)$ in an Xd-C program, the corresponding variable $x_m\in Dom(s)\setminus \{break, continue, return, RVal\}$ in an MSVL program, $b,b'\in \mathbb{Z}$, $j,j'\in N_0$ and $v,n\in D$, if $(G,E\vdash x_c, M \stackrel{l}{\Rightarrow} (b,j))$ and $(G,E \vdash x_c, M \Rightarrow v)$ in the Xd-C program, as well as $s^l(x_m)=(b',j')$ and $s^r(x_m)=n$ in the MSVL program, then $\alpha\vdash ptr(b,j)\sim ptr(b',j')$ and $\alpha\vdash v\sim n$.

% if $x\in Dom(E)\cup Dom(G)$ in the Xd-C program, then $x\in Dom(s)\setminus \{break, continue, return, RVal\}$ in the MSVL program and there exist $b\in \mathbb{Z}$, $j\in N_0$ and $v\in D$
%such that $(G,E\vdash x, M \stackrel{l}{\Rightarrow} (b,j))$, $(G,E \vdash x, M \Rightarrow v)$ in the Xd-C program, $\alpha\vdash ptr(b,j)\sim ptr(s^l(x))$ and $\alpha\vdash v\sim s^r(x)$.\\
%For any variable $x$, if $x\in Dom(s)\setminus \{break, continue, return, RVal\}$ in the MSVL program, then $x\in Dom(E)\cup Dom(G)$ in the Xd-C program and there exist $b\in \mathbb{Z}$, $j\in N_0$ and $v\in D$
%such that $(G,E\vdash x, M \stackrel{l}{\Rightarrow} (b,j))$, $(G,E \vdash x, M \Rightarrow v)$, $\alpha\vdash ptr(b,j)\sim ptr(s^l(x))$ and $\alpha\vdash v\sim s^r(x)$.

%(1) ($\Rightarrow$)
%If $x\in (Dom(E) \cup Dom(G))$ and $(G,E \vdash x, M \Rightarrow v)$, then $x\in Dom(s^r)$ and $\alpha\vdash v\equiv s^r(x)$, for any %$G\in \mathbb{G}$, $E\in \mathbb{E}$,
%$x \in \mathcal{V}$ and $v\in D$.\\%, $b\in \mathbb{Z}$, $b'\in \mathbb{Z}$, $\delta\in N_0$ and $\delta'\in N_0$.\\
%(2)($\Leftarrow$)
%If $x\in Dom(s^r)$, then
%$x\in (Dom(E) \cup Dom(G))$ and there exists $v\in D$ such that $(G,E \vdash x, M \Rightarrow v)$ and $\alpha\vdash v\equiv s^r(x)$, for any
% $x \in \mathcal{V}$.
\end{definition}

Definition \ref{se} defines the equivalence between states $M$ and $s$. Intuitively, $M$ is equivalent to $s$ means that for each variable in both Xd-C and MSVL programs, the locations and values of the variable are equivalent, respectively. The following is an example of state equivalence.

\textit{Example 1.}
In an Xd-C program, there are two declared variables $x_c$ and $y_c$.  At state $M$, the location of $x_c$ is (0xffff0000, 0) with value 1, while the location of $y_c$ is (0xffff1111, 0) with value $ptr(\mbox{0xffff0000},0)$. That is, $x_c=1$ and $y_c=\&x_c$. In the corresponding MSVL program, $x_m$ and $y_m$ are also declared variables.  At state $s$, the location of $x_m$ is (0xffffaaaa, 0) with value 1, while the location of $y_m$ is
(0xffff3333, 0) with value $ptr(\mbox{0xffffaaaa},0)$. We also have $x_m=1$ and $y_m=\&x_m$. Thus, for a given memory injection $\alpha$ satisfying $\alpha(\mbox{0xffff0000})=(\mbox{0xffffaaaa},0)$ and $\alpha(\mbox{0xffff1111})=(\mbox{0xffff3333},0)$, we have $\alpha \vdash ptr(\mbox{0xffff0000},0)\sim ptr(\mbox{0xffffaaaa},0)$, $\alpha\vdash 1\sim 1$ and
$\alpha \vdash ptr(\mbox{0xffff1111},0)\sim ptr(\mbox{0xffff3333},0)$.
Therefore, the locations and values of the two variables between $M$ and $s$ are equivalent, respectively. Hence, $\alpha \vdash M\sim s$.

%For a variable $x$, we define a function $var(x)$ as follows:
%$$var(x)=\left\{\begin{array}{ll}var(x')& \mbox{if $x=\&x'$ for some variable $x'$}\\ x &\mbox{otherwise}\end{array}\right .$$
%We define a matching relation $EnvMatch(\alpha, G, E, M, s)$ between an Xd-C environment $E$ and memory
%state $M$  and reference to  as follows:\\

\begin{definition}\label{lee} (Left-value Expression Equivalence)
For a given memory injection $\alpha$, an Xd-C left-value expression $e$ is equivalent to an MSVL left-value expression $a$, denoted by $\alpha\vdash e\sim_l a$, if and only if
for any $M$, $s$, $b$, $j$, $b'$, $j'$ and $\sigma$, if $\alpha\vdash M\sim s$,  $(G,E \vdash e, M \stackrel{l}{\Rightarrow} (b,j))$ in the Xd-C program and $(a, \sigma,s,|\sigma|+1)\stackrel{l}{\Rightarrow}  (b',j')$ in the MSVL program, then $\alpha \vdash ptr(b,j)\sim ptr(b',j')$.
\end{definition}

%Definition \ref{lee} defines the equivalence between left-value expressions in Xd-C and MSVL.
Intuitively, the equivalence between left-value expressions is merely that the equivalence of expression locations in Xd-C and MSVL. Further, if the locations of expressions are equivalent, the expressions stand for the same variable and their values are equivalent. For instance, the following example illustrates the situation.

\textit{Example 2.} %Let $e_c$ denote an Xd-C expression $e$ and $M(e)$ an MSVL expression $e$.
As in \textit{Example 1}, in an Xd-C program, $*y_c$ is a left-value expression and
$(G,E \vdash  *y_c, M \stackrel{l}{\Rightarrow} (\mbox{0xffff0000},0))$, while in the corresponding MSVL program, $( *y_m, \sigma,s,|\sigma|+1)\stackrel{l}{\Rightarrow}  (\mbox{0xffffaaaa},0)$. Since $\alpha\vdash ptr(\mbox{0xffff0000},0)\sim ptr(\mbox{0xffffaaaa},0)$, we have $\alpha\vdash  *y_c\sim_l  *y_m$.
Here, $*y_c$ and $*y_m$  stand for variables $x_c$ and $x_m$, respectively. In fact, the values of $*y_c$ and $*y_m$ are equivalent.

\begin{definition} \label{ree} (Right-value Expression Equivalence)
For a given memory injection $\alpha$, an Xd-C right-value expression $e$ is equivalent to an MSVL right-value expression $a$, denoted by $\alpha\vdash e\sim_r a$, if and only if
 for any $M$, $s$, $v$, $n$ and $\sigma$, if $\alpha\vdash M\sim s$, $(G,E \vdash e, M \Rightarrow v)$ in the Xd-C program and $(a,\sigma,s,|\sigma|+1) \Downarrow  n$ in the MSVL program, then $\alpha \vdash v\sim n$.
\end{definition}

As a matter of fact, the equivalence between right-value expressions is really the equivalence between  the expression values in Xd-C and MSVL under the condition of state equivalence. The following is an example of the equivalence relation.

\textit{Example 3.} In an Xd-C program, $(x_c>0)?2:3$ is a right-value expression with $x_c=1$, hence,
$(G,E \vdash (x_c>0)?2:3, M  \Rightarrow 2)$. Whereas in the corresponding MSVL program, $x_m=1$ and $(if(x_m>0)~then~2~else~3, \sigma,s,|\sigma|+1) \Downarrow  2$. Since $\alpha\vdash 2\sim 2$, we have $\alpha\vdash (x_c>0)?2:3\sim_r if(x_m>0)~then~2~else~3$.

\begin{definition}\label{ee}  (Expression Equivalence)
For a given memory injection $\alpha$, an Xd-C expression $e$ is equivalent to an MSVL expression $a$, denoted by $\alpha\vdash e\sim_e a$, if and only if either
  $e$ and $a$ are both left-value expressions and $\alpha\vdash e\sim_l a$, or $e$ and $a$ are both only right-value expressions and $\alpha\vdash e\sim_r a$.
\end{definition}

%For an Xd-C left-value expression $e$ and an MSVL left-value expression $a$, for any  $\alpha$,  $M$, $s$, $v$, $n$ and $\sigma$,
%if $\alpha\vdash M\sim s$, $e\sim a$, $(G,E \vdash e, M \Rightarrow v)$ in the Xd-C program and $(a,\sigma,s,|\sigma|+1) \Downarrow  n$  in the MSVL program, then $\alpha \vdash v\sim n$.

\begin{definition}\label{tse} (Terminating Statement Equivalence)
For a given memory injection $\alpha$, a terminating  statement $cs$ in an Xd-C  program is equivalent to a terminating statement $ms$ in an MSVL program, denoted by $\alpha\vdash cs\sim_t ms$, if and only if for any $M$, $s_i$, $out$ and $M'$, if $\alpha\vdash M\sim s_i$ and $(G,E \vdash cs, M\stackrel{t}{\Rightarrow}  out, M')$ in the Xd-C program, then there exists $\sigma \in \Gamma$ such that
%the following condition holds:\\
 $(ms, \sigma_{i-1}, s_i, i)\stackrel{*}{\rightarrow}(true, \sigma, \emptyset,|\sigma|+1)$ in the MSVL program and $\alpha \vdash M'\sim s_{|\sigma|}$.
\end{definition}

The equivalence between terminating statements in Xd-C and MSVL indicates that if the initial states are equivalent, then after executing the programs, the final states are equivalent. For instance, \textit{Example 4} shows the equivalence relation between Xd-C and MSVL statements.

\textit{Example 4.} As in \textit{Example 1}, in an Xd-C program, ``$x_c=2;$'' is a terminating statement and $(G,E \vdash x_c=2;, M  \Rightarrow out, M')$. At state $M'$, the value of $x_c$ is changed to 2. The locations of $x_c$ and $y_c$, and the value of $y_c$ are not changed. Whereas in the corresponding MSVL program, $s_i=s$ and
 $(x_m:=2, \sigma_{i-1}, s_i, i)\stackrel{*}{\rightarrow}(true, \sigma, \emptyset,|\sigma|+1)$. At state $s_{|\sigma|}$, $s^r_{|\sigma|}(x_m)=2$, $s^l_{|\sigma|}(x_m)=s^l_{i}(x_m)$, $s^l_{|\sigma|}(y_m)=s^l_{i}(y_m)$ and  $s^r_{|\sigma|}(y_m)=s^r_{i}(y_m)$.
 Hence, $\alpha\vdash M'\sim s_{|\sigma|}$. Consequently, we have $\alpha\vdash x_c=2;\sim_t x_m:=2$.

\begin{definition}\label{xd-cse}  (Xd-C Statement Equivalence)
 An Xd-C statement $cs$ is equivalent to $cs'$ executed from state $M$, denoted by $(cs,M)\cong (cs',M)$, if and only if
 two intervals $M_\sigma=(M_0,M_1,...)$ and $M'_\sigma=(M'_0,M'_1,...)$ generated by respectively executing $cs$  and $cs'$ from $M$ are equivalent, that is,  $M_i=M'_i$ for $i\geq 0$.
 \end{definition}

\begin{Lem}\label{Cloop} (\mbox{C\_loop})
In an Xd-C program, if $(G,E \vdash e, M  {\Rightarrow} true)$ and $(G,E \vdash cs, M \stackrel{t}{\Rightarrow} Normal,M_1)$,
then $(\mbox{\ttfamily while}(e)\{cs\},M)\cong (cs;\mbox{\ttfamily while}(e)\{cs\},M)$.
\end{Lem}

\begin{Proof}
Suppose $M_\sigma$ and $M'_\sigma$ are generated by executing $\mbox{\ttfamily while}(e)\{cs\}$ and ``$cs;\mbox{\ttfamily while}(e)\{cs\}$'' from $M$, respectively.
Since $(G,E \vdash e, M  {\Rightarrow} true)$ and $(G,E \vdash cs, M \stackrel{t}{\Rightarrow} Normal,M_1)$, a prefix $(M,M_1)$ (rule $T13$) of $M_\sigma$  is generated  while the same prefix $(M,M_1)$ (rule $T7$) of $M'_\sigma$  is also generated. %Further, $(\mbox{\ttfamily while}(e)\{cs\},M_1)\cong (\mbox{\ttfamily while}(e)\{cs\},M_1)$.
Therefore, $M_\sigma$ and $M'_\sigma$ are equivalent. According to Definition \ref{xd-cse}, $(\mbox{\ttfamily while}(e)\{cs\},M)\cong (cs;\mbox{\ttfamily while}(e)\linebreak\{cs\}, M)$.
\end{Proof}
\rightline{$\square$}

\begin{definition}\label{dse}  (Diverging Statement Equivalence)
For a given memory injection $\alpha$, a diverging statement $cs$ in an Xd-C program is equivalent to a diverging statement $ms$ in an MSVL program, denoted by $\alpha\vdash cs\sim_d ms$, if and only if for any $M$ and $s_i$, if $\alpha\vdash M\sim s_i$ and $(G,E \vdash cs, M\stackrel{T}{\Rightarrow} \infty)$ in the Xd-C program,
then $cs$ at state $M$ and $ms$ at state $s_i$ can both be rewritten as an infinite sequence of terminating statements $(cs,M)\cong (cs_1;cs_2;...,M)$ and
$(ms,\sigma_{i-1},s_i,i)\stackrel{*}{\rightarrowtail} (ms_1;ms_2;..., \sigma_{i-1},s_i,i)$, and $\alpha\vdash cs_j\sim_t ms_j$ for all $j\geq 1$.
%$p_0=p$ and $(p_k, \sigma_{i-1+k}, s_{i+k}, i+k){\rightarrow}(p_{k+1}, \sigma_{i+k}, s_{i+k+1}, i+k+1)$ in the MSVL program, for any $k\geq 0$.
\end{definition}

%The equivalence between diverging statements in Xd-C and MSVL means that
%if the initial states are equivalent, then  the execution of the Xd-C and MSVL programs cannot terminate.
The following is an example of diverging statement equivalence.

\textit{Example 5.} In an Xd-C program, the initial value of variable $x_c$ is 1. ``$\mbox{\ttfamily while}(x_c>0)\{x_c\mathit{++};\}$'' is a diverging statement and $(\mbox{\ttfamily while}(x_c>0)\{x_c\mathit{++};\},\linebreak M)\cong (x_c\mathit{++};x_c\mathit{++};...,M)$.
 Whereas in the corresponding MSVL program, the initial value of variable $x_m$ is also 1. $(\mbox{\ttfamily while}(x_m>0)\{x_m:=x_m+1\},\sigma_{i-1},s_i,i)\stackrel{*}{\rightarrowtail} (x_m:=x_m+1;x_m:=x_m+1;...,\sigma_{i-1},s_i,i)$.
 Since $\alpha \vdash x_c\mathit{++};\sim_t x_m:=x_m+1$, we have
 $\alpha\vdash \mbox{\ttfamily while}(x_c>0)\{x_c\mathit{++};\}\sim_d \mbox{\ttfamily while}(x_m>0)\{x_m:=x_m+1\}$.

\begin{definition}\label{ste}  (Statement Equivalence)
For a given memory injection $\alpha$, a statement $cs$ in an Xd-C program is equivalent to a statement $ms$ in an MSVL program, denoted by $\alpha\vdash cs\sim_s ms$, if and only if either $\alpha\vdash cs\sim_t ms$ or $\alpha\vdash cs\sim_d ms$.
\end{definition}

\subsection{Proof of expression equivalence}

\begin{Thm}\label{e-eq}
Suppose an Xd-C expression $e$ is transformed to an MSVL expression $a$ by Algorithm \ref{ExTr} along with converting an Xd-C program to an MSVL program.  That is, $a=ExTr(e)$.
For a given $\alpha$, any $M\in \mathbb{M}$ and $s\in \mathbb{S}$, if $\alpha\vdash M\sim s$, then $\alpha\vdash e\sim_e a$.
\end{Thm}

\begin{Proof}
The proof proceeds by induction on the structure of expressions.

 %Let $Exp$ denote the set of all Xd-C expressions. We need to prove that for all $e\in Exp$, $P(e)$ holds, where $P(e)$ is defined as follows:\\
% if $e$ is a left-value expression and for any $\alpha$, $M$, $s$, $b$ and $j$, if $\alpha\vdash M\sim s$ and  $(G,E \vdash e, M \stackrel{l}{\Rightarrow} (b,j))$, then there exist
%    $b'\in \mathbb{Z}$ and $j'\in N_0$ such that
%  $(ExTr(e), \sigma,s,|\sigma|+1)\stackrel{l}{\Rightarrow}  (b',j')$ and $\alpha \vdash ptr(b,j)\sim ptr(b',j')$ for any $\sigma\in \Gamma \cup \{\epsilon\}$; \\
%  if $e$ is a right-value expression rather than a left-value expression and for any $\alpha$,  $M$, $s$ and $v$,
%  if $\alpha\vdash M\sim s$ and $(G,E \vdash e, M \Rightarrow v)$, then there exist $n\in D$ such that  $(ExTr(e),\sigma,s,|\sigma|+1) \Downarrow  n$ and $\alpha \vdash v\sim n$ for any $\sigma\in \Gamma \cup \{\epsilon\}$.
%The proof is given as follows:

Base:%\\[-2ex]
\begin{itemize}
\item[1.] For a constant $c$, the conclusion is trivially true.

 \item[2.] For a variable $id_c$, $ExTr(id_c)=id_m$. Here, $id_c$ and $id_m$ are left-value expressions.
{\small\[
\begin{array}{lllr}
(1)&&\alpha\vdash M\sim s&\mbox{given condition}\\
(2)&\Longrightarrow&\multicolumn{2}{l}{\forall b,b',j,j'.(G,E\vdash id_c, M\stackrel{l}{\Rightarrow} (b,j))\wedge s^l(id_m)=(b',j')}\\
  &&\rightarrow \alpha\vdash ptr(b,j)\sim ptr(b',j')& \mbox{Definition}~ \ref{se},~(1)\\
(3)&&(id_m,\sigma,s,|\sigma|+1)\stackrel{l}{\Rightarrow}  s^l(id_m) &\mbox{L1}\\
%(4)&&\alpha\vdash M\sim s&\mbox{given condition}\\
(4)&\Longrightarrow&\multicolumn{2}{l}{\forall b,b',j,j'.(G,E\vdash id_c, M\stackrel{l}{\Rightarrow} (b,j))\wedge (id_m,\sigma,s,|\sigma|+1)\stackrel{l}{\Rightarrow}(b',j')}\\
  &&\rightarrow \alpha\vdash ptr(b,j)\sim ptr(b',j')& (1, 2 , 3)\\
(5)&\Longleftrightarrow& \alpha\vdash id_c\sim_l id_m&\mbox{Definition}~ \ref{lee},~(4)\\
(6)&\Longrightarrow& \alpha\vdash id_c\sim_e id_m&\mbox{Definition}~ \ref{ee},~(5)
\end{array}
\]}
Induction:
\item[3.] For $id_c[e]$ of type $\tau$, $ExTr(id_c[e])=id_m[ra]$, where $ra=ExTr(e)$, $id_c$, $id_m$, $id_c[e]$ and $id_m[ra]$ are all left-value expressions  while $e$ and $ra$ are both right-value expressions.
{\small\[
\begin{array}{llllr}
(1)&&(\alpha\vdash id_c\sim_e id_m)\wedge(\alpha\vdash e\sim_e ra) &\multicolumn{2}{r}{\mbox{hypothesis}}\\
(2)&& \alpha\vdash M\sim s&\multicolumn{2}{r}{\mbox{given condition}}\\
(3)&&id_c[e]\equiv *(id_c+e)&&C13\\
(4)&&id_c+e\equiv \&(*(id_c+e))&&C2,C6\\
(5)&&\multicolumn{2}{l}{ G,E\vdash id_c[e],M \stackrel{l}{\Rightarrow} (b,j)}&\mbox{assumption}\\
(6)&\Longrightarrow&G,E\vdash *(id_c+e),M \stackrel{l}{\Rightarrow} (b,j)&&(3,5)\\
(7)&\Longrightarrow&G,E\vdash id_c+e,M {\Rightarrow} ptr(b,j)&&C6,~(4,6)\\
(8)&\Longrightarrow&\multicolumn{3}{l}{(G,E\vdash id_c,M \stackrel{l}{\Rightarrow} (b,0))\wedge (G,E\vdash e,M\Rightarrow v_1)\wedge }\\
&&j=v_1\cdot\mathit{sizeof}(\tau)&&C1,~C5,~C8,~(7)\\
(9)&&\multicolumn{2}{l}{ (id_m[ra], \sigma,s,|\sigma|+1)\stackrel{l}{\Rightarrow} (b',j')}&\mbox{assumption}\\
(10)&\Longrightarrow&\multicolumn{3}{l}{((id_m, \sigma,s,|\sigma|+1)\stackrel{l}{\Rightarrow} (b',0))\wedge (ra, \sigma, s, |\sigma|+1) \Downarrow  n_1\wedge}\\
&&\multicolumn{2}{l}{j'=n_1\cdot\mathit{sizeof}(\tau)}&L2,~(9)\\
(11)&&\multicolumn{3}{l}{(G,E \vdash id_c, M \stackrel{l}{\Rightarrow} (b,0)) \wedge((id_m, \sigma,s,|\sigma|+1)\stackrel{l}{\Rightarrow} (b',0))\wedge}\\
&&\multicolumn{2}{l}{(G,E \vdash e, M \Rightarrow v_1)\wedge((ra, \sigma, s, |\sigma|+1) \Downarrow  n_1)}&(8,10)\\
(12)&\Longrightarrow&\multicolumn{3}{l}{\alpha\vdash ptr(b,0)\sim ptr(b',0)\wedge \alpha\vdash v_1\sim n_1}\\
&&\multicolumn{3}{r}{\mbox{Definition}~ \ref{lee},  ~\ref{ree}, ~ \ref{ee},~(1,2,11)}\\
(13)&\Longrightarrow&\alpha\vdash ptr(b,0)\sim ptr(b',0) \wedge v_1=n_1\\
&&\multicolumn{3}{r}{v_1 \mbox{ and } n_1 \mbox{ are both interger values},~(12)}\\
(14)&\Longrightarrow&\multicolumn{3}{l}{\alpha\vdash ptr(b,v_1\cdot\mathit{sizeof}(\tau))\sim ptr(b',n_1\cdot\mathit{sizeof}(\tau))}\\
&&\multicolumn{3}{r}{\mbox{Lemma}~ \ref{adre},(13)}\\
(15)&\Longrightarrow&\alpha \vdash ptr(b,j)\sim ptr(b',j')&\multicolumn{2}{r}{(8,10)}\\
%(16)&&\multicolumn{3}{l}{ (\alpha\vdash M\sim s )\wedge(G,E\vdash id_c[e],M \stackrel{l}{\Rightarrow} (b,j))\wedge}\\
% &&\multicolumn{2}{l}{ (id_m[ra], \sigma,s,|\sigma|+1)\stackrel{l}{\Rightarrow} (b',j')}&\mbox{hypothesis}\\
%&\Longrightarrow&\alpha \vdash ptr(b,j)\sim ptr(b',j')&\multicolumn{2}{r}{(5),~(6),~(11)}\\
(16)&\Longleftrightarrow& \alpha\vdash id_c[e]\sim_l id_m[ra] & \multicolumn{2}{r}{\mbox{Definition}~ \ref{lee},~(15)}\\
(17)&\Longrightarrow& \alpha\vdash id_c[e]\sim_e id_m[ra] & \multicolumn{2}{r}{\mbox{Definition}~ \ref{ee},~(16)}
\end{array}
\]}
In a similar way, it can be proved that the conclusion is true for $id_c[e_1][e_2]$.
%$\alpha\vdash id[e_1][e_2]\sim_e ExTr(id[e_1][e_2])$ can be proved.

\item[4.] For $le.x$ of type $\tau$, $ExTr(le.x)=la.y$, where $la=ExTr(le)$ and $y=ExTr(x)$, $le$, $la$, $le.x$ and $la.y$ are all left-value expressions.
    When translating a member $x$ of a struct variable to an expression in MSVL using $ExTr$, we do the following: if $x$ is the $k$th member of a struct $S_c$ with filed list $\varphi$ in the Xd-C program, it is translated to the $k$th member of a struct $S_m$ with filed list $\varphi$ in the MSVL program.
     We assume %that $x$ is the $m$th member of $le$, $y$ the $m$th member of $la$ and
     $\tau_i$ is the type of $i$th member of struct $S_c$ for $0\leq i< k$. Therefore,
    $\mathit{field\_offset}(x,\varphi)=\mathit{field\_offset}(y,\varphi)=\mathit{sizeof}(\tau_1)+\ldots+\mathit{sizeof}(\tau_{k-1})$. For convenience, we denote $\delta'=\mathit{field\_offset}(x,\varphi)$ in the following.\\[-6ex]

    {\small\[
\begin{array}{llllr}
(1)& & \alpha\vdash le\sim_e la  &\multicolumn{2}{r}{\mbox{hypothesis}}\\
(2)&&\alpha\vdash M\sim s&\multicolumn{2}{r}{\mbox{given condition}}\\
(3)&&\multicolumn{3}{l}{ (G,E\vdash le.x,M \stackrel{l}{\Rightarrow} (b,j))\wedge  (la.y,\sigma,s,|\sigma|+1)\stackrel{l}{\Rightarrow} (b',j')~~\mbox{assumption}}\\
(4)&\Longrightarrow&\multicolumn{3}{l}{(G,E\vdash le,M \stackrel{l}{\Rightarrow} (b,j_1))\wedge j=j_1+\delta'\wedge(la, \sigma,s,|\sigma|+1)\stackrel{l}{\Rightarrow} (b',j_1')\wedge}\\
 &&j'=j_1'+\delta'&\multicolumn{2}{r}{ C3, ~\mbox{L4},~(3)}\\
%(3)&&\multicolumn{3}{l}{(\alpha\vdash M\sim s)\wedge(G,E \vdash le, M \stackrel{l}{\Rightarrow} (b,j_1))\wedge((la, \sigma,s,|\sigma|+1)\stackrel{l}{\Rightarrow} (b',j_1'))}\\
(5)&\Longrightarrow& \multicolumn{3}{l}{\alpha\vdash ptr(b,j_1)\sim ptr(b',j_1')\wedge j=j_1+\delta' \wedge j'=j_1'+\delta'}\\
&&\multicolumn{3}{r}{\mbox{Definition}~ \ref{lee}, ~ \ref{ee},~(1,2,4)}\\
%(4)&\supset&\multicolumn{2}{l}{\alpha\vdash ptr(b,j_1+\delta')\sim ptr(b',j_1'+\delta')}&\\
(6)&\Longrightarrow&\alpha\vdash ptr(b,j)\sim ptr(b',j')&\multicolumn{2}{r}{\mbox{Lemma}~ \ref{adre},~(5)}\\
%(4)&&\multicolumn{3}{l}{\forall M,s,\sigma,b,j,b',j'.(\alpha\vdash M\sim s )\wedge(G,E\vdash le.x,M \stackrel{l}{\Rightarrow} (b,j))\wedge}\\
% &&\multicolumn{3}{l}{ (la.y, \sigma,s,|\sigma|+1)\stackrel{l}{\Rightarrow} (b',j')}\\
%&\supset&\multicolumn{2}{l}{\alpha\vdash ptr(b,j)\sim ptr(b',j')}&(1),~(5)\\
(7)&\Longleftrightarrow&  \alpha\vdash le.x\sim_l la.y  & \multicolumn{2}{r}{\mbox{Definition}~ \ref{lee},~(6)}\\
(8)&\Longrightarrow&  \alpha\vdash le.x\sim_e la.y  & \multicolumn{2}{r}{\mbox{Definition}~ \ref{ee},~(7)}
\end{array}
\]}\noindent
Similarly, we can prove that the conclusion is true for $le\rightarrow x$.
%$\alpha\vdash le\rightarrow x \sim_e ExTr(le)\rightarrow ExTr(x)$ can be proved.

\item[5.] For $*e$, $ExTr(*e)=*pt$, where $pt=ExTr(e)$, $*e$ and $*pt$ are both left-value expressions  while $e$ and $pt$ are both right-value expressions.
       {\small\[
\begin{array}{llllr}
(1)&&\alpha\vdash e\sim_e pt&\multicolumn{2}{r}{\mbox{hypothesis}}\\
(2)&&\alpha\vdash M\sim s&\multicolumn{2}{r}{\mbox{given condition}}\\
(3)&&\multicolumn{2}{l}{e\equiv \&(*e)\wedge pt\equiv\&(*pt)}&C2,~C6\\
(4)&&\multicolumn{2}{l}{(G,E\vdash *e,M \stackrel{l}{\Rightarrow}  (b, j))\wedge (*pt,\sigma,s,|\sigma|+1)\stackrel{l}{\Rightarrow}  (b',j')}&\mbox{assumption}\\
(5)&\Longrightarrow&\multicolumn{3}{l}{(G,E \vdash e, M  {\Rightarrow} ptr(b,j))\wedge(pt, \sigma,s,|\sigma|+1){\Downarrow} ptr(b',j')}\\
&&\multicolumn{3}{r}{C6,~\mbox{R3},~(3,4)}\\
(6)&\Longrightarrow& \alpha\vdash ptr(b,j)\sim ptr(b',j')&\multicolumn{2}{r}{\mbox{Definition}~\ref{lee},~ \ref{ree}, ~\ref{ee}, ~(1,2,5)}\\
(7)&\Longleftrightarrow& \alpha\vdash *e\sim_l *pt & \multicolumn{2}{r}{\mbox{Definition}~ \ref{lee},~(6)}\\
(8)&\Longrightarrow& \alpha\vdash *e\sim_e *pt & \multicolumn{2}{r}{\mbox{Definition}~ \ref{ee},~(7)}
\end{array}
\]}
\item[6.] For $\&le$, $ExTr(\&le)=\&la$, where $la=ExTr(le)$, $\&le$ and $\&la$ are both right-value expressions  while $le$ and $la$ are both left-value expressions.
      {\small\[
\begin{array}{llllr}
(1)&&\alpha\vdash le\sim_e la&\multicolumn{2}{r}{\mbox{hypothesis}}\\
(2)&&\alpha\vdash M\sim s&\multicolumn{2}{r}{\mbox{given condition}}\\
(3)&&\multicolumn{2}{l}{le\equiv *(\&le)\wedge la\equiv*(\& la)}&C2,~C6\\
(4)&&\multicolumn{2}{l}{(G,E\vdash \&le,M {\Rightarrow}  ptr(b, j))\wedge (\&la, \sigma,s,|\sigma|+1){\Downarrow}  ptr(b',j')}\\
&&\multicolumn{3}{r}{\mbox{assumption}}\\
(5)&\Longrightarrow&\multicolumn{3}{l}{ (G,E \vdash le, M  \stackrel{l}{\Rightarrow} (b,j))\wedge((la, \sigma,s,|\sigma|+1)\stackrel{l}{\Rightarrow} (b',j'))}\\
&&\multicolumn{3}{r}{C2,~\mbox{L6},~(3,4)}\\
(6)&\Longrightarrow& \alpha\vdash ptr(b,j)\sim ptr(b',j')&\multicolumn{2}{r}{\mbox{Definition} ~\ref{lee},~\ref{ee}, ~(1,2,5)}\\
(7)&\Longleftrightarrow& \alpha\vdash \&le\sim_r \& la & \multicolumn{2}{r}{\mbox{Definition}~ \ref{ree},~(6)}\\
(8)&\Longrightarrow& \alpha\vdash \&le\sim_e \& la & \multicolumn{2}{r}{\mbox{Definition}~ \ref{ee},~(7)}
\end{array}
\]}
 \item[7.] For $(\tau) e$, $ExTr((\tau) e)=(\tau) ra$,  where $ra=ExTr(e)$, $e$, $ra$, $(\tau)e$
   and $(\tau)ra$ are all right-value expressions.
    {\small\[
\begin{array}{llllr}
(1)& &\alpha\vdash e\sim_e ra&\multicolumn{2}{r}{\mbox{hypothesis}}\\
%(2)&\Longleftrightarrow &\alpha \vdash e\sim_l ra\vee\alpha\vdash e\sim_r ra&\multicolumn{2}{r}{\mbox{Definition}~ \ref{ee},~(1)}\\
(2)&&\alpha\vdash M\sim s&\multicolumn{2}{r}{\mbox{given condition}}\\
(3)&&\multicolumn{2}{l}{(G,E\vdash (\tau)e,M {\Rightarrow} v)\wedge((\tau)ra, \sigma,s,|\sigma|+1){\Downarrow}  n}&\mbox{assumption}\\
(4)&\Longrightarrow&\multicolumn{3}{l}{ (G,E \vdash e, M  {\Rightarrow} v_1)\wedge v=(\tau)v_1\wedge((ra, \sigma,s,|\sigma|+1){\Downarrow} n_1)\wedge n=(\tau)n_1 }\\
&&&\multicolumn{2}{r}{C11,~\mbox{R4},~(3)}\\
(5)&\Longrightarrow& \alpha\vdash v_1\sim n_1\wedge v=(\tau)v_1\wedge n=(\tau)n_1 &\multicolumn{2}{r}{\mbox{Definition} ~\ref{lee},~ \ref{ree},~\ref{ee},~ (1,2,4)}\\
(6)&\Longrightarrow& \multicolumn{3}{l}{v_1= n_1\wedge v=(\tau)v_1\wedge n=(\tau)n_1}\\
&&\multicolumn{3}{r}{\mbox{$v_1$ and $n_1$ are both non-pointer values,}~(5)}\\
(7)&\Longrightarrow&v=n&&(6)\\
(8)&\Longrightarrow& \alpha\vdash v\sim n&\multicolumn{2}{r}{\mbox{definition of } \alpha\vdash v\sim n,~(7)}\\
(9)&\Longleftrightarrow& \alpha\vdash (\tau)e\sim_r  (\tau)ra & \multicolumn{2}{r}{\mbox{Definition}~ \ref{ree},~(8)}\\
(10)&\Longrightarrow& \alpha\vdash (\tau)e\sim_e  (\tau)ra & \multicolumn{2}{r}{\mbox{Definition}~ \ref{ee},~(9)}
\end{array}
\]}\noindent
In a similar way, $\alpha\vdash op_1 ~e_1\sim_e op_1~ ExTr(e_1)$ can be proved.
 \item[8.] For $e_1~aop~e_2$, $ExTr(e_1 ~aop ~e_2)=ra_1 ~aop ~ra_2$, where $ra_1=ExTr(e_1)$, $ra_2=ExTr(e_2)$ and all expressions are right-value expressions. To prove the conclusion, the following three cases need to be taken into account.

     Case 1: $e_1$, $e_2$, $ra_1$ and $ra_2$ are of non-pointer type.
         {\small\[
\begin{array}{llllr}
(1)&&\alpha\vdash e_i\sim_e ra_i~~~(i=1,2)&\multicolumn{2}{r}{\mbox{hypothesis}}\\
(2)&&\alpha\vdash M\sim s&\multicolumn{2}{r}{\mbox{given condition}}\\
(3)&&\multicolumn{3}{l}{(G,E\vdash e_1 ~aop ~e_2,M {\Rightarrow} v)\wedge (ra_1 ~aop ~ra_2, \sigma,s,|\sigma|+1){\Downarrow}  n}\\
&&\multicolumn{3}{r}{\mbox{assumption}}\\
(4)&\Longrightarrow&\multicolumn{3}{l}{(G,E \vdash e_1, M  {\Rightarrow} v_1)\wedge(G,E \vdash e_2, M  {\Rightarrow} v_2)\wedge v=v_1~aop~v_2\wedge}\\
&&\multicolumn{3}{l}{((ra_1, \sigma,s,|\sigma|+1){\Downarrow} n_1)\wedge((ra_2, \sigma,s,|\sigma|+1){\Downarrow} n_2)\wedge n=n_1~aop~n_2}\\
&&&\multicolumn{2}{r}{C8,~\mbox{R6},~(3)}\\
(5)&\Longrightarrow&\multicolumn{3}{l}{ \alpha\vdash v_1\sim n_1\wedge \alpha\vdash v_2\sim n_2\wedge v=v_1~aop~v_2\wedge n=n_1~aop~n_2}\\ &&\multicolumn{3}{r}{\mbox{Definition}~\ref{lee},~ \ref{ree},~ \ref{ee},~ (1, 2, 4)}\\
(6)&\Longrightarrow& \multicolumn{3}{l}{v_1= n_1\wedge v_2=n_2\wedge v=v_1~aop~v_2\wedge n=n_1~aop~n_2}\\
&&\multicolumn{3}{r}{\mbox{$v_1$, $v_2$, $n_1$ and $n_2$ are all non-pointer values,}~(5)}\\
(7)&\Longrightarrow& v=n&&(6)\\
(8)&\Longrightarrow& \alpha\vdash v\sim n&\multicolumn{2}{r}{\mbox{definition of } \alpha\vdash v\sim n,~ (7)}\\
(9)&\Longleftrightarrow& \alpha\vdash e_1 ~aop ~e_2\sim_r  ra_1 ~aop ~ra_2 & \multicolumn{2}{r}{\mbox{Definition}~ \ref{ree},~(8)}\\
(10)&\Longrightarrow& \alpha\vdash e_1 ~aop ~e_2\sim_e  ra_1 ~aop ~ra_2 & \multicolumn{2}{r}{\mbox{Definition}~ \ref{ee},~(9)}
\end{array}
\]}\noindent
 Case 2: both $e_1$ and $ra_1$ are of pointer type $\tau*$ while both $e_2$ and $ra_2$ are of integer type. Here, $aop=+\mid -$.
{\small\[
\begin{array}{llllr}
(1)&&\alpha\vdash e_i\sim_e ra_i~~~(i=1,2)&\multicolumn{2}{r}{\mbox{hypothesis}}\\
(2)&&\alpha\vdash M\sim s&\multicolumn{2}{r}{\mbox{given condition}}\\
(3)&& \multicolumn{2}{l}{G,E\vdash e_1 ~aop ~e_2,M {\Rightarrow} ptr(b,j)}&\mbox{assumption}\\
(4)&\Longrightarrow&\multicolumn{2}{l}{(G,E \vdash e_1, M  {\Rightarrow} ptr(b,j_1))\wedge (G,E \vdash e_2, M  {\Rightarrow} v_2)\wedge}\\
&& j=j_1+v_2*\mathit{sizeof}(\tau)&\multicolumn{2}{r}{C8,~(3)}\\
(5)&&
(ra_1 ~aop ~ra_2, \sigma,s,|\sigma|+1){\Downarrow}  ptr(b',j')&\multicolumn{2}{r}{\mbox{assumption}}\\
(6)&\Longrightarrow&\multicolumn{3}{l}{(ra_1, \sigma,s,|\sigma|+1){\Downarrow} ptr(b',j_1')\wedge(ra_2, \sigma,s,|\sigma|+1){\Downarrow} n_2\wedge}\\
&& j'=j_1'+n_2*\mathit{sizeof}(\tau)&\multicolumn{2}{r}{\mbox{R6},~(5)}\\
(7)&&\multicolumn{2}{l}{(G,E \vdash e_1, M  {\Rightarrow} ptr(b,j_1))\wedge (G,E \vdash e_2, M  {\Rightarrow} v_2)\wedge}&\\
&&\multicolumn{2}{l}{(ra_1, \sigma,s,|\sigma|+1){\Downarrow} ptr(b',j_1')\wedge(ra_2, \sigma,s,|\sigma|+1){\Downarrow} n_2}&(4,6)\\
(8)&\Longrightarrow& \alpha\vdash ptr(b,j_1)\sim ptr(b',j_1)\wedge \alpha\vdash v_2\sim n_2\\
&&\multicolumn{3}{r}{\mbox{Definition}~ \ref{lee},~ \ref{ree},~ \ref{ee}, ~(1,2,7)}\\
(9)&\Longrightarrow& \multicolumn{3}{l}{\alpha\vdash ptr(b,j_1)\sim ptr(b',j_1)\wedge v_2= n_2}\\
&&\multicolumn{3}{r}{\mbox{$v_2$ and $n_2$ are both non-pointer values, (8)}}\\
(10)&\Longrightarrow& \multicolumn{3}{l}{\alpha\vdash ptr(b,j_1+v_2*\mathit{sizeof}(\tau))\sim ptr(b,j_1'+n_2*\mathit{sizeof}(\tau))}\\
&&\multicolumn{3}{r}{\mbox{Lemma } \ref{adre},(9)}\\
(11)&\Longrightarrow& \alpha \vdash ptr(b,j)\sim ptr(b',j')&\multicolumn{2}{r}{(4,6,10)}\\
%(12)&&\multicolumn{3}{l}{(\alpha\vdash M\sim s )\wedge(G,E\vdash e_1 ~aop ~e_2,M {\Rightarrow} ptr(b,j))\wedge}\\
 %&&(ra_1 ~aop ~ra_2, \sigma,s,|\sigma|+1){\Downarrow} ptr(b',j')&\multicolumn{2}{r}{ \mbox{(1),~(5)}}\\
%(13)&\Longrightarrow&\alpha\vdash ptr(b,j)\sim ptr(b',j')&\multicolumn{2}{r}{(1),~(2),~(7)}\\
(12)&\Longleftrightarrow&\alpha\vdash e_1 ~aop ~e_2\sim_r ra_1 ~aop ~ra_2&\multicolumn{2}{r}{\mbox{Definition}~ \ref{ree},~(11)} \\
(13)&\Longrightarrow& \alpha\vdash e_1 ~aop ~e_2\sim_e  ra_1 ~aop ~ra_2 & \multicolumn{2}{r}{\mbox{Definition}~ \ref{ee},~(12)}
\end{array}
\]}\noindent
Case 3: both $e_2$ and $ra_2$ are of pointer type $\tau*$ while both $e_1$ and $ra_1$ are of integer type. The proof is similar to Case 2.

    For $e_1 ~rop ~e_2$, $e_1 ~\mathit{\&\&} ~e_2$ and $e_1 ~\bitor\bitor ~e_2$, similar proofs can be given.

    % $\alpha\vdash e_1 ~rop ~e_2 \sim_e ExTr(e_1) ~rop~ ExTr(e_2)$, $\alpha\vdash e_1 ~\mathit{\&\&} ~e_2 \sim_e ExTr(e_1) ~\wedge~ ExTr(e_2)$ and $\alpha\vdash e_1 ~\bitor\bitor ~e_2 \sim_e ExTr(e_1) ~\vee~ ExTr(e_2)$  can be proved.

\item[9.] For $e_1?e_2:e_3$, $ExTr(e_1?e_2:e_3)=if(b)~then~ ra_2~ else ~ra_3$, where $b=ExTr(e_1)$, $ra_2=ExTr(e_2)$, $ra_3=ExTr(e_3)$ and all expressions are right-value expressions.
    {\small\[
\begin{array}{llllr}
(1)&&\multicolumn{2}{l}{(\alpha\vdash e_1\sim_e b)\wedge (\alpha\vdash e_2\sim_e ra_2)\wedge(\alpha\vdash e_3\sim_e ra_3)}&\mbox{hypothesis}\\
(2)&&\alpha\vdash M\sim s&\multicolumn{2}{r}{\mbox{given condition}}\\
(3)&&\multicolumn{3}{l}{(G,E\vdash e_1?e_2:e_3,M {\Rightarrow} v)\wedge(if(b)~then~ ra_2~ else ~ra_3, \sigma,s,|\sigma|+1){\Downarrow}  n }\\
&&\multicolumn{3}{r}{\mbox{assumption}}\\
(4)&\Longrightarrow&\multicolumn{3}{l}{  (G,E \vdash e_1, M  {\Rightarrow} v_1)\wedge(v_1= true\wedge(G,E \vdash e_2, M  {\Rightarrow} v)\vee }\\
&&\multicolumn{3}{l}{v_1=false\wedge(G,E \vdash e_3, M  {\Rightarrow} v))\wedge
(b, \sigma,s,|\sigma|+1){\Downarrow} t \wedge ( t=true \wedge
}\\
&&\multicolumn{3}{l}{ (ra_2, \sigma,s,|\sigma|+1){\Downarrow} n \vee t=false \wedge
(ra_3, \sigma,s,|\sigma|+1){\Downarrow} n)}\\
&&&\multicolumn{2}{r}{C9,~C10,~\mbox{R7, R8},~(3)}\\
(5)&\Longrightarrow&\multicolumn{3}{l}{v_1=t=true\wedge \alpha\vdash n\sim v \vee v_1=t=false \wedge \alpha\vdash n\sim v}\\ &&\multicolumn{3}{r}{\mbox{Definition}~\ref{lee},~ \ref{ree},~\ref{ee},~ (1,2,4)}\\
(6)&\Longrightarrow&\multicolumn{2}{l}{ \alpha\vdash v\sim n}&(5)\\
(7)&\Longleftrightarrow& \alpha\vdash e_1?e_2:e_3\sim_r  if(b)~then~ ra_2~ else ~ra_3 & \multicolumn{2}{r}{\mbox{Definition}~ \ref{ree},~(6)}\\
(8)&\Longrightarrow& \alpha\vdash e_1?e_2:e_3\sim_e  if(b)~then~ ra_2~ else ~ra_3 & \multicolumn{2}{r}{\mbox{Definition}~ \ref{ee},~(7)}
\end{array}
\]}
\item[10.] For a function call $x(e_1,...,e_k)$,  if $x$ points to a user-defined function, $ExTr(x(e_1,...,e_k))=ext~f(ra_1,...,ra_k, RVal)$ otherwise $ExTr(x(e_1,...,\linebreak e_k))=ext~f(ra_1,...,ra_k)$, where $ra_i=ExTr(e_i)$ for $1\leq i\leq k$, $f=ExTr(x)$ and all expressions are right-value expressions. The following two cases need to be considered.

    Case 1: $x(e_1,...,e_k)$ is a call of an external function $extern~ \tau~id(par)$, where $par=(\tau_1~ y_1, ..., \tau_k~ y_k)$.
    By induction hypothesis, $f$ is also a call of $extern~ \tau~id(par)$ but in the form $ext~ f(ra_1,...,ra_k)$.
    {\small\[
\begin{array}{llllr}
(1)&& \alpha\vdash e_i\sim_e ra_i~~~~~~~~(i=1,...,k)&\multicolumn{2}{r}{\mbox{hypothesis}}\\
(2)&&\alpha\vdash M\sim s&\multicolumn{2}{r}{\mbox{given condition}}\\
(3)&&G,E\vdash x(e_1,...,e_k),M {\Rightarrow} v&\multicolumn{2}{r}{\mbox{assumption}}\\
(4)&\Longrightarrow&G,E \vdash x(e_1,...,e_k), M  \stackrel{t}{\Rightarrow} v, M&\multicolumn{2}{r}{C12, (3)}\\
(5)&\Longrightarrow&\multicolumn{3}{l}{\bigwedge_{i=1}^{k}(G,E\vdash e_i,M\Rightarrow v_i)\wedge
(G\vdash id(v_1,...,v_k), M\stackrel{t}{\Rightarrow} v, M)}\\
&&\multicolumn{3}{r}{C36,(4)}\\
(6)&\Longrightarrow&\multicolumn{2}{l}{ \bigwedge_{i=1}^{k}(G,E\vdash e_i,M\Rightarrow v_i)\wedge v=id(v_1,...,v_k)}&C38,(5)\\
(7)&&(ext~f(ra_1,...,ra_k),\sigma,s,|\sigma|+1)\Downarrow n&\multicolumn{2}{r}{\mbox{assumption}}\\
(8)&\Longrightarrow&\multicolumn{2}{l}{\bigwedge_{i=1}^{k}(ra_i,s,\sigma,|\sigma|+1)\Downarrow n_i\wedge n=id(n_1,...,n_k)}&\mbox{R11},(7)\\
(9)&&\multicolumn{3}{l}{\bigwedge_{i=1}^{k}(G,E\vdash e_i,M\Rightarrow v_i)\wedge v=id(v_1,...,v_k)
\wedge}\\
&&\bigwedge_{i=1}^{k}(ra_i,s,\sigma,|\sigma|+1)\Downarrow n_i\wedge n=id(n_1,...,n_k)&\multicolumn{2}{r}{\mbox{(6,8)}}\\
(10)&\Longrightarrow&\multicolumn{3}{l}{\bigwedge_{i=1}^{k}\alpha\vdash v_i\sim n_i\wedge v=id(v_1,...,v_k) \wedge n=id(n_1,...,n_k)}\\
&&\multicolumn{3}{r}{\mbox{Definition}~\ref{lee}, ~\ref{ree}, ~ \ref{ee}, ~(1,2,9)}\\
(11)&\Longrightarrow&\multicolumn{2}{l}{\alpha\vdash v\sim n}&(10)\\
%(12)&&\multicolumn{3}{l}{(G,E\vdash x(e_1,...,e_m),M {\Rightarrow} v)\wedge} \\
%&&(f(ra_1,...,ra_m),\sigma,s,|\sigma|+1)\Downarrow n&\multicolumn{2}{r}{\mbox{hypothesis}}\\
%(13)&\Longrightarrow&\alpha\vdash v\sim n&\multicolumn{2}{r}{(3),~(4),~(7)}\\
(12)&\Longleftrightarrow&\alpha\vdash x(e_1,...,e_k)\sim_r ext~f(ra_1,...,ra_k)&\multicolumn{2}{r}{\mbox{Definition}~ \ref{ree},~(11)} \\
(13)&\Longrightarrow& \alpha\vdash x(e_1,...,e_k)\sim_e  ext~f(ra_1,...,ra_k) & \multicolumn{2}{r}{\mbox{Definition}~ \ref{ee},~(12)}
\end{array}
\]}\noindent
     Case 2: $x$ points to a user-defined function $\tau~id(par)\{dcl;cs\}$,  where $par=(\tau_1~ y_1, ..., \tau_k~ y_k)$ and $dcl;cs$ is the body of the function.
     By induction hypothesis, $f$ points to $function~id( \tau_1~ y_1, ..., \tau_k~ y_k, \tau ~RVal) \{\phi\}$, where $\phi$ is translated from $dcl;cs$ ($\phi=StmtTr(dcl;cs)$ see Algorithm \ref{StmtTr}).
       {\small\[
\begin{array}{llllr}
(1)&& \multicolumn{2}{l}{\alpha\vdash e_i\sim_e ra_i~~~~~~~~(i=1,...,k)}&\mbox{hypothesis}\\
(2)&&\alpha\vdash M\sim s&\multicolumn{2}{r}{\mbox{given condition}}\\
(3)&&\multicolumn{2}{l}{G,E\vdash x(e_1,...,e_k),M {\Rightarrow} v}&\mbox{assumption}\\
(4)&\Longrightarrow&\multicolumn{3}{l}{\bigwedge_{i=1}^{k}(G,E\vdash e_i,M\Rightarrow v_i)\wedge
(G\vdash id(v_1,...,v_k), M\stackrel{t}{\Rightarrow} v, M)}\\
&&\multicolumn{3}{r}{C12,~C36,~(3)}\\
(5)&\Longrightarrow&\multicolumn{3}{l}{ \bigwedge_{i=1}^{k}(G,E\vdash e_i,M\Rightarrow v_i)\wedge\mathit{alloc\_vars}(M,par\textbf{+}dcl,E)=(M_1,b^*)\wedge}\\
&&\multicolumn{3}{l}{ v_{arg}=(v_1,...,v_k)\wedge \mathit{bind\_params}(E,M_1,par, v_{args})=M_2\wedge}\\
&&\multicolumn{3}{l}{(G,E\vdash cs,M_2 \stackrel{t}{\Rightarrow} Return(v), M_3)\wedge  (Return(v),\tau\#v)}\\
&&&&C37,~(4)\\
&&\multicolumn{3}{l}{\mbox{\{here $par\textbf{+}dcl$, $alloc\_vars$,   $bind\_params$ and $\tau\#v$ are}}\\
 &&\multicolumn{3}{l}{\mbox{borrowed from \cite{blazy2009mechanized}, see Appendix A.\}}}\\
%&&\multicolumn{3}{l}{\mbox{$par$ and $dcl$, and concatenate them together.\}}}\\
(6)&&\multicolumn{2}{l}{(ext~f(ra_1,...,ra_k,RVal),\sigma,s,|\sigma|+1)\Downarrow n}&\mbox{assumption}\\
(7)&\Longrightarrow&\multicolumn{3}{l}{\bigwedge_{i=1}^{k}(ra_i,s,\sigma,|\sigma|+1)\Downarrow n_i\wedge}\\
&&\multicolumn{3}{l}{(\phi \wedge \bigwedge_{j=1}^{k} y_j\Leftarrow n_j, \epsilon, s'_0, 0)\stackrel{*}{\rightarrow}(true, \sigma', \emptyset, |\sigma'|+1)}\\
&&\multicolumn{2}{l}{\wedge s'_0=s \wedge n=s'^r_{|\sigma'|}(RVal)}&\mbox{R10},~(6)\\
(8)&&\multicolumn{3}{l}{(G,E\vdash x(e_1,...,e_k),M {\Rightarrow} v)\wedge} \\
&&\multicolumn{2}{l}{(ext~f(ra_1,...,ra_k,RVal),\sigma,s,|\sigma|+1)\Downarrow n}&(3,6)\\
(9)&\Longrightarrow&\alpha\vdash v\sim n&\multicolumn{2}{r}{\mbox{proved in Section } \ref{eqs},~(2,5,7)}\\
(10)&\Longleftrightarrow&\multicolumn{2}{l}{\alpha\vdash x(e_1,...,e_k)\sim_r ext~f(ra_1,...,ra_k, RVal)}&\mbox{Definition}~ \ref{ree},~(9) \\
(11)&\Longrightarrow& \multicolumn{2}{l}{\alpha\vdash x(e_1,...,e_k)\sim_e  ext~f(ra_1,...,ra_k, RVal)} &\mbox{Definition}~ \ref{ee},~(10)
\end{array}
\]}
\end{itemize}
\end{Proof}
\rightline{$\square$}

\subsection{Proof of statement equivalence} \label{eqs}
\begin{Thm}\label{s-eq}
Suppose an Xd-C statement $cs$ is transformed to an MSVL statement $ms$ by Algorithm \ref{StmtTr} along with transforming an Xd-C program to an MSVL program. That is, $ms=StmtTr(cs)$. For a given $\alpha$, any $M\in \mathbb{M}$ and $s_i\in \mathbb{S}$, if $\alpha\vdash M\sim s_i$, then $\alpha\vdash cs\sim_s ms$.
\end{Thm}

\begin{Proof}
The proof proceeds by induction on the rules of operational semantics of statements.
To do so, we consider two cases: terminating and diverging statement equivalences.
That is, \\[4ex]
(1) Terminating case:
{\small $$G,E\vdash cs,M \stackrel{t}{\Rightarrow} (out, M')\Longrightarrow P(cs,M,M',out)$$}\noindent
where
{\small
$$\begin{array}{lll}
P(cs,M,M',out)\Longleftrightarrow& (\alpha\vdash M\sim s_i\Longrightarrow \\
&(ms,\sigma_{i-1},s_i,i)\stackrel{*}{\rightarrow} (true,\sigma,\emptyset, |\sigma|+1)\wedge\\
&\alpha\vdash M'\sim s_{|\sigma|})&\mbox{TER}
\end{array}
$$}\noindent
(2) Diverging case:
{\small $$  G,E \vdash cs, M \stackrel{T}{\Rightarrow} \infty \Longrightarrow P'(cs,M)$$}\noindent
%where
%{\small
%$$\begin{array}{ll}
%P'(cs,M)\Leftrightarrow& (\forall s_i.(\alpha\vdash M\sim s_i)\Rightarrow \\
%&\exists \sigma. (StmtTr(cs),\sigma_{i-1},s_i,i)\stackrel{*}{\rightarrow} (true,\sigma,\emptyset, |\sigma|+1)\wedge\\
%&\alpha\vdash M'\sim s_{|\sigma|})
%\end{array}
%$$}
where
{\small
$$\begin{array}{lll}
P'(cs,M)\Longleftrightarrow& ( \alpha\vdash M\sim s_i \Longrightarrow (cs,M)\cong (cs_1;cs_2;...,M) \wedge\\
& (ms,\sigma_{i-1},s_i,i)\stackrel{*}{\rightarrowtail} (ms_1;ms_2;..., \sigma_{i-1},s_i,i)\\
&\wedge \bigwedge_{j=1}^{\infty}(\alpha\vdash cs_j\sim_t ms_j))&\mbox{DV}
\end{array}
$$}\noindent

%The proof of the terminating statement equivalence is given as follows:
Case 1:\\[-6ex]

Base:
\begin{itemize}
  \item [1.] For rule $T1$ w.r.t. a null statement `$;$',  the conclusion is trivially true.
  %In an Xd-C program, for any $M$ and $M'$, $(G,E\vdash ;,M \stackrel{\epsilon}{\Rightarrow} Normal, M )$ (rule $C13$) while in an MSVL program, for any $s_i$, $(\mbox{\ttfamily empty}, \sigma_{i-1},s_i,i)\rightarrow (true, \sigma_i,\emptyset,i+1)$ (rule TR2), where $\mbox{\ttfamily empty}=StmtTr(;)$. For any $\alpha$ and $s_i$, if $\alpha\vdash M\sim s_i$, then $\alpha\vdash M\sim s_{|\sigma_i|}$ due to $|\sigma_i|=i$. Therefore, $P(;,M,M',Normal)$ holds.
  \item [2.] For rule $T2$ w.r.t. ``{\ttfamily break}$;$'', $StmtTr(\mbox{\ttfamily break};)=$``$break:=1$''.
  {\small\[
  \begin{array}{lllr}
 (1)&&(break:=1, \sigma_{i-1}, s_i, i)\\
    &  & \rightarrowtail (\bigcirc (break\Leftarrow 1 \wedge \mbox{\ttfamily empty}), \sigma_{i-1}, s_{i}, i)& \mbox{UASS}\\
    &  & \rightarrow (break\Leftarrow 1 \wedge \mbox{\ttfamily empty}, \sigma_i, s_{i+1}, i+1)& \mbox{TR1}\\
     &  & \rightarrowtail ( \mbox{\ttfamily empty}, \sigma_i, (s_{i+1}^l,s_{i+1}^r[1/break]), i+1)&\mbox{MIN1}\\
     & &\rightarrow(true, \sigma_{i+1}, \emptyset, i+2)&\mbox{TR2}\\

 %(2)&\vdash& \forall y\in Dom(s_i)\setminus \{break, continue, return, RVal\}.\\
% &&(\bigcircp y,\sigma_{i},s_{i+1},i+1)\Downarrow  s_i^r(y)&\mbox{R9}\\
 (2)&&\multicolumn{2}{l}{\forall y\in Dom(s_i)\setminus \{break, continue, return, RVal\}}\\
  & &( \mbox{\ttfamily empty}, \sigma_i, s_{i+1}, i+1)\\
  &&\rightarrowtail( \mbox{\ttfamily empty}, \sigma_i,  (s_{i+1}^l,s^r_{i+1}[s_i^r(y)/y]), i+1)& \mbox{MIN2}\\
  (3)&\Longrightarrow&\multicolumn{2}{l}{\forall y\in Dom(s_{i+1})\setminus \{break, continue, return, RVal\} }\\
   && s_{i+1}^r(y) =s_i^r(y)\wedge s_{i+1}^l(y)=s_i^l(y)&(2)\\
   (4)& &\alpha\vdash M\sim s_i&\mbox{given condition}\\
    (5) &\Longrightarrow& \alpha\vdash M\sim s_{i+1}&\mbox{Definition} ~\ref{se},~(3,4)\\
   (6)%&&\forall M. (G,E\vdash \mbox{\ttfamily break};,M \stackrel{t}{\Rightarrow} Break, M)\\
   &\Longleftrightarrow& P(\mbox{\ttfamily break};,M,M,Break)&\mbox{TER},~(1 , 5)
  \end{array}
  \]}\noindent
  Note that, $s_i\neq s_{i+1}$ in the above proof even so $\alpha\vdash M\sim s_i$ and
$\alpha\vdash M\sim s_{i+1}$ under the consideration without variables $break$, $continue$, $return$ and $RVal$.

   In the same way, we can prove that the conclusions are all true for rules $T3$, $T4$ and $T5$ w.r.t. ``{\ttfamily continue}$;$'',
   ``{\ttfamily return}$;$'' and ``{\ttfamily return} $e;$'', respectively.
 \item[3.] For rule $T6$ w.r.t. ``$le=e;$'', $StmtTr(le=e;)=$``$la:=ra$'', where $la=ExTr(le)$ and $ra=ExTr(e)$.
 {\small\[
  \begin{array}{llllr}
  (1)&&\alpha\vdash le\sim_e la&\multicolumn{2}{r}{\mbox{Theorem}~\ref{e-eq}}\\
  (2)&&\alpha\vdash e\sim_e ra&\multicolumn{2}{r}{\mbox{Theorem}~\ref{e-eq}}\\
  (3)&&\alpha\vdash M\sim s_i &\multicolumn{2}{r}{\mbox{given condition}}\\
  (4)&&\multicolumn{2}{l}{ (G,E\vdash le,M \stackrel{l}{\Rightarrow} (b,j))\wedge(la, \sigma_{i-1},s_i,i)\stackrel{l}{\Rightarrow} (b',j')}&\mbox{assumption}\\
(5)&\Longrightarrow&\alpha \vdash ptr(b,j)\sim ptr(b',j')&\multicolumn{2}{r}{\mbox{Definition}~\ref{lee}, ~\ref{ee},~(1,3,4)}\\
  (6)&&\multicolumn{2}{l}{(G,E\vdash e,M  {\Rightarrow} v)\wedge(ra, \sigma_{i-1},s_i,i) {\Downarrow} n}&\mbox{assumption}\\
(7)&\Longrightarrow&\alpha \vdash v\sim n&\multicolumn{2}{r}{\mbox{Definition}~\ref{ree}, ~\ref{ee},~(2,3,6)}\\
 \multicolumn{5}{l}{\mbox{In the MSVL program, we have}}\\
 (8)&& s_i^l(x_m)=(b',j')&\multicolumn{2}{r}{\mbox{assumption}}\\
 (9)&& \multicolumn{3}{l}{(la:=ra, \sigma_{i-1}, s_i, i)} \\
 &&\multicolumn{2}{l}{\rightarrowtail(\bigcirc (x_m\Leftarrow n \wedge \mbox{\ttfamily empty}), \sigma_{i-1}, s_{i}, i)}&\mbox{UASS, (8)}\\
      && \multicolumn{2}{l}{\rightarrow ((x_m\Leftarrow n \wedge \mbox{\ttfamily empty}), \sigma_i, s_{i+1}, i+1)}& \mbox{TR1} \\
       &&\multicolumn{2}{l}{\rightarrowtail ( \mbox{\ttfamily empty}, \sigma_i, (s_{i+1}^l,s_{i+1}^r[n/x_m]), i+1)}& \mbox{MIN1} \\
 %    & &\rightarrow(true, \sigma_{i+1}, \emptyset, i+2)&\multicolumn{2}{r}{\mbox{TR2}}\\
 (10) %&&UA\wedge s_i^l(x)=(b',j')& \multicolumn{2}{r}{\mbox{hypothesis}}\\
 &\Longrightarrow& \multicolumn{2}{l}{s_{i+1}^l(x_m)=(b',j') \wedge s_{i+1}^r(x_m)=n} &(9)\\
\multicolumn{5}{l}{\mbox{Whereas in the Xd-C program, we have}}\\
 (11) &&\multicolumn{3}{l}{(\alpha\vdash M\sim s_i) \wedge s_i^l(x_m)=(b',j')\wedge(\alpha\vdash ptr(b,j)\sim ptr(b',j'))}\\
 &&\multicolumn{3}{r}{ (3,5,8)}\\
 (12)&\Longrightarrow&G,E\vdash x_c, M \stackrel{l}{\Rightarrow} (b,j)&\multicolumn{2}{r}{\mbox{Definition}~\ref{se},~\alpha\mbox{ is an injective function}}\\
%AS&\DEF&\multicolumn{3}{l}{(G,E \vdash le, M \stackrel{l}{\Rightarrow} (b,j))\wedge (G,E \vdash e, M \Rightarrow v) \wedge}\\ &&\multicolumn{3}{l}{storeval(type(le),M,(b,j),v)= M'}\\
% &&\multicolumn{3}{l}{(G,E \vdash le=e;, M \stackrel{\epsilon}{\Rightarrow} Normal,  M')}\\
 %&&\multicolumn{3}{l}{\forall y.(G,E \vdash y, M \stackrel{l}{\Rightarrow} (b_y,j_y)) \wedge (G,E\vdash x, M {\Rightarrow} v_y)}\\
(13)&&\multicolumn{2}{l}{storeval(type(le),M,(b,j),v)= M'}&\mbox{assumption} \\
(14)&\Longrightarrow&\multicolumn{3}{l}{(G,E\vdash x_c, M'\stackrel{l}{\Rightarrow} (b,j))\wedge loadval(type(x_c),M',(b,j))=v}\\ &&\multicolumn{3}{r}{T6,~(12,13)}\\
&&\multicolumn{3}{l}{\mbox{\{here $storeval$ and  $loadval$ are borrowed from \cite{blazy2009mechanized}, see Appendix A.\}}}\\
 %&&\multicolumn{3}{l}{\mbox{\}}}\\
(15)&\Longrightarrow&\multicolumn{2}{l}{(G,E\vdash x_c, M'\stackrel{l}{\Rightarrow} (b,j))\wedge (G,E\vdash x_c, M'{\Rightarrow} v)}&C5,(14)\\
\multicolumn{5}{l}{\mbox{For other variables in the Xd-C and MSVL programs, their locations and}}\\
\multicolumn{5}{l}{\mbox{variables are not changed. Thus,}}\\
 (16)%&&\multicolumn{2}{l}{(\alpha\vdash M\sim s_i)\wedge AS }&\mbox{hypothesis}\\
  &\Longrightarrow&\multicolumn{2}{l}{(la:=ra, \sigma_{i-1}, s_i, i)\stackrel{*}{\rightarrow}(true, \sigma_{i+1}, \emptyset, i+2)}&\mbox{TR2},(9)\\
  (17)&\Longrightarrow& \alpha\vdash M'\sim s_{i+1}& \multicolumn{2}{r}{(3,5,7,10,15)}\\
 (18)&\Longleftrightarrow& P(le=e;,M,M', Normal)&\multicolumn{2}{r}{\mbox{TER},~(16,17)}
  \end{array}
  \]}\noindent
  Note that, (16) tells us the MSVL program eventually terminates and (17) indicates that
  final states $M'$ and $s_{i+1}$ are equivalent.

 Induction:
  \item [4.] For rule $T7$ w.r.t. ``$cs_1;cs_2$'',
  $StmtTr(cs_1;cs_2)=$``$ms_1;ms_2$'', if there is no {\ttfamily break}, {\ttfamily return} or {\ttfamily continue} in $cs_1$, where $ms_1=StmtTr(cs_1)$ and $ms_2=StmtTr(cs_2)$.
  {\small\[
  \begin{array}{llllr}
  (1)&&\alpha\vdash M\sim s_i &\multicolumn{2}{r}{\mbox{given condition}}\\
 (2)& &\multicolumn{3}{l}{ (G,E\vdash cs_1,M \stackrel{t_1}{\Rightarrow} Normal,M_1)\wedge P(cs_1,M,M_1,Normal)\wedge}\\
 && \multicolumn{2}{l}{ (G,E \vdash cs_2, M_1 \stackrel{t_2}{\Rightarrow} out,  M_2)\wedge P(cs_2,M_1,M_2, out) }&\mbox{hypothesis}\\
 &&\multicolumn{3}{l}{\mbox{\{here $Normal$ and $out$ are borrowed from \cite{blazy2009mechanized}, see Fig.\ref{semele}.\}}}\\
(3) &\Longrightarrow  & \multicolumn{3}{l}{ (ms_1,\sigma_{i-1},s_i,i)\stackrel{*}{\rightarrow} (true,\sigma_j,\emptyset, j+1)\wedge \alpha\vdash M_1\sim s_{j} \wedge}\\
 &&  \multicolumn{2}{l}{(ms_2,\sigma_{j-1},s_j,i)\stackrel{*}{\rightarrow} (true,\sigma,\emptyset, |\sigma|+1)\wedge \alpha\vdash M_2\sim s_{|\sigma|} }&\mbox{TER},~(1,2)\\
  (4)&\Longrightarrow&  (ms_1;ms_2, \sigma_{i-1}, s_i, i)\\
      & &\stackrel{*}{\rightarrow}(\mbox{\ttfamily empty};ms_2, \sigma_{j-1}, s_j, j)&&(3)\\
      & &\rightarrowtail (ms_2, \sigma_{j-1}, s_j, j)&&\mbox{CHOP}\\
      & &\stackrel{*}{\rightarrow}(true, \sigma, \emptyset,|\sigma|+1))\wedge (\alpha\vdash M_2\sim s_{|\sigma|} )&&(3)\\
  (5)&\Longleftrightarrow& P(cs_1;cs_2,M,M_2,out)&&\mbox{TER},~(4)
  \end{array}
  \]}\noindent
  For this rule, if there are {\ttfamily break}, {\ttfamily return} and {\ttfamily continue} in $cs_1$, the conclusions can be proved in a similar way.
\item[5.] For rule $T8$ w.r.t. ``$cs_1;cs_2$'', $StmtTr(cs_1;cs_2)=$``$ ms_1;${\ttfamily if}$(break=0)$ {\ttfamily then} $\{ms_2\}$ {\ttfamily else \{empty\}}'',
    if there is a ``{\ttfamily break}$;$'' statement in $cs_1$, where $ms_1=StmtTr(cs_1)$ and $ms_2=StmtTr(cs_2)$.
    {\small\[
  \begin{array}{llllr}
  (1)&&\alpha\vdash M\sim s_i&\multicolumn{2}{r}{\mbox{given condition}}\\
 (2)& &\multicolumn{2}{l}{(G,E\vdash cs_1,M \stackrel{t}{\Rightarrow} Break,M')\wedge P(cs_1,M,M',Break)}&\mbox{hypothesis}  \\
 (3)&\Longleftrightarrow  & \multicolumn{2}{l}{(ms_1,\sigma_{i-1},s_i,i)\stackrel{*}{\rightarrow} (true,\sigma_j,\emptyset, j+1)\wedge\alpha\vdash M'\sim s_{j}}&\mbox{TER},~(1,2)\\
  (4)&\Longrightarrow& \multicolumn{3}{l}{ ( (\mbox{\ttfamily $ms_1;$if$(break=0)$then$\{ms_2\}$else\{empty\}}, \sigma_{i-1}, s_i, i)}\\
      & &\multicolumn{3}{l}{\stackrel{*}{\rightarrow}(\wedge \{\mbox{\ttfamily empty}, break\Leftarrow 1 \};\mbox{\ttfamily if$(break=0)$then$\{ms_2\}$else\{empty\}},}\\
      && \sigma_{j-1}, s_j, j)&&(3,4)\\
     &  &\multicolumn{3}{l}{\rightarrowtail (\mbox{\ttfamily empty};\mbox{\ttfamily if$(break=0)$then$\{ms_2\}$else\{empty\}},\sigma_{j-1}, s_j[1/break], }\\
     & &~~~~j)&\multicolumn{2}{r}{\mbox{L1, R1, MIN1}}\\
     && \multicolumn{2}{l}{ \rightarrowtail (\mbox{\ttfamily if$(break=0)$then$\{ms_2\}$else\{empty\}}, \sigma_{j-1}, s_j, j)}& \mbox{CHOP}\\
       &&\multicolumn{3}{l}{\rightarrowtail((break=0 \wedge ms_2) \vee (\neg break=0 \wedge \mbox{\ttfamily empty}), \sigma_{j-1}, s_j, j)~~~~~~~~~~ \mbox{IF}}\\
     &&   \rightarrowtail(\mbox{\ttfamily empty},\sigma_{j-1}, s_j, j)&\multicolumn{2}{r}{\mbox{B3, B4, F1, T1, F2}}\\
      && \rightarrow (true, \sigma_j,\emptyset,j+1))
      \wedge (\alpha\vdash M'\sim s_{j})& \multicolumn{2}{r}{\mbox{TR2},~(3)}\\
  (5)&\Longleftrightarrow& P(cs_1;cs_2,M,M',Break)&&\mbox{TER},~(4)
  \end{array}
  \]}\noindent
   In a similar way, it can be proved that
   {\small\[\begin{array}{ll}
   &(G,E\vdash cs_1,M \stackrel{l}{\Rightarrow} out,M')\wedge out\in\{Return, Return(v), Continue\}\wedge\\
   & P(cs_1,M,M',out)\\
   \Longrightarrow& P(cs_1;cs_2,M,M',out)
   \end{array}
   \]}
%  if $(G,E\vdash cs_1,M \stackrel{t}{\Rightarrow} out, M')$, $out\in\{Return, Return(v), Continue\}$ and $P(cs_1,M,M',Break)$, then $P(cs_1;cs_2,M,M',out)$ holds.
  \item [6.] For rule $T9$ w.r.t. ``{\ttfamily if$(e)\{cs_1\}$else$\{cs_2\}$}'', $StmtTr(${\ttfamily if$(e)\{cs_1\}$else}$\{cs_2\})=$ ``{\ttfamily if$(b)$then$\{ms_1\}$else$\{ ms_2\}$}'', where $b=ExTr(e)$, $ms_1=StmtTr(cs_1)$ and $ms_2=StmtTr(cs_2)$.
       {\small\[
  \begin{array}{llllr}
  (1)& &\alpha\vdash e \sim_e b &\multicolumn{2}{r}{\mbox{Theorem}~ \ref{e-eq}}\\
  (2)& &\alpha\vdash M\sim s_i &\multicolumn{2}{r}{\mbox{given condition}}\\
 (3)& &\multicolumn{3}{l}{(G,E\vdash e,M {\Rightarrow} true)\wedge(G,E\vdash cs_1,M \stackrel{t}{\Rightarrow} out,M')\wedge } \\
 &&P(cs_1,M,M',out)&\multicolumn{2}{r}{\mbox{hypothesis}}\\
 (4)&\Longleftrightarrow  &   \multicolumn{2}{l}{(ms_1,\sigma_{i-1},s_i,i)\stackrel{*}{\rightarrow} (true,\sigma,\emptyset, |\sigma|+1)\wedge \alpha\vdash M'\sim s_{|\sigma|}}&\mbox{TER},~(2,3)\\
  (5)&\Longrightarrow&   (b,\sigma_{i-1},s_i,i)\Downarrow true & \multicolumn{2}{r}{\mbox{Definition}~\ref{ree}, ~\ref{ee},~(1,2,3)}\\
  (6)&\Longrightarrow& ( (\mbox{\ttfamily if$(b)$then$\{ms_1\}$else$\{ms_2\}$}, \sigma_{i-1}, s_i, i)&\\
     & & \rightarrowtail((b\wedge ms_1)\vee (\neg b \wedge ms_2), \sigma_{i-1}, s_{i}, i) &&\mbox{IF}\\
     &&  \rightarrowtail (ms_1, \sigma_{i-1}, s_{i}, i)&\multicolumn{2}{r}{\mbox{B4, T1, F1, F2, (5)}}\\
    &&   \multicolumn{2}{l}{\stackrel{*}{\rightarrow}(true, \sigma, \emptyset,|\sigma|+1))
      \wedge (\alpha\vdash M'\sim s_{|\sigma|})}&(4)\\
  (7)&\Longleftrightarrow& P(\mbox{\ttfamily if$(e)\{cs_1\}$else$\{cs_2\}$},M,M',out)&&\mbox{TER},~(6)
  \end{array}
  \]}\noindent
   In a similar way, for rule $T10$ w.r.t. ``{\ttfamily if$(e)\{cs_1\}$else$\{cs_2\}$}'', it can be proved that
   {\small\[\begin{array}{ll}
   & (G,E\vdash e,M {\Rightarrow} false)\wedge(G,E\vdash cs_2,M \stackrel{t}{\Rightarrow} out,M')\wedge P(cs_2,M,M',out)\\
   \Longrightarrow& P(\mbox{\ttfamily if$(e)\{cs_1\}$else$\{cs_2\}$},M,M',out)
   \end{array}
   \]}
  \item [7.] For rule $T11$ w.r.t. ``{\ttfamily while}$(e)\{cs\}$'', $StmtTr(\mbox{{\ttfamily while}}(e)\{cs\})=$``$\mbox{{\ttfamily while}}(b)\{ms\}$'', where $b=ExTr(e)$ and $ms=StmtTr(cs)$, if there is no {\ttfamily break}, {\ttfamily return} or {\ttfamily continue} in $cs$.
   {\small\[
  \begin{array}{llllr}
  (1)&&\alpha\vdash e \sim_e b &\multicolumn{2}{r}{\mbox{Theorem}~ \ref{e-eq}}\\
  (2)&&\alpha\vdash M\sim s_i &\multicolumn{2}{r}{\mbox{given condition}}\\
 (3)& &\multicolumn{2}{l}{G,E\vdash e,M {\Rightarrow} false}&\mbox{assumption}\\
  (4)&\Longrightarrow& (b,\sigma_{i-1},s_i,i)\Downarrow false & \multicolumn{2}{r}{\mbox{Definition}~\ref{ree},  ~\ref{ee},~(1,2,3)}\\
  (5)&\Longrightarrow&( (\mbox{\ttfamily while}(b)\{ms\}, \sigma_{i-1}, s_i, i)\\
      &&\multicolumn{3}{l}{\rightarrowtail(\mbox{\ttfamily if} (b) \mbox{\ttfamily then} \{ms\wedge more;\mbox{\ttfamily while}(b)\{ms\}\}\mbox{\ttfamily else} \{\mbox{\ttfamily empty}\},\sigma_{i-1}, s_i, i)} \\
      &&\multicolumn{3}{r}{\mbox{WHL}}\\
       & &\multicolumn{3}{l}{{\rightarrowtail} ((b\wedge (ms\wedge more;\mbox{\ttfamily while}(b)\{ms\}))\vee (\neg b \wedge \mbox{\ttfamily empty}), \sigma_{i-1}, s_{i}, i)~~\mbox{IF}}\\
       &&\rightarrowtail (\mbox{\ttfamily empty}, \sigma_{i-1}, s_{i}, i)&\multicolumn{2}{r}{\mbox{B4, F1, T1, F2, (4)}}\\
      &&\rightarrow(true, \sigma_i, \emptyset, i+1) )\wedge (\alpha\vdash M\sim s_{i})&\multicolumn{2}{r}{\mbox{TR2},~(2)}\\
  (6)&\Longleftrightarrow& \multicolumn{2}{l}{P(\mbox{{\ttfamily while}}(e)\{cs\},M,M,Normal)}&\mbox{TER},~(5)
  \end{array}
  \]}\noindent
 If there are {\ttfamily break}, {\ttfamily return} and {\ttfamily continue} in $cs_1$, the conclusions can similarly be proved.

  \item[8.]  For rule $T12$ w.r.t. ``{\ttfamily while}$(e)\{cs\}$'', $StmtTr(\mbox{{\ttfamily while}}(e)\{cs\})=$``{\ttfamily while}$(b ~and\linebreak  break=0)\{ms\};break:=0$'',  where $b=ExTr(e)$ and $ms=StmtTr(cs)$, if there is a ``{\ttfamily break}$;$'' statement in $cs$. Note that, the initial value of $break$ is 0.
   {\small\[
  \begin{array}{llllr}
  (1)&&\alpha\vdash e \sim_e b &\multicolumn{2}{r}{\mbox{Theorem}~ \ref{e-eq}}\\
  (2)&& \alpha\vdash M\sim s_i&\multicolumn{2}{r}{\mbox{given condition}}\\
 (3)& &\multicolumn{3}{l}{(G,E\vdash e,M {\Rightarrow} true)\wedge(G,E\vdash cs,M \stackrel{t}{\Rightarrow} Break,M')\wedge}\\
 &&\multicolumn{2}{l}{ P(cs,M,M',Break)\wedge Break\stackrel{loop}{\rightsquigarrow}Normal} &\mbox{hypothesis}\\
  &&\multicolumn{3}{l}{\mbox{\{here $\stackrel{loop}{\rightsquigarrow}$ is borrowed from \cite{blazy2009mechanized}, see Appendix A.\}}}\\
 (4)&\Longleftrightarrow & %&\multicolumn{3}{l}{(G,E\vdash e,M {\Rightarrow} true)\wedge (G,E\vdash cs,M \stackrel{t}{\Rightarrow} Break,M')\wedge}\\
 \multicolumn{2}{l}{ (ms,\sigma_{i-1},s_i,i)\stackrel{*}{\rightarrow} (true,\sigma_j,\emptyset, j+1)\wedge \alpha\vdash M'\sim s_{j} }&\mbox{TER},~(2,3)\\
  (5)&\Longrightarrow&   (b,\sigma_{i-1},s_i,i)\Downarrow true& \multicolumn{2}{r}{ \mbox{Definition}~\ref{ree}, ~\ref{ee},~(1,2,3)}\\
  (6)&\Longrightarrow&(b\wedge break=0,\sigma_{i-1},s_i,i)\Downarrow true & \multicolumn{2}{r}{\mbox{B3, B5, (5)}}\\

  (7)&\Longrightarrow&\multicolumn{3}{l}{( (\mbox{\ttfamily while$(b ~and ~break=0)\{ms\};break:=0$}, \sigma_{i-1}, s_i, i)}\\
    &&  \multicolumn{3}{l}{\stackrel{*}{\rightarrowtail}((b\wedge break=0 \wedge (ms\wedge more;\mbox{\ttfamily while$(b ~and ~break=0)\{ms\}$})\vee }\\
     &  &\multicolumn{2}{l}{~~~~\neg (b \wedge break=0) \wedge \mbox{\ttfamily empty});break:=0, \sigma_{i-1}, s_{i}, i)}&\mbox{WHL, IF}\\
    &&  \multicolumn{3}{l}{ \rightarrowtail (ms\wedge more;\mbox{\ttfamily while$(b ~and ~break=0)\{ms\};$}break:=0, \sigma_{i-1}, s_{i}, i)}\\
     &  &\multicolumn{3}{r}{\mbox{B4, T1, F1, F2, (6)}}\\
      && \multicolumn{3}{l}{\stackrel{*}{\rightarrow}(\wedge \{\mbox{\ttfamily empty}, break\Leftarrow 1\};\mbox{\ttfamily while}(b ~and ~break=0)\{ms\};break:=0,}\\
   &    & ~~~~\sigma_{j-1},s_{j}, j)&&(3,4)\\
    &&   \multicolumn{3}{l}{\rightarrowtail(\mbox{\ttfamily empty};\mbox{\ttfamily while}(b ~and ~break=0)\{ms\};break:=0, \sigma_{j-1},}\\
    &   & ~~~~s_{j}[1/break],j)& \multicolumn{2}{r}{\mbox{MIN1}}\\
      && \multicolumn{3}{l}{\rightarrowtail(\mbox{\ttfamily while$(b ~and ~break=0)\{ms\};$}
       break:=0, \sigma_{j-1}, s_{j}, j)~~~~~~\mbox{CHOP}}\\
     &&  \multicolumn{3}{l}{\stackrel{*}{\rightarrowtail} ((b\wedge break=0 \wedge (ms\wedge more; \mbox{\ttfamily while$(b ~and ~break=0)\{ms\}$})\vee } \\
     &  &\multicolumn{2}{l}{~~~~\neg (b \wedge break=0) \wedge \mbox{\ttfamily empty});break:=0, \sigma_{j-1}, s_{j}, j)}&\mbox{WHL, IF}\\
    &   &  \rightarrowtail(\mbox{\ttfamily empty}; break:=0,\sigma_{j-1}, s_{j}, j ) &\multicolumn{2}{r}{\mbox{F1, T1, F2}}\\
    & &  \multicolumn{2}{l}{\rightarrowtail(break:=0,\sigma_{j-1}, s_{j}, j ))\wedge (\alpha\vdash M'\sim s_{j})} &\mbox{CHOP, (4)}\\
   (8)&\Longrightarrow& \multicolumn{3}{l}{(break:=0,\sigma_{j-1}, s_{j}, j ) \stackrel{*}{\rightarrow}(true, \sigma_{j+1}, \emptyset, j+2)\wedge}\\
      &&(\alpha\vdash M'\sim s_{j+1})
       &\multicolumn{2}{r}{\mbox{proved in Step 2.}}\\
  (9)&\Longleftrightarrow& P(\mbox{{\ttfamily while}}(e)\{cs\},M,M', Normal)&&\mbox{TER},~(7,8)
  \end{array}
  \]}\noindent
 In a similar way, it can be proved that
   {\small\[\begin{array}{ll}
   &(G,E\vdash e, M \Rightarrow true)\wedge(G,E \vdash cs, M \stackrel{t}{\Rightarrow}  out, M')\wedge\\
   &out\in \{ Return, Return(v)\} \wedge P(cs,M,M',out)\wedge out\stackrel{loop}{\rightsquigarrow}out'\\
   \Longrightarrow& P(\mbox{{\ttfamily while}}(e)\{cs\},M,M',out')
   \end{array}
   \]}
\item[9.] For rule $T13$ w.r.t. ``{\ttfamily while}$(e)\{cs\}$'', $StmtTr(\mbox{{\ttfamily  while}}(e)\{cs\}) =$ ``{\ttfamily  while}$(b)\{ms\}$'', where $b=ExTr(e)$ and $ms=StmtTr(cs)$, if there is no {\ttfamily break}, {\ttfamily return} or {\ttfamily continue} in $cs$.
{\small\[
  \begin{array}{lllr}
  (1)& &\alpha\vdash e \sim_e b &\mbox{Theorem}~ \ref{e-eq}\\
  (2)&& \alpha\vdash M\sim s_i&\mbox{given condition}\\
 (3)& &\multicolumn{2}{l}{ (G,E\vdash e,M {\Rightarrow} true)\wedge(G,E\vdash cs,M \stackrel{t_1}{\Rightarrow} Normal,M_1)\wedge }\\
 &&\multicolumn{2}{l}{ P(cs,M,M_1,Normal)\wedge(G,E \vdash \mbox{{\ttfamily  while}}(e) \{cs\}, M_1 \stackrel{t_2}{\Rightarrow}  out, M_2)\wedge } \\
 &&  P(\mbox{{\ttfamily  while}}(e)\{cs\},M_1,M_2,out) &\mbox{hypothesis}\\
 (4)&\Longleftrightarrow&  \multicolumn{2}{l}{(   (ms,\sigma_{i-1},s_i,i)\stackrel{*}{\rightarrow} (true,\sigma_j,\emptyset, j+1)\wedge\alpha\vdash M_1\sim s_{j})\wedge } \\
  &&\multicolumn{2}{l} { (\mbox{{\ttfamily  while}}(b)\{ms\},\sigma_{j-1},s_j,j)\stackrel{*}{\rightarrow} (true,\sigma,\emptyset, |\sigma|+1)\wedge }\\
  &&\alpha\vdash M_2\sim s_{|\sigma|})&\mbox{TER},~(3)\\
  (5)&\Longrightarrow&  (b,\sigma_{i-1},s_i,i)\Downarrow true & \mbox{Definition}~\ref{ree},~\ref{ee},~(1)\\
  (6) &\Longrightarrow&\multicolumn{2}{l}{(
   (\mbox{\ttfamily  while}(b)\{ms\}, \sigma_{i-1}, s_i, i)}\\
       & &\multicolumn{2}{l}{\stackrel{*}{\rightarrowtail} ((b\wedge (ms\wedge more;\mbox{\ttfamily  while}(b)\{ms\}))\vee (\neg b \wedge \mbox{\ttfamily empty}),\sigma_{i-1}, s_{i}, i)}\\
   &     &&\mbox{WHL, IF}\\
    &&   \rightarrowtail (ms\wedge more;\mbox{\ttfamily  while}(b)\{ms\}, \sigma_{i-1}, s_{i}, i)&\mbox{B4, T1, F1, F2, (5)}\\
     &&  \stackrel{*}{\rightarrow}(\mbox{\ttfamily empty};\mbox{\ttfamily  while}(b)\{ms\}, \sigma_{j-1}, s_{j}, j)&(4)\\
     &&  \rightarrowtail (\mbox{\ttfamily  while}(b)\{ms\}, \sigma_{j-1}, s_{j}, j) &\mbox{CHOP}\\
   &&  \stackrel{*}{\rightarrow} (true,\sigma,\emptyset, |\sigma|+1))\wedge(\alpha\vdash M_2\sim s_{|\sigma|})&(4)\\
  (7)&\Longleftrightarrow& P(\mbox{{\ttfamily while}}(e)\{cs\},M,M_2, out)&\mbox{TER},~(6)
  \end{array}
  \]}\noindent
 In a similar way, it can be proved that
   {\small\[\begin{array}{ll}
   & (G,E\vdash e, M \Rightarrow true)\wedge(G,E \vdash cs, M \stackrel{t_1}{\Rightarrow}  Continue, M_1)\wedge\\
   & P(cs,M,M_1,Continue)\wedge(G,E \vdash \mbox{{\ttfamily  while}}(e)\{cs\}, M_1 \stackrel{t_2}{\Rightarrow}  out, M_2)\wedge\\
   & P(\mbox{{\ttfamily  while}}(e)\{cs\},M_1,M_2,out)\\
   \Longrightarrow& P(\mbox{\ttfamily  while}(e)\{cs\},M,M_2,out)
   \end{array}
   \]}
     \item [10.]  For rules $T24$ and $T25$ w.r.t. a function call ``$x(e_{1},...,e_{m});$'', $StmtTr(x(e_{1},...,\linebreak e_{m});)=f(ra_1,...,ra_m,RVal)$, where $f=ExTr(x)$ and $ra_k=ExTr(e_k)$ for all $1\leq k\leq m$, if $x$ points to a user-defined function with a return value.
         We assume $x$ points to $\tau~id(par)\{ dcl;cs\}$, where $par=(\tau_1~ y_1, ..., \tau_m~ y_m)$. Thus, $f$ points to $function~id(\tau_1~ y_1, ..., \tau_m~ y_m, \tau ~RVal)\{ mdcl;ms\}$ translated from $\tau~id(par)\{ dcl;cs\}$, where $mdcl=DecTr(dcl)$ and $ms=StmtTr(cs)$.
          {\small\[
  \begin{array}{llllr}
  (1)& &\alpha\vdash e_k \sim_e ra_k ~~~~~(1\leq k\leq m)&\multicolumn{2}{r}{\mbox{Theorem}~ \ref{e-eq}}\\
  (2)&&  \alpha \vdash M\sim s_i & \multicolumn{2}{r}{\mbox{given condition}}\\
  (3)&&  \multicolumn{3}{l}{ \bigwedge_{k=1}^{m}(G,E\vdash e_k,M {\Rightarrow} v_k)\wedge v_{args}=(v_1,...,v_m)\wedge}\\
     &&\multicolumn{3}{l}{ \mathit{alloc\_vars}(M,par\textbf{+}dcl,E)=(M_1,b^*)\wedge}\\
    && \multicolumn{3}{l}{\mathit{bind\_params}(E,M_1, par, v_{args})=M_2\wedge}\\
          &&\multicolumn{3}{l}{(G,E\vdash cs,M_2 \stackrel{t}{\Rightarrow} Return(v), M_3)\wedge(Return(v),\tau\#v)\wedge}\\
          && \multicolumn{2}{l}{M_4=free(M_3,b^*)\wedge P(cs,M_2,M_3, Return(v))} &\mbox{hypothesis}\\
            &&\multicolumn{3}{l}{\mbox{\{here $alloc\_vars$,  $bind\_params$, $Return(v)$ and $free$ are borrowed}}\\
            &&\multicolumn{3}{l}{\mbox{ from \cite{blazy2009mechanized}, see Appendix A.\}}}\\
 (4)&&\alpha \vdash M_2\sim s_t&\multicolumn{2}{r}{\mbox{assumption}}\\
 (5)&\Longrightarrow& \multicolumn{2}{l}{(ms,\sigma_{t-1},s_t,t)\stackrel{*}{\rightarrow} (true,\sigma_j,\emptyset, j+1)\wedge(\alpha\vdash M_3\sim s_j)} &\mbox{TER},~(3,4)\\
 (6)& &\multicolumn{3}{l}{(f(ra_1, . . . , ra_m,RVal), \sigma_{i-1}, s_i, i)}\\
    & & \multicolumn{3}{l}{\rightarrowtail(id(ra_1, . . . , ra_m,RVal), \sigma_{i-1}, s_i, i)}\\
    & &  \multicolumn{3}{l}{\rightarrowtail (
       (\wedge_{k=1}^m \tau_k~ y_k\Leftarrow ra_k\wedge mdcl);ms; \bigcirc (ext~mfree(y_1,...,y_m,mdcl)} \\
     &&\multicolumn{2}{l}{ ~~~~ \wedge\mbox{\ttfamily empty}),\sigma_{i-1}, s_{i}, i)}&\mbox{FUN}\\        &&\multicolumn{3}{l}{\mbox{ \{here $mfree$ is defined in Section \ref{function}.\}}}\\
     && \multicolumn{3}{l}{\stackrel{*}{\rightarrow} (
       ms; \bigcirc (ext~mfree(y_1,...,y_m,mdcl)\wedge\mbox{\ttfamily empty}), \sigma_{t-1}, s_{t}, t)}\\
       &&\multicolumn{3}{r}{\mbox{MIN1, TR1}}\\
       \end{array}\]}\noindent
       In the Xd-C program, the memory state transfers from $M$ to $M_2$. Compared to $M$, $M_2$ allocates memory blocks to variables $y_1,...,y_m$ and variables in $dcl$, and assigns $v_k$ to $y_k$ for $1\leq k\leq m$. Whereas in the MSVL program,
       the state transfers from $s_i$ to $s_t$. We assume $(ra_k,\sigma_{i-1},s_i,i)\Downarrow n_k$ for all $1\leq k\leq m$. Thus,
       $\alpha\vdash v_k\sim n_k$ due to (1) and (2). Compared to $s_i$, $s_t$ also allocates memory blocks to variables $y_1,...,y_m$ and variables in $mdcl$, and assigns $n_k$ to $y_k$ for $1\leq k\leq m$.
        Hence, $\alpha\vdash M_2\sim s_t$.
       Since $(G,E\vdash cs,M_2 \stackrel{t}{\Rightarrow} Return(v), M_3)$, ``{\ttfamily return $e;$}'' must be executed at the last state of executing $cs$, and $(G,E\vdash e, M_3\Rightarrow v)$. Thus, $RVal\Leftarrow ra$ must be executed at the last state of executing $ms$, where $ra=ExTr(e)$. We assume $(ra,\sigma_{j-1},s_j,j)\Downarrow n$.
       Accordingly, the reduction continues as follows:
     {\small\[\begin{array}{lllr}
     (7)&~~~~& \multicolumn{2}{l}{(
       ms; \bigcirc (ext~mfree(y_1,...,y_m,mdcl)\wedge\mbox{\ttfamily empty}), \sigma_{t-1}, s_{t}, t)}\\
       && \multicolumn{2}{l}{\stackrel{*}{\rightarrow} (RVal\Leftarrow ra \wedge \mbox{\ttfamily empty};\bigcirc (ext~mfree(y_1,...,y_m,mdcl)\wedge \mbox{\ttfamily empty}),} \\
       && \sigma_{j-1}, s_j, j)&(5)\\
       &&{\rightarrowtail} (\mbox{\ttfamily empty};\bigcirc (ext~mfree(y_1,...,y_m,mdcl)\wedge \mbox{\ttfamily empty}),
        \sigma_{j-1}, \\
        &&(s_j^l, s_j^r[n/RVal]), j)&\mbox{MIN1}\\
        &&{\rightarrow} (ext~mfree(y_1,...,y_m,mdcl)\wedge \mbox{\ttfamily empty},\sigma_{j}, s_{j+1}, j+1)&\mbox{CHOP, TR1} \\
        &&{\rightarrow} (true, \sigma_{j+1}, \emptyset, j+2)&\mbox{EXT3, TR2}\\
      \end{array}\]}\noindent
       According to (5),  we have $\alpha\vdash M_3\sim s_j$. Further, $\alpha\vdash v\sim s_j^r(RVal)$ due to $\alpha \vdash e\sim_e ra$.
      % For any variable $z$ in the Xd-C and MSVL programs, $b,b'\in \mathbb{Z}$, $j,j'\in N_0$ and $v_z,n_z\in D$, if $(G,E\vdash z, M_3 \stackrel{l}{\Rightarrow} (b,j))$ and $(G,E \vdash z, M_3 \Rightarrow v_z)$ in the Xd-C program, as well as $s_j^l(z)=(b',j')$ and $s_j^r(z)=n_z$ in the MSVL program, then $\alpha\vdash ptr(b,j)\sim ptr(b',j')$ and $\alpha\vdash v_z\sim n_z$.
       $M_4$ just removes $y_1,...,y_m$ and variables in $dcl$ from $M_3$ while $s_{j+1}$ just removes $y_1,...,y_m$ and variables in $mdcl$ from $s_j$.
       %Thus, for any variable $z$, if $z\in Dom(G)\cup Dom(E)$, then $z\in Dom(s_{j+1})\setminus \{break, continue, return, RVal\}$ and if
      % $z\in Dom(s_{j+1})\setminus \{break, continue, return, RVal\}$, then $z\in Dom(G)\cup Dom(E)$.
     Thus, the locations and values of variables are not changed and $\alpha\vdash M_4\sim s_{j+1}$.
   %  $(G,E\vdash z, M_4 \stackrel{l}{\Rightarrow} (b,j))$ and $(G,E \vdash z, M_4 \Rightarrow v_z)$.
      % Similar to (2), we have $s_{j+1}^l(z)=s_{j}^l(z)=(b',j')$ and $s_{j+1}^r(z)=s_{j}^r(z)=n_z$. Therefore, $\alpha \vdash M_4\sim s_{j+1}$ and
      Therefore,
      {\small$$ \begin{array}{lllr}
        (8)& \Longrightarrow&(f(ra_1, . . . , ra_m,RVal), \sigma_{i-1}, s_i, i)\stackrel{*}{\rightarrow} (true, \sigma_{j+1}, \emptyset, j+2)\wedge\\
         &&\alpha\vdash M_4\sim s_{j+1}&(6,7)\\
         (9)&\Longleftrightarrow& P(x(e_1,...,e_m);,M,M_4,out )&\mbox{TER},~(8)
         \end{array}
      $$}\noindent
     In a similar way, if $x$ points to a user-defined function without a return value, it can be proved that $P(x(e_{1},...,e_{m});,M,M_4,out)$ holds.

  \item[11.]  For rules $T24$ and $T26$ w.r.t. a function call ``$x(e_{1},...,e_{m});$'', $StmtTr(x(e_{1},...,\linebreak e_{m});)=ext ~f( ra_1,...,ra_m)$, where $f=ExTr(x)$ and $ra_k=ExTr(e_k)$ for all $1\leq k\leq m$ and $x$ points to an external function $extern~[\tau\mid void] ~id(par)$, where $par=(\tau_1~ y_1, ..., \tau_m~ y_m)$.
     % We define $EFUN$ in the Xd-C program as
         {\small\[\begin{array}{lllr}
       %  (1)&&\alpha\vdash M\sim s_i&\mbox{given condition}\\
       %  (2)& & \multicolumn{2}{l}{\bigwedge_{k=1}^{m}(G,E\vdash e_k,M {\Rightarrow} v_k)\wedge v_{args}=(v_1,...,v_m)\wedge v=id(v_{args})\wedge}\\
       %  && v_t=``id(v_{args},v)"\wedge \bigwedge_{k=1}^{m} (ra_k,\sigma_{i-1},s_i,i)\Downarrow n_k&\mbox{assumption}\\
      (1)&&(\wedge\{\bigcirc \mbox{\ttfamily empty},ext~ f (ra_1, . . . , ra_m)\},\sigma_{i-1}, s_i, i)\\
      &&\rightarrow(\mbox{\ttfamily empty},\sigma_{i}, s_{i+1}, i+1)&\mbox{EXT2}\\
       &&\rightarrow (true,\sigma_{i+1}, \emptyset, i+2)\wedge s_i=s_{i+1}&\mbox{TR2}\\
        (2)& \Longrightarrow&( (\wedge\{\bigcirc \mbox{\ttfamily empty},ext~ f (ra_1, . . . , ra_m)\},\sigma_{i-1},s_i, i)\\
        & &\stackrel{*}{\rightarrow}(true,\sigma_{i+1}, \emptyset, i+2)\wedge \alpha\vdash M\sim s_{i+1})&(1)\\
        (3)& \Longleftrightarrow& P(x(e_1,...,e_m),M,M,out)&\mbox{TER},~(2)
       \end{array}\]}\noindent
      Note that, (1) indicates that $s_i=s_{i+1}$ and $\langle s_i \rangle$ is a model of $ext~f(n_1, . . . , n_m)$.

\end{itemize}

Similar to Step 6, we can prove that for the rules of {\ttfamily switch} statements, the conclusions are true, and
similar to Steps 7, 8 and 9, we can prove for the rules of {\ttfamily for} loops, the conclusions are true.

%The proof of the diverging statement equivalence is given as follows:
Case 2:

Base:
\begin{itemize}
\item[1.] For rule $D1$ w.r.t. ``{\ttfamily while}(e)\{cs\}'', $StmtTr(\mbox{\ttfamily while}(e)\{cs\})=$ ``{\ttfamily  while}$(b)\{ms\}$'',  where $b=ExTr(e)$ and $ms=StmtTr(cs)$, if there is no {\ttfamily break}, {\ttfamily return} or {\ttfamily continue} in $cs$. $G,E\vdash \mbox{\ttfamily while}(e)\{cs\},M\stackrel{T}{\Rightarrow} \infty$.
    {\small\[
  \begin{array}{llllr}
  (1)&&\alpha\vdash M\sim s_i&\multicolumn{2}{r}{\mbox{given condition}}\\
  (2)&&\alpha\vdash e\sim_e b&\multicolumn{2}{r}{\mbox{Theorem}~ \ref{e-eq}}\\
% (3)& & \multicolumn{3}{l}{\forall M,M'(G,E \vdash e, M  {\Rightarrow} true \wedge (G,E \vdash cs, M \stackrel{t}{\Rightarrow} Normal, M')}  \\
% &&\multicolumn{3}{l}{\rightarrow G,E~\vdash~e,~M'  {\Rightarrow} v' \wedge is\_true(v',type(e)))\wedge}\\
% &&G,E~\vdash~e,~M  {\Rightarrow} true&\multicolumn{2}{r}{\mbox{assumption}}\\
 (3)& & \multicolumn{3}{l}{\forall i\in N_0(G,E \vdash e, M_i  {\Rightarrow} true \wedge (G,E \vdash cs, M_i \stackrel{t}{\Rightarrow} Normal, M_{i+1})}  \\
 &&\multicolumn{2}{l}{\rightarrow G,E~\vdash~e,~M_{i+1}  {\Rightarrow} true ) \wedge M_0=M} & \mbox{(infinite loop) assumption} \\
 (4)&\Longrightarrow&  \multicolumn{3}{l}{   (\mbox{\ttfamily while}(e)\{cs\},M)\cong (cs;\mbox{\ttfamily while}(e)\{cs\},M)\cong(cs;cs;...,M)}\\
 && \multicolumn{3}{r}{\mbox{Lemma \ref{Cloop}},~(3)}\\
  (5)&\Longrightarrow&(b, \sigma_{i-1}, s_i, i) \Downarrow true& \multicolumn{2}{r}{\mbox{Definition}~\ref{ree}, ~\ref{ee},~(1,2,3)} \\
 (6)&\Longrightarrow& \multicolumn{3}{l}{(\mbox{\ttfamily while}(b)\{ms\},\sigma_{i-1},s_i,i)}\\
 &&\multicolumn{2}{l}{\stackrel{*}{\rightarrowtail} (ms;\mbox{\ttfamily while}(b)\{ms\}, \sigma_{i-1},s_i,i)}&\mbox{WHL, IF, (5)} \\
 &&  \multicolumn{2}{l}{\stackrel{*}{\rightarrowtail} (ms;ms;..., \sigma_{i-1},s_i,i) } &\mbox{WHL, IF, }(1,2,3)\\
 (7)&\Longrightarrow&\alpha\vdash cs\sim_t ms&&\mbox{Case 1}\\
  (8)&\Longleftrightarrow& P'(cs;cs',M)&&\mbox{DV},~(4,6,7)
  \end{array}
  \]}\noindent
  If there are {\ttfamily break}, {\ttfamily return} and {\ttfamily continue} in $cs$,  the conclusions can similarly be proved.

Induction:
\item[2.] For rule $D2$ w.r.t. ``$cs;cs'$'', $StmtTr(cs;cs')=$``$ms;ms'$'', where $ms=StmtTr(cs)$ and $ms'=StmtTr(cs')$, if there is no {\ttfamily break}, {\ttfamily return} or {\ttfamily continue} in $cs$. $(G,E \vdash cs, M \stackrel{T}{\Rightarrow} \infty)$ and $(G,E \vdash cs;cs', M \stackrel{T}{\Rightarrow} \infty)$.
   {\small\[
  \begin{array}{lllr}
  (1)&&\alpha\vdash M\sim s_i&\mbox{given condition}\\
 (2)& & (G,E \vdash cs, M \stackrel{T}{\Rightarrow} \infty) \wedge P'(cs,M) &\mbox{hypothesis}\\
 (3)&\Longleftrightarrow&  \multicolumn{2}{l}{   (cs,M)\cong (cs_1;cs_2;...,M)\wedge (ms,\sigma_{i-1},s_i,i)\stackrel{*}{\rightarrowtail} (ms_1;ms_2;..., \sigma_{i-1}, }\\
 && s_i,i)\wedge \bigwedge_{j=1}^{\infty}(\alpha\vdash cs_j\sim_t ms_j) &\mbox{DV, (2)}\\
  (4)&\Longrightarrow&\multicolumn{2}{l}{ (cs;cs',M)\cong (cs_1;cs_2;...,M) \wedge}\\
   &&\multicolumn{2}{l}{(ms;ms',\sigma_{i-1},s_i,i)\stackrel{*}{\rightarrowtail} (ms_1;ms_2;..., \sigma_{i-1},s_i,i)\wedge
    }\\
    && \bigwedge_{j=1}^{\infty}(\alpha\vdash cs_j\sim_t ms_j) &E1,~(3)\\
  (5)&\Longleftrightarrow& P'(cs;cs',M)&\mbox{DV},~(4)
  \end{array}
  \]}\noindent
If there are {\ttfamily break}, {\ttfamily return} and {\ttfamily continue} in $cs$,  the conclusions can be proved in a similar way.
\item[3.] For rule $D3$ w.r.t. ``$cs;cs'$'', $StmtTr(cs;cs')=$``$ms;ms'$'',  where $ms=StmtTr(cs)$ and $ms'=StmtTr(cs')$, if there is no {\ttfamily break}, {\ttfamily return} or {\ttfamily continue} in $cs$. $G,E \vdash cs, M \stackrel{t}{\Rightarrow} Normal,M_1$, $G,E \vdash cs', M_1 \stackrel{T}{\Rightarrow} \infty$ and $G,E \vdash cs;cs', M_1 \stackrel{T}{\Rightarrow} \infty$.
    %Then $\alpha\vdash cs\sim_s ms$.
       {\small\[
  \begin{array}{llllr}
  (1)&& \alpha\vdash M\sim s_i&\multicolumn{2}{r}{\mbox{given condition}}\\
 (2)& &\multicolumn{3}{l}{(G,E \vdash cs, M \stackrel{t}{\Rightarrow} Normal,M_1)\wedge(G,E \vdash cs', M_1 \stackrel{T}{\Rightarrow} \infty)\wedge }\\
 &&P'(cs',M_1)&\multicolumn{2}{r}{\mbox{hypothesis}}\\
 (3)&\Longrightarrow&   P(cs,M,M_1,Normal)&\multicolumn{2}{r}{\mbox{Terminating Statement Equivalence, (2)}}\\
  (4)&\Longrightarrow&\multicolumn{3}{l}{(ms,\sigma_{i-1},s_i,i)\stackrel{*}{\rightarrow}(true, \sigma_t,\emptyset, t+1)\wedge\alpha\vdash M_1\sim s_t\wedge }\\
 &&\multicolumn{3}{l}{ (cs',M_1)\cong (cs_1;cs_2;...,M_1) \wedge}\\
  &&\multicolumn{3}{l}{(ms',\sigma_{t-1},s_t,t)\stackrel{*}{\rightarrowtail} (ms_1;ms_2;..., \sigma_{t-1},s_t,t) \wedge
    } \\
    &&\bigwedge_{j=1}^{\infty}(\alpha\vdash cs_j\sim_t ms_j)&&\mbox{TER, DV},~(2,3)\\
  (5)&\Longrightarrow&\multicolumn{3}{l}{  (cs;cs',M)\cong (cs;cs_1;cs_2;...,M) \wedge}\\
   &&\multicolumn{3}{l}{(ms;ms',\sigma_{i-1},s_i,i)\stackrel{*}{\rightarrowtail} (ms;ms_1;ms_2;..., \sigma_{i-1},s_i,i) \wedge}\\
    && \multicolumn{2}{l}{\bigwedge_{j=1}^{\infty}(\alpha\vdash cs_j\sim_t ms_j)\wedge \alpha\vdash cs\sim_t ms} & E2,~(3,4) \\
  (6)&\Longleftrightarrow& P'(cs;cs',M)&&\mbox{DV},~(5)
  \end{array}
  \]}\noindent
If there are {\ttfamily break}, {\ttfamily return} and {\ttfamily continue} in $cs$,  the conclusions are also true. % in a similar way.
%$\frac{G,E~\vdash~cs_1,~M \stackrel{t}{\Rightarrow} Normal,~ M_1~~~G,E~\vdash~cs_2,~M \stackrel{T}{\Rightarrow} \infty }{G,E~\vdash~cs_1;cs_2,~M \stackrel{t.T}{\Rightarrow} \infty}~{\scriptstyle (C40)}\\$

\item[4.] For rule $D4$ w.r.t. ``{\ttfamily if$(e)\{cs\}$else$\{cs'\}$}'', $StmtTr(\mbox{{\ttfamily if$(e)\{cs\}$else$\{cs'\}$}})=$ ``{\ttfamily if$(b)$then$\{ms\}$else$\{ms'\}$}'',  where $b=ExTr(e)$, $ms=StmtTr(cs)$ and $ms'=StmtTr(cs')$. $(G,E \vdash e, M \Rightarrow true)$, $(G,E \vdash cs, M \stackrel{T}{\Rightarrow} \infty)$ and $(G,E \vdash \mbox{\ttfamily if$(e)\{cs\}$else$\{cs'\}$}, M \stackrel{T}{\Rightarrow} \infty)$.
       {\small\[
  \begin{array}{llllr}
  (1)&&\alpha\vdash M\sim s_i&\multicolumn{2}{r}{\mbox{given condition}}\\
  (2)&&\alpha \vdash e\sim_e b &\multicolumn{2}{r}{\mbox{Theorem} ~\ref{e-eq}}\\
 (3)& &\multicolumn{3}{l}{(G,E \vdash e, M \Rightarrow true)\wedge(G,E \vdash cs, M \stackrel{T}{\Rightarrow} \infty) \wedge P'(cs,M)}\\
 &&&&\mbox{hypothesis}\\
 (4)&\Longrightarrow&  \multicolumn{3}{l}{(cs,M)\cong (cs_1;cs_2;...,M) \wedge}\\
 &&\multicolumn{3}{l}{(ms,\sigma_{i-1},s_i,i)\stackrel{*}{\rightarrowtail} (ms_1;ms_2;..., \sigma_{i-1},s_i,i)} \\
 &&\wedge \bigwedge_{j=1}^{\infty}(\alpha\vdash cs_j\sim_t ms_j)&&\mbox{DV, }(3)\\
  (5)&\Longrightarrow& \multicolumn{2}{l}{(\mbox{\ttfamily if$(e)\{cs\}$else$\{cs'\}$},M)\cong(cs,M) \cong(cs_1;cs_2;...,M)}&E3,~(3,4)\\
   (6)&\Longrightarrow&(b, \sigma_{i-1}, s_i, i) \Downarrow true&\multicolumn{2}{r}{ \mbox{Definition}~\ref{ree}, ~\ref{ee},~(1,2,3)}\\
   (7)&\Longrightarrow&\multicolumn{2}{l}{(\mbox{\ttfamily if$(b)$then$\{ms\}$else$\{ms'\}$},\sigma_{i-1},s_i,i)}\\
  && \stackrel{*}{\rightarrowtail} (ms, \sigma_{i-1},s_i,i)&&\mbox{IF},~(6)\\
   &&\multicolumn{2}{l}{\stackrel{*}{\rightarrowtail} (ms_1;ms_2;..., \sigma_{i-1},s_i,i)\wedge
    \bigwedge_{j=1}^{\infty}(\alpha\vdash cs_j\sim_t ms_j)}&(4)\\
  (8)&\Longleftrightarrow& P'(\mbox{\ttfamily if$(e)\{cs\}$else$\{cs'\}$},M)&&\mbox{DV, }(5,7)
  \end{array}
  \]}\noindent
For rule $D5$ w.r.t. ``{\ttfamily if$(e)\{cs\}$else$\{cs'\}$}'' and $D6$ w.r.t. ``{\ttfamily while}(e)\{cs\}'', we can similarly prove that
     {\small\[\begin{array}{lll}
(1)&& (G,E \vdash e, M \Rightarrow false)\wedge (G,E \vdash cs', M \stackrel{T}{\Rightarrow} \infty)\wedge P'(cs',M)\\
 &\Longrightarrow &P'(\mbox{{\ttfamily if$(e)\{cs\}$else$\{cs'\}$}},M)\\
(2)&& (G,E \vdash e, M \Rightarrow true)\wedge (G,E \vdash cs, M \stackrel{T}{\Rightarrow} \infty)\wedge P'(cs,M)\\
 &\Longrightarrow &P'(\mbox{\ttfamily while}(e)\{cs\},M)
\end{array}\]}\noindent

\item[5.] For rule $D7$ w.r.t. ``{\ttfamily while}(e)\{cs\}'', $StmtTr(\mbox{\ttfamily while}(e)\{cs\})=$ ``{\ttfamily  while}$(b)\{ms\}$'',  where $b=ExTr(e)$ and $ms=StmtTr(cs)$, if there is no {\ttfamily break}, {\ttfamily return} or {\ttfamily continue} in $cs$.
          {\small\[
  \begin{array}{llllr}
  (1)&& \alpha\vdash M\sim s_i &\multicolumn{2}{r}{\mbox{given condition}}\\
  (2)&&\alpha \vdash e\sim_e b &&\mbox{Theorem} ~\ref{e-eq}\\
 (3)& &\multicolumn{3}{l}{(G,E \vdash e, M \Rightarrow true)\wedge(G,E \vdash cs, M \stackrel{t}{\Rightarrow} Normal, M_1) \wedge}\\
 &&\multicolumn{3}{l}{(G,E \vdash \mbox{\ttfamily while}(e)\{cs\}, M_1 \stackrel{T}{\Rightarrow} \infty)\wedge P'(\mbox{\ttfamily while}(e)\{cs\},M_1)~~\mbox{hypothesis}}\\
 (4)&\Longrightarrow&P(cs,M,M_1,Normal)& \multicolumn{2}{r}{\mbox{Terminating Statement Equivalence, }(1,3)}\\
   (5)&\Longrightarrow&  \multicolumn{2}{l}{(ms,\sigma_{i-1},s_i,i)\stackrel{*}{\rightarrow}(true, \sigma_t,\emptyset, t+1)\wedge \alpha\vdash M_1\sim s_t }&\mbox{TER},~(1,4)\\
  (6)&\Longrightarrow&\multicolumn{3}{l}{ (\mbox{\ttfamily while}(e)\{cs\},M_1)\cong (cs_1;cs_2;...,M_1) \wedge} \\
 &&\multicolumn{3}{l}{ (\mbox{\ttfamily while}(b)\{ms\},\sigma_{t-1},s_t,t)\stackrel{*}{\rightarrowtail} (ms_1;ms_2;..., \sigma_{t-1},s_t,t)\wedge}\\
 && \bigwedge_{j=1}^{\infty}(\alpha\vdash cs_j\sim_t ms_j)&&\mbox{DV, }(3)\\
   (7)&\Longrightarrow&\multicolumn{2}{l}{(\mbox{\ttfamily while}(e)\{cs\},M) \cong(cs;cs_1;cs_2;...,M)}&E2,~E5,~(3,6)\\
 (8)&\Longrightarrow&(b, \sigma_{i-1}, s_i, i) \Downarrow true& \multicolumn{2}{r}{\mbox{Definition}~\ref{ree}, ~\ref{ee},~(1,2,3)}\\
  (9) &\Longrightarrow&\multicolumn{3}{l}{(\mbox{\ttfamily while}(b)\{ms\},\sigma_{i-1},s_i,i)}\\
  &&\multicolumn{2}{l}{\stackrel{*}{\rightarrowtail} (ms\wedge more;\mbox{\ttfamily while}(b)\{ms\}, \sigma_{i-1},s_i,i)}&\mbox{WHL, IF, (8)}\\
   &&\multicolumn{3}{l}{\stackrel{*}{\rightarrowtail} (ms;ms_1;ms_2;..., \sigma_{i-1},s_i,i)\wedge\bigwedge_{j=1}^{\infty}(\alpha\vdash cs_j\sim_t ms_j)\wedge}\\
   &&\alpha\vdash cs\sim_t ms&\multicolumn{2}{r}{\mbox{Case }1,~(6)}\\
  (10)&\Longleftrightarrow& P'(\mbox{\ttfamily while}(e)\{cs\},M)&&\mbox{DV, }(7,9)
  \end{array}
  \]}\noindent
    If there are {\ttfamily break}, {\ttfamily return} and {\ttfamily continue} in $cs$, the conclusions are also true.
 In a similar way, we can prove that
 {\small\[
 \begin{array}{ll}
 &(G,E \vdash e, M \Rightarrow true)\wedge (G,E \vdash cs, M \stackrel{t}{\Rightarrow} Continue, M_1)\wedge\\
 &(G,E \vdash \mbox{{\ttfamily  while}}(e)\{cs\}, M_1 \stackrel{T}{\Rightarrow} \infty) \wedge P'(\mbox{{\ttfamily  while}}(e)\{cs\},M_1)\\
 \Longrightarrow& P'(\mbox{\ttfamily  while}(e)\{cs\},M)
 \end{array}
 \]}
\item[6.] For rules $D8$ and $D9$ w.r.t. ``$x(e_{1},...,e_{m});$'' , $StmtTr(x(e_{1},...,e_{m});)=$ ``$f(ra_1,..., ra_m,RVal)$'', where $ra_k=ExTr(e_k)$ for all $1\leq k\leq m$, if $x$ points to a user-defined function with a return value $\tau~id(par)\{ dcl;cs\}$, where $par=(\tau_1~ y_1, ..., \tau_m~ y_m)$. Thus, $f$ points to $function~id(\tau_1~ y_1, ..., \tau_m\linebreak y_m, \tau ~RVal)\{mdcl;ms\}$ translated from $\tau~id(par)\{ dcl;cs\}$, where $mdcl=DecTr(dcl)$ and $ms=StmtTr(cs)$.
         {\small\[
  \begin{array}{lllr}
  (1)&&\alpha\vdash M\sim s_i&\mbox{given condition}\\
  (2)&&\alpha \vdash e_k\sim_e ra_k (1\leq k\leq m) &\mbox{Theorem} ~\ref{e-eq}\\
 (3)& &\multicolumn{2}{l}{\mathit{alloc\_vars}(M,par\textbf{+}dcl,E)=(M_1,b^*) \wedge v_{args}=(v_1,...,v_m)\wedge}\\
 &&\multicolumn{2}{l}{bind\_params(E,M_1, par, v_{args})=M_2 \wedge(G,E\vdash cs,M_2 \stackrel{t}{\Rightarrow} \infty)\wedge}\\
  && P'(cs,M_2)&\mbox{hypothesis}\\
 (4)&\Longrightarrow&  \multicolumn{2}{l}{ ((f(ra_1, . . . , ra_m,RVal), \sigma_{i-1}, s_i, i)\stackrel{*}{\rightarrow}( ms;}\\
 &&  \multicolumn{2}{l}{\bigcirc (ext~mfree(y_1,...,y_m,mdcl)\wedge \mbox{\ttfamily empty}),
        \sigma_{t-1}, s_t, t)\wedge \alpha \vdash M_2\sim s_t)}\\
  &&\multicolumn{2}{l}{\wedge(\alpha\vdash \tau_1~ y_1=e_1;...;\tau_m~ y_m=e_m;dcl; \sim_t \bigwedge_{k=1}^m \tau_k~y_k\Leftarrow ra_k\wedge mdcl)} \\
   && \multicolumn{2}{r}{\mbox{proved in Step 10 of Case 1}}\\
  (5)&\Longrightarrow&\multicolumn{2}{l}{(cs,M_2)\cong (cs_1;cs_2;...,M_2)\wedge}\\
  &&\multicolumn{2}{l}{(ms,\sigma_{t-1},s_t,t)\stackrel{*}{\rightarrowtail} (ms_1;ms_2;..., \sigma_{t-1},s_t,t)\wedge \bigwedge_{j=1}^{\infty}(\alpha\vdash cs_j\sim_t ms_j)}\\
&&&\mbox{DV, }(3,4)\\
 (6)&\Longrightarrow& \multicolumn{2}{l}{(x(e_1,...,e_m);,M)\cong (\tau_1~ y_1=e_1;...;\tau_m~ y_m=e_m;dcl;cs_1;cs_2;...,M)}\\
 &&&E7,~(3,5)\\
 (7)&\Longrightarrow&\multicolumn{2}{l}{(f(ra_1, . . . , ra_m,RVal), \sigma_{i-1}, s_i, i)\stackrel{*}{\rightarrowtail}}\\
  &&\multicolumn{2}{l}{(\bigwedge_{k=1}^m \tau_k~y_k\Leftarrow ra_k\wedge mdcl;ms_1;ms_2;...,\sigma_{i-1}, s_i, i)\wedge}\\
 &&\multicolumn{2}{l}{(\alpha\vdash \tau_1~ y_1=e_1;...;\tau_m~ y_m=e_m;dcl; \sim_t \bigwedge_{k=1}^m \tau_k~y_k\Leftarrow ra_k\wedge mdcl)\wedge} \\
 && \bigwedge_{j=1}^{\infty}(\alpha\vdash cs_j\sim_t ms_j))&\mbox{FUN},~(4,5) \\
  (8)&\Longleftrightarrow& P'( x(e_{1},...,e_{m});,M)&\mbox{DV, }(6,7)
  \end{array}
  \]}\noindent
      If $x$ points to a user-defined function without a return value, the conclusion can be proved in a similar way.
%
%$\frac{\begin{array}{c}\scriptstyle G,E~\vdash~e,~M \Rightarrow v~~~is\_true(v,type(e))\\
%  \scriptstyle G,E~\vdash~cs,~M \stackrel{t}{\Rightarrow} (Normal\mid Continue),~M_1~~~G,E~\vdash~\mbox{\ttfamily\scriptsize while}(e)\{cs\},~M_1 \stackrel{T}{\Rightarrow} \infty \end{array}}
%  {G,E~\vdash~\mbox{\ttfamily\scriptsize while}(e)\{cs\},~M \stackrel{t.T}{\Rightarrow} \infty}~~{\scriptstyle (C42)}$
%
\end{itemize}
\end{Proof}
\rightline{$\square$}

%We have proved equivalence between Xd-C statements,

% Once an Xd-C program $P$ is transformed into an MSVL program $Q$ by Algorithm \ref{PrgmTr}, $P$ and $Q$ are semantically equivalent. So we have the following theorem:

Based on the equivalences between expressions and statements in Xd-C and MSVL,
 we have the following theorem:

\begin{Thm}\label{p-eq}
If an Xd-C program $P$ is transformed to an MSVL program $Q$ by Algorithm \ref{PrgmTr}, then $P$ is semantically equivalent to $Q$, denoted by $P\sim_p Q$.
\end{Thm}

\begin{Proof}
  Suppose an Xd-C program $P$ is composed of $k_1$ expressions and $k_2$ statements, where $k_1$ and $k_2$ are constants. When Xd-C program $P$ is transformed to MSVL program $Q$ by Algorithm \ref{PrgmTr}, we have $Q=PrgmTr(P)$. Actually, translating $P$ to $Q$ is merely translating expression $e_i$ ($0\leq i \leq k_1$) and statement $cs_j$ ($0\leq j\leq k_2$) in $P$ to $a_i$ and $ms_j$ in $Q$ in one-to-one manner by Algorithm \ref{ExTr} and \ref{StmtTr}, respectively. That is, $a_i=ExTr(e_i)$ ($0\leq i \leq k_1$) and $ms_j=StmtTr(cs_j)$ ($0\leq j\leq k_2$).
 Let $M$ and $s_0$ be the initial states of $P$ and $Q$, respectively.
 According to Theorem \ref{e-eq} and \ref{s-eq}, for a given $\alpha$, if $\alpha\vdash M \sim s_0$, then $\alpha \vdash e_i\sim_e a_i$ and
 $\alpha \vdash cs_j\sim_s ms_j$ for all $0\leq i \leq k_1$ and $0\leq j\leq k_2$. As a result, $P$ is equivalent to $Q$, that is, $P\sim_p Q$.

\end{Proof}
\rightline{$\square$}

\subsection{Time Complexity}
Let $t$ be the number of declarations, $n$ the number of statements in an Xd-C program and $m$ the average number of expressions in a statement. Correspondingly, it is not difficult to prove that
the time complexity of the translation algorithm is $O(m\cdot n+t)$. Normally, the number of expressions in a statement is no more than a constant $k_1$ and the number of declarations in an Xd-C program is also no more than a constant $k_2$. As a result, the time complexity is $O(n)$.

\section{Implementation} \label{imp}
\begin{figure}[htb!]
\centering
\includegraphics[width=4.5in]{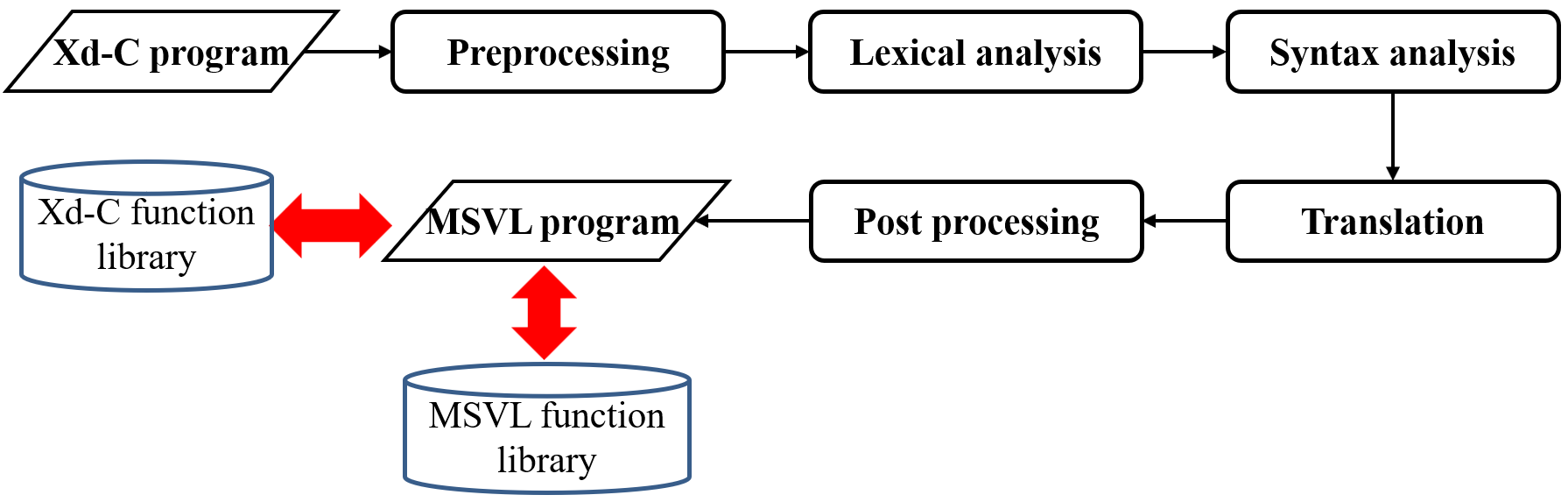}
\caption{Architecture of $C2M$}
\label{Implementation}
\end{figure}

We have implemented the proposed approach in  a tool named $C2M$.
The architecture of the tool is shown in Fig. \ref{Implementation}. An Xd-C program is first preprocessed. In this phase, {\ttfamily \#include} statements are removed by merging all Xd-C files in a project into a file according to their invoking relationships. Macro definitions such as {\ttfamily \#ifdef}, {\ttfamily \#define} and {\ttfamily \#undef} are processed using MinGW \cite{mingw} to generate an Xd-C program without them.
Then, lexer and parser of Xd-C programs based on Parser Generator (PG) are employed to do the lexical analysis and syntax analysis, respectively. Further, a syntax tree of an Xd-C program is generated and translated to an MSVL program using the algorithms presented before. Finally, post processing adjusts the format of the generated MSVL program and outputs it to a file with a suffix of ``.m''. Since a generated MSVL program may invoke MSVL and Xd-C library functions, we have built our libraries of Xd-C and MSVL functions.

In order to show the usability and scalability of our tool in translating
real-world Xd-C programs to MSVL programs, we have applied $C2M$ on 13 programs
from industry whose sizes range from $\approx$ 0.5k  to $\approx$ 17k lines as shown in Table \ref{benchmark}.
In this benchmark, Xd-C programs from RERS P14 to RERS P19 are taken from RERS Grey-Box Challenge 2012 (RERS) \cite{howar2012rers}.
\emph{LTLNFBA} \cite{ltlnfba} is a software for translating an LTL formula to a B\"{u}chi automaton. Program \emph{carc} \cite{Car} is a license plate recognition system. The other 5 programs \emph{bzip2}, \emph{mcf}, \emph{art}, \emph{gzip} and \emph{twolf} %\emph{gap}
are from
SPEC2000 \cite{SPEC}.
The experiments have been carried out on a 64-bit Windows
7 PC with a 4.00GHz Intel(R) Core(TM) i7 processor and
64GB memory.

Table \ref{benchmark} shows the
experimental results on the benchmark. Column ``Program'' represents names of programs. Column ``LOC'' shows sizes of Xd-C programs and column ``LOM'' lists sizes of MSVL programs translated from Xd-C programs.
Column ``Time'' shows the time consumed for accomplishing the translation tasks.
Experimental results in Table \ref{benchmark} show that for all the programs, our tool can effectively output the translation results and the size of the generated MSVL programs is about 2.6 times of Xd-C Programs.

\begin{table}[htb!]
\centering
\caption{ Results of \emph{C2M }on real-world programs}
\label{benchmark}
\begin{tabular}{|c|c||c|c|} \hline
Program &LOC & LOM & Time(s) \\
\hline \hline
RERS P14&514&2261 &0.46 \\ \hline
RERS P15&1353&5016 &2.04\\ \hline
RERS P16&1304&5271 &2.18\\ \hline
RERS P17&2100&7753 &4.38\\ \hline
RERS P18&3306&12677 &11.81\\ \hline
RERS P19&8079&28332 &63.83\\ \hline
LTLNFBA & 3296& 9113&0.76 \\ \hline
carc & 2170 & 4027 & 0.59  \\ \hline
bzip2 & 2320 & 4976 &0.55 \\ \hline
mcf   &  1322& 2124 &0.36 \\ \hline
art   & 886 &1514& 0.28 \\ \hline
gzip  & 3773& 8189& 0.80  \\ \hline
twolf & 17452& 33114 & 7.11 \\ \hline\hline
Total&47875&124376&95.15\\\hline
\end{tabular}
\end{table}

\section{Conclusion}\label{conc}
In order to verify safety, reliability and security properties of C programs by means of a runtime verification tool UMC4M \cite{Wang2017Full,8531789} based on MSVL and its compiler MC \cite{yang2018compiler}, we need to translate C programs to MSVL programs automatically. In this paper, we first present an approach to translating programs written in Xd-C to MSVL programs. Then we prove the equivalence between an original Xd-C program and the translated MSVL program. For doing so, we inductively prove the  equivalences  between expressions and statements, respectively involved in the Xd-C and MSVL programs.
 Further, we have developed a translator $C2M$ based on the proposed algorithms. Moreover, to evaluate usability and scalability of $C2M$, a benchmark of experiments including 13 programs from industry has been conducted. The results show that $C2M$ works effectively. However, since Xd-C is only a fragment of ANSI-C, to translate a general C program to an MSVL program, we have to manually translate the C program to an Xd-C program first.

  In the near future, we plan  to further optimize the translating approach and tool so as to improve efficiency of transformation. Further, we will build a library of functions for MSVL so that a large scale MSVL program can effectively be built.
  In addition, to verify properties of safety, reliability and security of C programs more effectively, we will further investigate techniques of runtime verification at code level by using translator $C2M$,  MSVL compiler MC and verifier UMC4M of the toolkit MSV \cite{DBLP:journals/chinaf/ZhangDT16} such that C or MSVL programs in large scale can effectively be  verified.

\section*{References}
\bibliographystyle{elsarticle-num}
\bibliography{wangmeng}

\section*{Appendix}
\centerline{Appendix A: Operational Semantics of Xd-C}
%
%\[
%\begin{array}{ll}
%\langle e,G,E,M \rangle \xrightarrow{l} \ell& \mbox{(evaluation of expressions in left-value position)}\\
%\langle e,G,E,M \rangle \rightarrow c& \mbox{(evaluation of expressions in right-value poisition)}\\
%\langle s,G,E,M \rangle \xrightarrow{t} out, M' & \mbox{(evaluation of statements, terminating case)}\\
%\langle sw,G,E,M \rangle \xrightarrow{t} out, M'& \mbox{(evaluation of the cases of a switch, terminating case)}\\
%\langle funct,G,M \rangle \xrightarrow{t} v, M'& \mbox{(evaluation of function invocations, terminating case)}\\
%\langle s,G,E,M \rangle \xrightarrow{T} \infty & \mbox{(evaluation of statements, diverging case)}\\
%\langle sw,G,E,M \rangle \xrightarrow{T} \infty& \mbox{(evaluation of the cases of a switch, diverging case)}\\
%\langle funct,G,M \rangle \xrightarrow{T} \infty& \mbox{(evaluation of function invocations, diverging case)}\\
%\langle P \rangle \rightarrow B& \mbox{(execution of whole programs)}\\
%\end{array}
%\]

The following is operational semantics of expressions in Xd-C.

Expressions in left-value position:
\[\begin{array}{l}
\frac{E(id)=b  ~or~  (id \notin Dom(E) ~ and ~ symbol(G,id)=\lfloor b\rfloor)}{
G,E~ \vdash~ id,~ M \stackrel{l}{\Rightarrow} (b,0)}{\scriptstyle(C1)} ~~~~~~~~\frac{G,E~\vdash~ e, ~M\Rightarrow ptr(\ell)}{G,E ~\vdash~ *e,M \stackrel{l}{\Rightarrow} \ell} {\scriptstyle(C2)} \\
  \frac{G,E~\vdash~ le,~ M \stackrel{l}{\Rightarrow} (b,\delta)~~type(le)=struct~ id'\{\varphi\}~~ \mathit{field\_offset}(x,\varphi)=\lfloor \delta' \rfloor}{G,E~\vdash~ le.x,~M  \stackrel{l}{\Rightarrow} (b,\delta+\delta')}{\scriptstyle(C3)}
\end{array}
\]
where $Dom(E)$ is the domain of $E$; $ptr(\ell)$ is a pointer value pointing to $\ell$; $type(e)$ returns the type of $e$;
$struct~ id'\{\varphi\}$ is a struct type where $id'$ is the name and a list $\varphi$ is its fields; $\mathit{field\_offset}(x,\varphi)$ returns the byte offset of the field named $x$ in a struct whose field list is $\varphi$.

In order to access and store values in memory locations, $loadval(\tau, M, (b,\delta))$  and $storeval(\tau,M,(b,\delta),v)$ are respectively defined as follows:
{\small
\[
\begin{array}{l}
\begin{array}{llll}
\mbox{Access modes:}&\mu &::= By\_value(\kappa) & \mbox{access by value}\\
&&~~\mid By\_reference & \mbox{access by reference}\\
%&&~~\mid By\_struct & \mbox{access by struct}\\
&&~~\mid By\_nothing & \mbox{no access}
\end{array}\\
\begin{array}{l}
\mbox{Associating acess modes to Xd-C types:}\\
\begin{array}{l}
A(signed~char)=A(char)=By\_value(int8signed)\\
A(unsigned~ char)=By\_value(int8unsigned)\\
A(signed~short ~int)=A(short~int)=By\_value(int16signed)\\
A(unsigned ~short ~int)=By\_value(int16unsigned)\\
A(signed ~int)=A(unsigned~int)=A(int)=By\_value(int32)\\
A(float)=By\_value(float32)\\
A(double)=By\_value(float64)\\
A(long~ double)=By\_value(float64)\\
A(\tau*)=By\_value(int32)\\
A(array)=By\_reference\\
A(struct)=By\_nothing\\
A(void)=By\_nothing
\end{array}\\
\mbox{Accessing or updating a value of type $\tau$ at location $(b,\delta)$ in memory state $M$:}\\
\begin{array}{rll}
loadval(\tau, M, (b,\delta))&= load(\kappa, M, b,\delta)& \mbox{if } A(\tau)=By\_value(\kappa)\\
loadval(\tau, M, (b,\delta))&=\lfloor ptr(b,\delta) \rfloor & \mbox{if }  A(\tau)=By\_reference\\
loadval(\tau, M, (b,\delta))&=\emptyset&  \mbox{if }  A(\tau)=By\_nothing\\
storeval(\tau,M,(b,\delta),v)&=store(\kappa, M,b,\delta,v) &\mbox{if } A(\tau)=By\_value(\kappa)\\
storeval(\tau,M,(b,\delta),v)&=\emptyset&  \mbox{otherwise}
\end{array}
\end{array}
\end{array}\]
}
%$loadval(type(le),M',\ell)$ reads consecutive bytes at $\ell$ in memory state $M'$ and if successful returns the contents of these bytes as value $v$,
$loadval(\tau, M, (b,\delta))$ reads consecutive bytes at $(b,\delta)$ in memory state $M$ and returns the contents of these bytes as value $v$ if successful
while $storeval(\tau,M,(b,\delta),\linebreak v)$ stores value $v$ at $(b,\delta)$ in memory state $M$ and returns an updated memory state.

Expressions in right-value position:
\[\begin{array}{l}
 {\scriptstyle G,E ~\vdash~ c,~M \Rightarrow c ~~(C4)} ~~~~~~~~~~~\frac{G,E~ \vdash ~le,~M \stackrel{l}{ \Rightarrow} \ell~~loadval(type(le),M,\ell)=\lfloor v \rfloor}{G,E~ \vdash~ le,~ M \Rightarrow  v }~~{\scriptstyle(C5)} \\
\frac{G,E~\vdash~ le,~ M \stackrel{l}{ \Rightarrow} \ell}{G,E~\vdash~ \&le,~M \Rightarrow ptr(\ell)} {\scriptstyle(C6)}~~~~
  \frac{G,E~\vdash~e_1,~M \Rightarrow v_1~~eval\_unop(op_1,v_1, type(e_1))=\lfloor v \rfloor}{G,E~\vdash~op_1 ~e_1,~M \Rightarrow v }{\scriptstyle(C7)} \\[1.5ex]
\frac{G,E~\vdash~e_1,~M \Rightarrow v_1~~G,E~\vdash~e_2,~M \Rightarrow v_2~~eval\_binop(op_2,v_1, type(e_1),v_2,type(e_2))=\lfloor v \rfloor}{G,E~\vdash~e_1~ op_2 ~e_2,~M \Rightarrow  v }{\scriptstyle(C8)}\\[1.5ex]
\frac{G,E~\vdash~e_1,~M \Rightarrow v_1~~is\_true(v_1,type(e_1))~~G,E~\vdash~e_2,~M \Rightarrow v_2}{G,E~\vdash~e_1? e_2:e_3,~M \Rightarrow v_2}{\scriptstyle(C9)}\\[1.5ex]
\frac{G,E~\vdash~e_1,~M \Rightarrow v_1~~is\_false(v_1,type(e_1))~~G,E~\vdash~e_3,~M \Rightarrow v_3}{G,E~\vdash~e_1? e_2:e_3,~M \Rightarrow v_3}{\scriptstyle(C10)}\\[1.5ex]
  \frac{G,E~\vdash~e,~M \Rightarrow v_1~~cast(v_1, type(e), \tau)=\lfloor v \rfloor}{G,E~\vdash~(\tau)e,~M \Rightarrow  v }{\scriptstyle(C11)} \\
  \frac{G,E~\vdash~ e_{fun}(e_{args}),~M \stackrel{t}{\Rightarrow} v,~M}
  {G,E~\vdash~ e_{fun}(e_{args}),~M {\Rightarrow} v} {\scriptstyle(C12)}
\end{array}
\]
 where $eval\_unop(op_1,v_1, type(e_1))$ describes a unary operation and returns the value of $op_1~ v_1$; $eval\_binop(op_2,v_1, type(e_1),v_2,type(e_2))$ returns the value of $v_1~ op_2~ v_2$; $is\_true(v_1,type(e_1))$ and $is\_false(v_1,type(e_1))$ determine the truth value of $v_1$, depending on its type, and the truth value of $v_1$ is false if $v_1$ equals 0 and true otherwise;  $cast(v_1, type(e), \tau)$ converts $v_1$ from its natural type $type(e)$ to the expected type $\tau$.

Taking binary addition as an example, the two argument expressions of types $\tau_1$ and $\tau_2$ are evaluated and
their values $v_1$ and $v_2$ are combined using the the $eval\_binop$ function.
The cases corresponding to binary addition are shown in Table \ref{binadd}.

\begin{table}[t]
\small
\centering
\caption{Binary addition}
\label{binadd}
\begin{tabular}{lllll} \hline
$\tau_1$&$\tau_2$&$v_1$&$v_2$&$eval\_binop(+,v_1,\tau_1,v_2,\tau_2)$\\
\hline\\[-1.5ex]
$int$&$int$&$n_1$&$n_2$&$n_1+n_2$\\
$float$&$float$&$f_1$&$f_2$&$f_1+f_2$ \\
$double$&$double$&$d_1$&$d_2$&$d_1+d_2$\\
$\tau*$&$int$&$ptr(b,\delta)$&$n$&$ptr(b,\delta+n*sizeof(\tau))$\\
$int$&$\tau*$&$n$&$ptr(b,\delta)$&$ptr(b,\delta+n*sizeof(\tau))$\\
\multicolumn{4}{c}{otherwise}&$\emptyset$\\
\hline
\end{tabular}
\end{table}

Some forms of C expressions are omitted but can be
expressed as syntactic sugar:
\[\begin{array}{ll}
\mbox{array access:}& id[e]\equiv *(id+e)~~~C13\\
 & id[e_1][e_2]\equiv *(id+e_1*n+e_2), \mbox{where $n$ is the number }\\
& \mbox{of elements in each row of $id[e_1][e_2]$.}~~~C14\\
\mbox{indirect field access:}& e\rightarrow x\equiv *(e.x)~~~C15\\
\end{array}
\]

Operational semantics for Xd-C statements (other than loops and {\ttfamily switch} statements):
\[\begin{array}{c}
 {\scriptstyle G,E~\vdash~;,~M \stackrel{\epsilon}{\Rightarrow} Normal, ~M ~~(T1)} ~~~~~~~~~~~
 {\scriptstyle G,E~\vdash~\mbox{{\ttfamily\scriptsize break}};,~M \stackrel{\epsilon}{\Rightarrow} Break, ~M ~~(T2)}\\
  {\scriptstyle G,E~\vdash~\mbox{{\ttfamily\scriptsize continue}};,~M \stackrel{\epsilon}{\Rightarrow} Continue, ~M~~(T3) ~~~~~~~~~~~
  G,E~\vdash~\mbox{{\ttfamily\scriptsize return}};,~M \stackrel{\epsilon}{\Rightarrow} Return, ~M
  ~~(T4)}\\[1ex]
 \frac{ G,E~\vdash~e,~M \Rightarrow v,~M'}{G,E~\vdash~\mbox{{\ttfamily\scriptsize return }} e;,~M \stackrel{\epsilon}{\Rightarrow} Return(v), M'} ~~
 ~~{\scriptstyle (T5)}%~~~~~~~~~~~~~\frac{ G,E~\vdash~e,~M \Rightarrow v,~M'}{G,E~\vdash~e;,~M \stackrel{\epsilon}{\Rightarrow} Normal, M'}~~~~{\scriptstyle (C17)}
 \\[1.5ex]
 %\frac{G,E~\vdash~e_1,~M \Rightarrow v_1,~M_1~~G,E~\vdash~e_2,~M_1 \Rightarrow v_2,~M_2~~eval\_binop(op_2,v_1, type(e_1),v_2,type(e_2))=\lfloor v \rfloor}{G,E~\vdash~e_1~ op_2 ~e_2,~M \Rightarrow v,~M_2}{\scriptstyle(8)}\\[1.5ex]
\frac{ G,E~\vdash~le,~M \stackrel{l}{\Rightarrow} \ell~~~~G,E~\vdash~e,~M \Rightarrow v~~~~storeval(type(le),M,\ell,v)=\lfloor M'\rfloor}{G,E~\vdash~le=e,~M \stackrel{\epsilon}{\Rightarrow} Normal, ~M'} {\scriptstyle(T6)}\\[1.5ex]
  \frac{G,E~\vdash~cs_1,~M \stackrel{t_1}{\Rightarrow} Normal, ~M_1~~G,E~\vdash~cs_2,~M_1 \stackrel{t_2}{\Rightarrow} out, ~M_2}{G,E~\vdash~(cs_1;cs_2),~M \stackrel{t_1.t_2}{\Rightarrow} out, M_2}{\scriptstyle(T7)} \\[1.5ex]
  \frac{G,E~\vdash~cs_1,~M \stackrel{t}{\Rightarrow} out, ~M'~~ out\neq Normal}{
  G,E~\vdash~(cs_1;cs_2),~M \stackrel{t}{\Rightarrow} out, M'}{\scriptstyle(T8)} \\[1.5ex]
   \frac{G,E~\vdash~e,~M \Rightarrow v~~~is\_true(v,type(e))~~~
 G,E~\vdash~cs_1,~M \stackrel{t}{\Rightarrow}  out,~M'
 }{ G,E~\vdash~\mbox{{\ttfamily\scriptsize if$(e)\{cs_1\}$else$\{cs_2\}$}},~M \stackrel{t}{\Rightarrow}  out,~M'}~~{\scriptstyle (T9)}\\[1.5ex]
    \frac{G,E~\vdash~e,~M \Rightarrow v~~~is\_false(v,type(e))~~~
 G,E~\vdash~cs_2,~M \stackrel{t}{\Rightarrow}  out,~M'
 }{ G,E~\vdash~\mbox{{\ttfamily\scriptsize if$(e)\{cs_1\}$else$\{cs_2\}$}},~M \stackrel{t}{\Rightarrow}  out,~M'}~~{\scriptstyle (T10)}
\end{array}
\]
$le\mathit{++}$ and $le\mathit{--}$ are omitted but can be expressed as follows:
\[\begin{array}{ll}
le\mathit{++}\equiv (le=le+1)\\
le\mathit{--}\equiv (le=le-1)
\end{array}
\]

The following rules define the execution of {\ttfamily while} and  {\ttfamily for} loops. The rules describing the execution of {\ttfamily do} loops resemble the rules for {\ttfamily while} loops and
are omitted in this paper.\\
Outcome updates (at the end of a loop execution):
$${\scriptstyle Break~\stackrel{loop}{\rightsquigarrow} ~Normal~~~Return\stackrel{loop}{\rightsquigarrow} Return ~~~Return(v)\stackrel{loop}{\rightsquigarrow} Return(v)}$$
{\ttfamily while} loops:
\[\begin{array}{c}
 \frac{G,E~\vdash~e,~M \Rightarrow v~~~is\_false(v,type(e))}{
 G,E~\vdash~ \mbox{{\ttfamily\scriptsize while}}(e)\{cs\},~M \stackrel{\epsilon}{\Rightarrow} Normal,~M}
 ~~{\scriptstyle (T11)} \\ [1ex]
 \frac{G,E~\vdash~e,~M \Rightarrow v~~~is\_true(v,type(e))~~~
 G,E~\vdash~cs,~M \stackrel{t}{\Rightarrow}  out,~M'~~~out\stackrel{loop}{\rightsquigarrow}out'
 }{ G,E~\vdash~\mbox{{\ttfamily\scriptsize while}}(e)\{cs\},~M \stackrel{t}{\Rightarrow}  out',~M'}~~{\scriptstyle (T12)} \\ [2ex]
  \frac{\begin{array}{c}\scriptstyle
  G,E~\vdash~e,~M \Rightarrow v~~~is\_true(v,type(e))~~~\\
  \scriptstyle G,E~\vdash~cs,~M \stackrel{t_1}{\Rightarrow}  (Normal|Continue),~M_1~~~
   G,E~\vdash~\mbox{{\ttfamily\scriptsize while}}(e)\{cs\},~M_1 \stackrel{t_2}{\Rightarrow}  out', M_2\end{array}}{
    G,E~\vdash~\mbox{{\ttfamily\scriptsize while}}(e)\{cs\},~M \stackrel{t_1. t_2}{\Rightarrow}  out', M_2}~~{\scriptstyle (T13)}
\end{array}
\]
{\ttfamily for} loops:
\[\begin{array}{c}
\frac{cs_1\neq ;~~~\scriptstyle G,E~\vdash~cs_1,~M \stackrel{t_1}{\Rightarrow}  Normal,~M_1~~~ G,E~\vdash~\mbox{\ttfamily\scriptsize for}(;e;cs_2)\{cs\},~M_1 \stackrel{t_2}{\Rightarrow}  out,~M_2}{
G,E~\vdash~\mbox{\ttfamily\scriptsize for}(cs_1;e;cs_2)\{cs\},~M \stackrel{t_1. t_2}{\Rightarrow}  out,~M_2}~~{\scriptstyle (T14)} \\ [2ex]
 \frac{ G,E~\vdash~e,~M \Rightarrow v~~~is\_false(v,type(e))}{
 G,E~\vdash~\mbox{\ttfamily\scriptsize for}(;e;cs_2)\{cs\},~M \stackrel{\epsilon}{\Rightarrow}  Normal, ~M}~~{\scriptstyle (T15)}\\[2ex]
  \frac{\begin{array}{c}\scriptstyle G,E~\vdash~e,~M \Rightarrow v~~~is\_true(v,type(e))~~~
  \scriptstyle G,E~\vdash~cs,~M \stackrel{t}{\Rightarrow}  out_1,~M'~~~out_1 \stackrel{loop}{\rightsquigarrow} out\end{array}}{
   G,E~\vdash~\mbox{\ttfamily\scriptsize for}(;e;cs_2)\{cs\},~M \stackrel{t}{\Rightarrow}  out,~ M'}~~{\scriptstyle (T16)}\\[2ex]
  \frac{\begin{array}{c}\scriptstyle G,E~\vdash~e,~M \Rightarrow v~~~is\_true(v,type(e))~~~
  G,E~\vdash~cs,~M \stackrel{t_1}{\Rightarrow} (Normal|Continue), ~M_1\\
 \scriptstyle G,E~\vdash~cs_2,~M_1 \stackrel{t_2}{\Rightarrow} Normal,~ M_2 ~~~ G,E~\vdash~\mbox{\ttfamily\scriptsize for}(;e;cs_2)\{cs\},~M_2 \stackrel{t_3}{\Rightarrow}  out,~ M_3
  \end{array}}{ G,E~\vdash~\mbox{\ttfamily\scriptsize for}(;e;cs_2)\{cs\},~M \stackrel{t_1. t_2 . t_3}{\Longrightarrow}  out,~ M_3}~~{\scriptstyle (T17)}
\end{array}
\]
%We omit the rules for {\ttfamily switch$(e)$} $sw$ statements, which are standard. Based on
%the integer value of $e$, the appropriate case of $sw$ is selected, and the corresponding
%suffix of $sw$ is executed like a sequence, therefore implementing the ``fall-through''
%behavior of switch cases. A $Break$ outcome for one of the cases terminates the
%switch normally.\\
Outcome updates (at the end of a switch execution):
$${
 \begin{array}{l}\scriptstyle Normal~\stackrel{switch}{\rightsquigarrow} ~Normal~~~Continue~\stackrel{switch}{\rightsquigarrow} ~Continue~~~ Break~\stackrel{switch}{\rightsquigarrow} ~Normal\\
 \scriptstyle Return\stackrel{switch}{\rightsquigarrow} Return ~~~~~~Return(v)\stackrel{switch}{\rightsquigarrow} Return(v)\end{array}}$$
{\ttfamily switch$(e)$\{case $n_0:cs_0;...;$case $n_m:cs_m;$default$:cs;$\}}:
\[\begin{array}{c}
 \frac{ G,E~\vdash~cs,~M \stackrel{t}{\Rightarrow} out,~M'}{G,E~\vdash~\mbox{{\ttfamily\scriptsize default:}}cs;,~M \stackrel{t}{\Rightarrow} out, M'} ~~
 ~~{\scriptstyle (T18)}~~~~~~~~~~~~~ \frac{ G,E~\vdash~cs,~M \stackrel{t}{\Rightarrow} out,~M'}{G,E~\vdash~\mbox{{\ttfamily\scriptsize case:}}cs;,~M \stackrel{t}{\Rightarrow} out, M'}~~~~{\scriptstyle (T19)}\\[1.5ex]
  \frac{\begin{array}{c}\scriptstyle G,E~\vdash~\mbox{\ttfamily\scriptsize case }n_i:cs_i;,~M \stackrel{t_1}{\Rightarrow} Normal, ~M_1\\
 \scriptstyle G,E~\vdash~\mbox{\ttfamily\scriptsize case $n_{i+1}:cs_{i+1};...;$case $n_m:cs_m;$defalut$:cs;$},~M \stackrel{t_2}{\Rightarrow} out, ~M_2\end{array}}{G,E~\vdash~\mbox{\ttfamily\scriptsize case $n_{i}:cs_{i};...;$case $n_m:cs_m;$defalut$:cs;$},~M \stackrel{t_1. t_2}{\Rightarrow} out, M_2}{\scriptstyle(T20)} \\[1.5ex]
  \frac{G,E~\vdash~\mbox{\ttfamily\scriptsize case }n_i:cs_i;,~M \stackrel{t}{\Rightarrow} out, ~M'~~ out\neq Normal}{
  G,E~\vdash~\mbox{\ttfamily\scriptsize case $n_{i}:cs_{i};...;$case $n_m:cs_m;$defalut$:cs;$},~M \stackrel{t}{\Rightarrow} out, M'}{\scriptstyle(T21)} \\[1.5ex]
    \frac{\begin{array}{c}\scriptstyle G,E~\vdash~e,~M \Rightarrow v~~~v==n_i\\
 \scriptstyle G,E~\vdash~\mbox{\ttfamily\scriptsize case $n_{i}:cs_{i};...;$case $n_m:cs_m;$defalut$:cs;$},~M \stackrel{t}{\Rightarrow} out_1, M'~~~out_1 \stackrel{switch}{\rightsquigarrow} out\end{array}}{G,E~\vdash~\mbox{\ttfamily\scriptsize switch($e$)\{case $n_{0}:cs_{0};...;$case $n_m:cs_m;$defalut$:cs;$\}},~M \stackrel{t}{\Rightarrow} out, M'}{\scriptstyle(T22)} \\[1.5ex]
     \frac{\begin{array}{c}\scriptstyle G,E~\vdash~e,~M \Rightarrow v~~~v!=n_0~~~v!=n_1~~~...~~~v!=n_m\\
 \scriptstyle G,E~\vdash~\mbox{\ttfamily\scriptsize defalut$:cs;$},~M \stackrel{t}{\Rightarrow} out_1, M'~~~out_1 \stackrel{switch}{\rightsquigarrow} out\end{array}}{G,E~\vdash~\mbox{\ttfamily\scriptsize switch($e$)\{case $n_{0}:cs_{0};...;$case $n_m:cs_m;$defalut$:cs;$\}},~M \stackrel{t}{\Rightarrow} out, M'}{\scriptstyle(T23)} \\[1.5ex]
\end{array}
\]
Function calls:
\[\begin{array}{c}
  \frac{\begin{array}{c}\scriptstyle
  G,E~\vdash~ e_{fun},~M \Rightarrow ptr(b,0)~~~G,E~\vdash~ e_{args},~M \Rightarrow v_{args}\\
  \scriptstyle \mathit{functdef}(G,b)=\lfloor funct \rfloor~~~ type\_of\_fundef(funct)=type(e_{fun}) ~~~\\
  \scriptstyle G~\vdash~ funct(v_{args}),~M \stackrel{t}{\Rightarrow} v_{res},~M'\end{array}}
  {G,E~\vdash~ e_{fun}(e_{args}),~M \stackrel{t}{\Rightarrow} v_{res},~M'}~~{\scriptstyle (T24)}%\\[2ex]
%   \frac{\begin{array}{c}\scriptstyle G,E~\vdash~ le,~M \stackrel{l}{\Rightarrow} \ell,~M_1
%   ~~~G,E~\vdash~ e_{fun},~M_1 \Rightarrow ptr(b,0),~M_2~~~G,E~\vdash~ e_{args},~M_2 \Rightarrow v_{args},~M_3\\
%  \scriptstyle \mathit{functdef}(G,b)=\lfloor funct \rfloor~~~ type\_of\_fundef(funct)=type(e_{fun}) ~~~G~\vdash~ funct(v_{args}),~M_3 \stackrel{t}{\Rightarrow} v_{res},~M_4 \\
%  \scriptstyle storeval(type(e),M_4, ptr(\ell),v_{res}=\lfloor M_5 \rfloor) \end{array}}
%  {G,E~\vdash~ le=e_{fun}(e_{args}),~M \stackrel{t}{\Rightarrow} v_{res},~M_5}~~{\scriptstyle (28)}\\[2ex]
\end{array}
\]
where $e_{fun}$ is a function pointer or a function name,  $e_{args}$ a list of arguments of the function and $v_{args}$ a list of values of arguments; $type\_of\_fundef(funct)$ returns the type of function $funct$ including the return type and types of parameters;
$v_{res}$ is the return value of  $funct(v_{args})$.\\
Compatibility between values, outcomes and return types:
$${\scriptstyle Normal, void \#undef~~~~Return, void\#undef~~~~Return(v),\tau\#v~\mbox{\scriptsize when }\tau\neq void}$$
Function invocations:
\[\begin{array}{c}
  \frac{\begin{array}{c}\scriptstyle funct=[\tau\mid void]~id(par)\{dcl;cs\}~~~alloc\_vars(M,par\textbf{+}dcl,E)=(M_1,b^*)\\
  \scriptstyle bind\_params(E,M_1,par,v_{args})=M_2~~~
  G,E~\vdash~cs,~M_2 \stackrel{t}{\Rightarrow} out, M_3~~~out,\tau\#v_{res} \end{array}}
  {G~\vdash funct(v_{args}),~M \stackrel{t}{\Rightarrow}v_{res},~ free(M_3,b^*)}~~{\scriptstyle (T25)}\\[2ex]
   \frac{funct=extern~[\tau\mid void] ~id(par)~~~v_{res}=id(v_{args})~~~v=``id(v_{args},v_{res})"}
  {G\vdash funct(v_{args}),~M \stackrel{v}{\Rightarrow} v_{res}, M}~~{\scriptstyle (T26)}
\end{array}
\]
where $dcl$ is a list of declarations ($dcl=(Pd;)^*$);
$alloc\_vars(M,par\textbf{+}dcl,E)$ allocates the memory
required for storing the formal parameters $par$ and the local variables $dcl$; %where each variable $\tau~ x$ is allocated one block with lower bound 0 and upper bound $sizeof(\tau)$, using the $alloc$ primitive of the memory model;
%These blocks initially contain undef values.
$bind\_params(E,M_1,par,v_{args})$ iterates the $storeval$ function in order to initialize formal parameters $par$ to
the values of the corresponding arguments $v_{args}$; $v_{res}=id(v_{args})$ obtains the return value of the function call $id(v_{args})$ and if the return type is $void$, the value of $v$ is $\emptyset$. An input/output event $v$ recorded in the trace is generated by a call to an external function. \\
Operational semantics for divergence:
\[\begin{array}{c}
\frac{\begin{array}{c}
\scriptstyle \forall i\in N_0(G,E~\vdash~e,~M_i  {\Rightarrow} v \wedge is\_true(v,type(e)) \wedge G,E~\vdash~cs,~M_i \stackrel{t}{\Rightarrow} (Normal\mid Continue),~M_{i+1}\\
\scriptstyle\rightarrow G,E~\vdash~e,~M_{i+1}  {\Rightarrow} v' \wedge is\_true(v',type(e)))\wedge M_0=M
\end{array} }{G,E~\vdash~\mbox{\ttfamily\scriptsize while}(e)\{cs\},~M \stackrel{T}{\Rightarrow} \infty}~{\scriptstyle (D1)}\\
  \frac{G,E~\vdash~cs_1,~M \stackrel{T}{\Rightarrow} \infty }{
  G,E~\vdash~cs_1;cs_2,~M \stackrel{T}{\Rightarrow} \infty}~{\scriptstyle (D2)}~~~~~
   \frac{G,E~\vdash~cs_1,~M \stackrel{t}{\Rightarrow} Normal,~ M_1~~~G,E~\vdash~cs_2,~M_1 \stackrel{T}{\Rightarrow} \infty }{G,E~\vdash~cs_1;cs_2,~M \stackrel{t.T}{\Rightarrow} \infty}~{\scriptstyle (D3)}\\
     \frac{G,E~\vdash~e,~M \Rightarrow v~~~is\_true(v,type(e))~~~
 G,E~\vdash~cs_1,~M \stackrel{T}{\Rightarrow}  \infty
 }{ G,E~\vdash~\mbox{{\ttfamily\scriptsize if$(e)\{cs_1\}$else$\{cs_2\}$}},~M \stackrel{T}{\Rightarrow}  \infty}~~{\scriptstyle (D4)}\\[1.5ex]
    \frac{G,E~\vdash~e,~M \Rightarrow v~~~is\_false(v,type(e))~~~
 G,E~\vdash~cs_2,~M \stackrel{T}{\Rightarrow}  \infty
 }{ G,E~\vdash~\mbox{{\ttfamily\scriptsize if$(e)\{cs_1\}$else$\{cs_2\}$}},~M \stackrel{T}{\Rightarrow} \infty}~~{\scriptstyle (D5)}\\
    \frac{G,E~\vdash~e,~M \Rightarrow v~~~is\_true(v,type(e))~~~G,E~\vdash~cs,~M \stackrel{T}{\Rightarrow} \infty}
    {G,E~\vdash~\mbox{\ttfamily\scriptsize while}(e)\{cs\},~M \stackrel{T}{\Rightarrow} \infty}~{\scriptstyle (D6)}\\
    \frac{\begin{array}{c}\scriptstyle G,E~\vdash~e,~M \Rightarrow v~~~is\_true(v,type(e))\\
  \scriptstyle G,E~\vdash~cs,~M \stackrel{t}{\Rightarrow} (Normal\mid Continue),~M_1~~~G,E~\vdash~\mbox{\ttfamily\scriptsize while}(e)\{cs\},~M_1 \stackrel{T}{\Rightarrow} \infty \end{array}}
  {G,E~\vdash~\mbox{\ttfamily\scriptsize while}(e)\{cs\},~M \stackrel{t.T}{\Rightarrow} \infty}~~{\scriptstyle (D7)}\\[2ex]
  \frac{\begin{array}{c}
  \scriptstyle G,E~\vdash~ e_{fun},~M \Rightarrow ptr(b,0)~~~G,E~\vdash~ e_{args},~M \Rightarrow v_{args}\\
  \scriptstyle \mathit{functdef}(G,b)=\lfloor funct \rfloor~~~ type\_of\_fundef(funct)=type(e_{fun})\\
   \scriptstyle G~\vdash~ funct(v_{args}),~M \stackrel{T}{\Rightarrow} \infty\end{array}}
  {G,E~\vdash~ e_{fun}(e_{args}),~M \stackrel{T}{\Rightarrow} \infty}{\scriptstyle (D8)}\\[2ex]
  \frac{\begin{array}{c}\scriptstyle F=\tau~id(par)\{dcl;cs\}~~~alloc\_vars(M,par\textbf{+}dcl,E)=(M_1,b^*)\\
  \scriptstyle bind\_params(E,M_1,par,v_{args})=M_2~~~
  G,E~\vdash~cs,~M_2 \stackrel{T}{\Rightarrow} \infty \end{array}}
  {G\vdash F(v_{args}),~M \stackrel{T}{\Rightarrow}\infty}~~{\scriptstyle (D9)}
\end{array}
\]
Observable behaviors of programs:
\[\begin{array}{c}
  \frac{\begin{array}{c}\scriptstyle G=globalenv(P)~~~M=initmem(P)\\
  \scriptstyle symbol(G,main(P))=\lfloor b \rfloor~~~\mathit{functdef}(G,b)=\lfloor f\rfloor~~~ (f(nil),G,E,M)\stackrel{t}{\Rightarrow} n,~M' \end{array}}
  {\vdash~ P \Rightarrow terminates(t,n)}~~{\scriptstyle (P1)}\\[2ex]
    \frac{\begin{array}{c}\scriptstyle G=globalenv(P)~~~M=initmem(P)\\
  \scriptstyle symbol(G,main(P))=\lfloor b \rfloor~~~\mathit{functdef}(G,b)=\lfloor f\rfloor~~~ G,E~\vdash~f(nil),~M \stackrel{T}{\Rightarrow} \infty \end{array}}
  {\vdash~ P \Rightarrow diverges(T)}~~{\scriptstyle (P2)}
\end{array}
\]
A global environment $G$ and a memory state $M$ are computed for $P$. If the main function invocation terminates with trace $t$ and result value $n$, the observed behavior of P is $terminates(t, n)$. If the function invocation diverges with a possibly
infinite trace $T$, the observed behavior is $diverges(T)$.

Based on the operational semantics, some semantic equivalence rules can be proved similarly to Lemma \ref{Cloop} and are given as follows:
{\small$$
\begin{array}{ll}
E1& (cs,M)\cong(cs_1;cs_2,M)\\
&\Longrightarrow (cs;cs',M)\cong(cs_1;cs_2;cs',M)\\
E2& (cs',M')\cong(cs_1;cs_2,M')\wedge (G,E\vdash cs,M\stackrel{t}{\Rightarrow}out,M')\\
&\Longrightarrow(cs;cs',M)\cong(cs;cs_1;cs_2,M)\\
E3& G,E\vdash e,M\Rightarrow true\\
& \Longrightarrow(\mbox{\ttfamily if}(e)\{cs\}\mbox{\ttfamily else}\{cs'\},M)\cong (cs,M)\\
E4& G,E\vdash e,M\Rightarrow false\\
& \Longrightarrow(\mbox{\ttfamily if}(e)\{cs\}\mbox{\ttfamily else}\{cs'\},M)\cong (cs',M)\\
E5& (G,E\vdash e,M\Rightarrow true)\wedge (G,E\vdash cs, M\stackrel{t}{\Rightarrow}(Normal|Continue),M_1)\\
& \Longrightarrow(\mbox{\ttfamily while}(e)\{cs\},M)\cong (cs;\mbox{\ttfamily while}(e)\{cs\},M)\\
E6& (G,E\vdash e,M\Rightarrow true)\wedge (G,E\vdash cs, M\stackrel{T}{\Rightarrow}\infty)\\
& \Longrightarrow(\mbox{\ttfamily while}(e)\{cs\},M)\cong (cs,M)\\
E7&  (G,E~\vdash~ e_{fun},~M \Rightarrow ptr(b,0))\wedge \mathit{functdef}(G,b)=\lfloor funct \rfloor \wedge \\
 & type\_of\_fundef(funct)=type(e_{fun})\wedge\\
 &funct=[\tau\mid void]~id(\tau_1~y_1,...,\tau_m~y_m)\{dcl;cs\}\wedge  (G,E\vdash cs,M\stackrel{T}{\Rightarrow}\infty)\\
 &\Longrightarrow (e_{fun}(e_1,...,e_m),M)\cong (\tau_1~ y_1=e_1;...;\tau_m~y_m=e_m;dcl;cs,M)
\end{array}
$$}

\centerline{Appendix B: Semantics of MSVL statements}
Table \ref{MSVLsynSem} shows that all MSVL statements are defined by PTL formulas.

\begin{table}[htb!]
\small
\centering
\caption{Syntax and Semantics of MSVL statements}
\label{MSVLsynSem}
\begin{tabular}{ll} \hline
{\sc Syntax}&{\sc Semantics}\\
\hline\\[-1.5ex]
{\ttfamily empty}&$\DEF\varepsilon$\\
{\ttfamily skip}& $\DEF \bigcirc \varepsilon$ \\
$la\Leftarrow ra$&$\DEF la=ra\wedge p_{la}$\\
$la:=ra$&$\DEF \bigcirc (la=ra\wedge p_{la})\wedge \bigcirc \varepsilon$\\
$ms_1$ {\ttfamily and} $ms_2$&$\DEF ms_1\wedge ms_1$\\
{\ttfamily next} $ms$&$\DEF\bigcirc ms$\\
$ms_1;ms_2$&$\DEF ms_1;ms_2$\\
{\ttfamily if$(b)$then$\{ms_1\}$else$\{ms_2\}$}& $\DEF(b\rightarrow ms_1)\wedge(\neg b\rightarrow ms_2)$\\
{\ttfamily while$(b)\{ms\}$}&$\DEF(b\wedge ms)^{\ast}\wedge\Box(\varepsilon\rightarrow\neg b)\vee (b\wedge ms)^\omega$\\
\hline
\end{tabular}
\end{table}

\end{document}